\documentclass[12pt,a4paper]{article}
\usepackage{amsmath}
\usepackage{amsfonts}
\usepackage{graphicx}
\usepackage{subcaption}

\usepackage[english]{babel}
\usepackage{hyperref}

\allowdisplaybreaks 

\newcommand{\md}{\mathrm{d}}

\newcommand{\av}[1]{\langle #1 \rangle}
\newcommand{\acom}[2]{\left\{ #1, #2 \right\} }
\newcommand{\com}[2]{\left[ #1, #2 \right]}
\newcommand{\ident}[1]{\mathrm{I}_{#1}}
\newcommand{\Tr}[1]{\text{Tr}\left(#1\right)} 
\DeclareMathOperator\tr{Tr} 
\DeclareMathOperator\Err{Err} 
\newcommand{\g}[1]{\gamma^{#1}}
\newcommand{\te}{\otimes}

\newcommand{\pd}[2]{\frac{\partial #1}{\partial #2}}
\newcommand{\R}{{\mathbb R}}   
\newcommand{\C}{{\mathbb C}}   

\title{Monte Carlo simulations of random non-commutative geometries}
\date{}

\author{John W. Barrett, Lisa Glaser
\\ \\
School of Mathematical Sciences\\
University of Nottingham\\
University Park\\
Nottingham NG7 2RD, UK\\
\\
\tt\small john.barrett@nottingham.ac.uk, \tt\small lisa.glaser@nottingham.ac.uk
\\ \\}
\date{May 11th, 2016}

\begin{document}
\maketitle

\begin{abstract}Random non-commutative geometries are introduced by integrating over the space of Dirac operators that form a spectral triple with a fixed algebra and Hilbert space. The cases with the simplest types of Clifford algebra are investigated using Monte Carlo simulations to compute the integrals. Various qualitatively different types of behaviour of these random Dirac operators are exhibited. Some features are explained in terms of the theory of random matrices but other phenomena remain mysterious. Some of the models with a quartic action of symmetry-breaking type display a phase transition. Close to the phase transition the spectrum of a typical Dirac operator shows manifold-like behaviour for the eigenvalues below a cut-off scale.
\end{abstract}

\section{Introduction}
A spectral triple is a way of encoding a geometry using a Dirac operator \cite{ConnesBook}. There is a Dirac operator $D$ acting on a Hilbert space $\mathcal H$  and an algebra $\mathcal A$ that acts on the same space. Examples with a commutative algebra are given by Riemannian manifolds, where the algebra is the algebra of functions on the manifold and the Dirac operator is the usual one acting on spinor fields. However, the point of spectral triples is that the algebra is allowed to be non-commutative, leading to a generalisation of the notion of geometry. 

A random geometry is a class of geometries $\mathcal G$ that fluctuates according to a probability measure. In this article, the probability measure is taken to be a constant times
\begin{equation}e^{-S(D)}\md D
\end{equation}
using a real-valued `action' $S(D)$ and a standard measure $\md D$ on the space $\mathcal G$ of Dirac operators. 

To make this computable, the class of geometries is taken to be the Dirac operators on a fixed finite-dimensional Hilbert space $\mathcal H$; thus $\mathcal G$ is a space of matrices. It turns out that the axioms for $D$ for these finite spectral triples are all linear and so $\mathcal G$ is a vector space  \cite{Barrett:2015naa}. Therefore one can take $\md D$ to be its Lebesgue measure, which is unique up to an overall constant. Thus the object of study is a random matrix model where the matrices are constrained to be Dirac operators.

The algebra $\mathcal A$ in this construction is also fixed, and is taken to be the algebra of $n\times n$ matrices, $M(n,\C)$. Spectral triples with this algebra are known as fuzzy spaces  \cite{Madore:1991bw} and are the simplest type of non-commutative spectral triple. Allowing the algebra to be non-commutative is important because it allows a new type of finite-dimensional approximation to a manifold. Staying within the realm of commutative algebras would lead to the algebra of functions on a finite set of points, which is a lattice approximation to a manifold; simple examples of such random commutative spectral triples are studied in \cite{Holfter:2002ka,deAlbuquerque:2003kq}. The fuzzy spaces are not lattice approximations and so the study of these is complementary to the study of random lattices. Intuitively one can think of the algebra $\mathcal A$ as consisting of the functions on a space with a certain minimum wavelength that is determined by $n$; this picture is known to be accurate for the most-studied example of the fuzzy two-sphere \cite{Grosse:1994ed}.

The purpose of this paper is to study the simplest examples of random fuzzy spaces by computing the statistics of the eigenvalues of $D$ using a Markov Chain Monte Carlo algorithm. The examples are determined by the value of $n$ and the type of gamma matrices used in the Dirac operator (explicit formulas are given in section \ref{sec:dirac}). A type $(p,q)$ geometry is one in which there are $p$ gamma matrices that square to $1$ and $q$ gamma matrices that square to $-1$. The examples studied here are types $(1,0)$, $(0,1)$, $(2,0)$, $(1,1)$, $(0,2)$ and $(0,3)$. 
It would be interesting to go to higher types but these types already exhibit an interesting set of different behaviours, some aspects of which are not yet understood from a theoretical point of view. 

The form of the action $S(D)$ has not yet been specified. In this paper it is assumed to be spectral, which means it is of the form
\begin{equation} S(D)=\tr V(D)=\sum_i V(\lambda_i)
\end{equation}
for some potential function $V$, with $V\ge b$ for some $b \in \mathbb{R}$ and $\lambda_i$ denoting the eigenvalues of the self-adjoint operator $D$. The Connes-Chamseddine spectral action \cite{chamseddine_spectral_1997,chamseddine_uncanny_2010}, in which $V(x)\to0$ as $x\to\infty$, is not suitable because the integral for the partition function
\begin{equation}Z=\int_{\mathcal G}e^{-S(D)}\md D
\end{equation}
does not converge. This is because $S(\mu D)$, for $\mu\in\R$,  converges to a finite constant as $\mu\to\infty$. 

In fact, it is necessary that $V(x)\to\infty$ instead. The simplest cases are investigated here, namely 
\begin{equation}S(D)=\tr \left(g_2 D^2+g_4 D^4\right)
\end{equation}
with $g_4>0$, or $g_4=0$ and $g_2>0$. By a simple change of variables in the integral, $D\to\mu D$, one can assume that either $g_4=1$, or $g_4=0$ and $g_2=1$, so one need only study these cases. Note that one could choose other spectral actions and it is possible that the results obtained here might motivate the study of other choices.

When $g_4>0$ and $g_2<0$ the potential has a symmetry-breaking double well form. In random matrix models it is known that this potential leads to a phase transition \cite{Cicuta:2000xz} and so this possibility is investigated here. It is shown here that, at least in some of the random Dirac operator models, there is good numerical evidence for the existence of a phase transition.  

The eigenvalue distribution is plotted for several values of the coupling constants and matrix sizes to exhibit the typical behaviours. On a (commutative) Riemannian manifold of dimension $m$ the eigenvalue distribution, or density of states, for the Dirac operator is approximately 
the density of flat space
\begin{equation}\label{eq:dos}\rho_0(\lambda)=|\lambda|^{m-1}
\end{equation}
when the eigenvalues are large enough. Most of the plots for random non-commutative geometries presented here look nothing like this, except that some of them are approximately constant ($m=1$) for some range of eigenvalues. 

The exception is close to the phase transition, where the distribution does indeed look very much like \eqref{eq:dos} for a range of eigenvalues below a high-energy cutoff. Thus as far as the eigenvalues are concerned, the random non-commutative geometries are behaving something like random Riemannian geometries in this regime.  The results presented here show that this is a promising area for future investigation. A phase transition to a geometric phase in a multi-matrix model with a rather different Yang-Mills-type action has also been observed in \cite{Azuma:2004zq,O'Connor:2013rla}.

The motivation for the present work comes from the close relation between random geometries and models of quantum gravity, though one does not need to know anything about quantum gravity to understand the results. Most approaches to random geometry have been stimulated by work on quantum gravity but some of them (e.g. dynamical triangulations or Liouville gravity) have found wider application. It is possible that random Dirac operators will also find other applications beside quantum gravity. In quantum gravity the maximum eigenvalue has a ready interpretation as a natural cutoff to gravitational phenomena at the Planck scale. However in a wider context it can be interpreted simply as a finite limit to the resolution to which a geometry is defined.

There are features in common with other models of quantum or random geometry, most notably the existence of a phase transition, which is also evident in dynamical triangulations \cite{Ambjorn:1997di} and lattice simulations \cite{Hamber:2009mt}. There are however some features of our system that are quite different from those of other models.
One point is that the requirement that the action has to have a compact global minimum in the non-compact $\mathcal G$ is a non-trivial constraint on the model. In other theories of discrete geometry, like causal dynamical triangulations \cite{ambjorn_nonperturbative_2012} or causal set theory \cite{henson_onset_2015}, the space of geometries explored in Monte Carlo is a combinatorial space of a finite number of elements and the code is guaranteed to reach equilibrium after a finite, although possibly long, time. 

Another interesting feature is the freedom to rescale $g_2$ and $g_4$ using the change of variables mentioned above. Monte Carlo simulations show that the rescaling does not change the qualitative features or our system, e.g. relative differences between eigenvalues will remain unchanged. This can be explained by the use of the Lebesgue measure, which does not distinguish any particular scale of energy. This is in marked contrast to systems such as the Ising model, or the causal set model of 2d orders \cite{surya_evidence_2012}.

The non-commutativity also distinguishes this approach from most others. Using a finite-dimensional commutative algebra necessarily leads to a lattice model of quantum geometry defined on a finite set of points. The use of non-commutative geometry allows a more general set of finite-dimensional models where the algebra is an algebra of matrices. Thus one can construct perfectly computable models of random geometry that are not lattice models. Moreover, the standard model of particle physics has a non-commutative geometry using exactly the same framework \cite{Barrett:2006qq, Connes:2006qv}, so the hope is it will be easy to combine the two into a unified model of gravity and particle physics.

The technical details of the Dirac operators, observable functions and Monte Carlo method are given in section \ref{sec:details}. The results for the action $\tr D^2$ are given in section \ref{sec:D2}, where it is explained how the results relate to the standard theory of Gaussian random matrices. Actions including a $\tr D^4$ term are studied in section \ref{sec:D4}, with particular attention paid to the symmetry-breaking case which exhibits a phase transition. Section \ref{sec:conc} discusses the interpretation of the results. The expansions of the action in terms of the constituent matrices of the Dirac operators are given in detail in appendix \ref{sec:GeoAndAc}.



\section{Technical details}\label{sec:details}

\subsection{The Dirac operators}\label{sec:dirac}
The spectral triples considered here are `real spectral triples', which consist of a finite-dimensional Hilbert space $\mathcal H$ together with some operators acting in $\mathcal H$. These are an algebra $\mathcal A$, a chirality operator $\Gamma$, an antilinear `real structure' $J$ and a self-adjoint Dirac operator $D$. For a given random geometry model $\mathcal H,\mathcal A,\Gamma$ and $J$ are fixed but $D$ is allowed to vary, subject to the axioms of non-commutative geometry. 

The axioms are solved in  \cite{Barrett:2015naa} to give explicit forms of the Dirac operator in terms of  $n\times n$  Hermitian matrices $H$ and $n\times n$ anti-Hermitian traceless matrices $L$ according to the formulas below. There are no other constraints on these $n\times n$ matrices, so these are the freely-specifiable data for the Dirac operator.

The Dirac operator acts on  $\mathcal H=V\otimes M(n,\C)$, with $V=\C^k$ the space on which the gamma matrices act. The gamma matrices are assumed to form an irreducible representation of the Clifford algebra, which implies that the chirality operator is trivial for $d=p+q$ odd. The dimension of $V$ is $k=2^{d/2}$ for $d$ even and  $k=2^{(d-1)/2}$ for $d$ odd. In the first two cases the sole gamma matrix is just $1$ or $-i$ respectively. In the remaining cases the gamma matrices are $2\times 2$ matrices, distinct gamma matrices anti-commuting.  As usual, $[\,\cdot\,,\,\cdot\,]$ denotes the commutator and $\{\,\cdot\,,\,\cdot\,\}$ the anti-commutator of matrices.

\begin{description}
\item[Type (1,0)]
\begin{equation}\label{dirac10} D=\{H,\cdot\,\} \end{equation}
\item [Type (0,1)]
\begin{equation}\label{dirac01} D=-i\;[L,\cdot\,]\end{equation}
\item [Type (2,0)] $(\gamma^1)^2=(\gamma^2)^2=1$.
\begin{equation}\label{dirac20} D=\gamma^1\otimes\{H_1,\cdot\,\}+\gamma^2\otimes\{H_2,\cdot\,\}\end{equation}
\item [Type (1,1)]  $(\gamma^1)^2=1, (\gamma^2)^2=-1$.
\begin{equation}\label{dirac11} D=\gamma^1\otimes\{H,\cdot\,\}+\gamma^2\otimes[L,\cdot\,]\end{equation}
\item [Type (0,2)] $(\gamma^1)^2=(\gamma^2)^2=-1$.
\begin{equation} \label{dirac02} D=\gamma^1\otimes[L_1,\cdot\,]+\gamma^2\otimes[L_2,\cdot\,]\end{equation}
\item [Type (0,3)] $(\gamma^1)^2=(\gamma^2)^2=(\gamma^3)^2=-1$.
\begin{equation}\label{dirac03} D=\{H,\cdot\,\}+\gamma^1\otimes[L_1,\cdot\,]+\gamma^2\otimes[L_2,\cdot\,]+\gamma^3\otimes[L_3,\cdot\,]
\end{equation}
\end{description}

A type $(p,q)$ geometry has a signature $s=(q-p) \mod 8$ which determines some of the characteristics of the spectrum of $D$. These properties are well-known, holding also for the case of a Riemannian geometry in dimension $d$, which is a type $(0,d)$ spectral triple with signature $s=d \mod 8$. The properties can be seen in the Monte Carlo simulations below. 

\begin{description}\item[Symmetry] For $s\ne 3,7$, if $\lambda$ is an eigenvalue then so is $-\lambda$.
\item[Doubling] For $s=2,3 \text{ or } 4$, each eigenvalue appears with an even multiplicity.
\end{description}
The proof of these is given briefly here. For even $s$, the chirality operator $\Gamma$ is non-trivial. It is Hermitian and has eigenvalues $\pm1$. The Dirac operator changes the chirality,
$D \Gamma= - \Gamma D$.
If $v$ is an eigenvector of eigenvalue $\lambda$ then $\Gamma v$ is an eigenvector with eigenvalue $-\lambda$.
As a result, the spectrum of the Dirac operator is symmetric around $0$. A similar argument holds for $s=1,5$ using the fact that $DJ=-JD$ so that $v$ and $Jv$ have opposite eigenvalues. 

For the doubling property, if $s=2,3 \text{ or } 4$ then $J^2=-1$ (it is `quaternionic'). Since $DJ=JD$ in these cases, if $v$ is an eigenvector then so is $Jv$ with the same eigenvalue. Moreover, one can check that $v$ and $Jv$ must be linearly independent: suppose the eigenvectors are proportional to each other, i.e., $Jv =c v$, with $c \in \mathbb{C}$, then
\begin{align}
-v=J^2 v &= J (c v) = \bar{c} J v = c \bar{c} v 
 \,,
\end{align}
which is a contradiction.


\subsection{A Monte Carlo algorithm for matrix geometries}
An observable $f(D)$ is a real- or complex-valued function of Dirac operators. The expectation value of $f$ is defined to be
	\begin{align}
	\av{f}= \frac1Z{\int f(D) e^{ -S(D)} \md D}.
	\end{align}
The integral can be approximated as a sum over a discrete ensemble $\{D_j, j=1,\ldots,N\}$.
\begin{align}
\av{f}_N&=\frac{\sum_{j} f(D_j) e^{-S(D_j)} }{\sum_{j}^{\mathstrut} e^{-S(D_j)}}
\end{align}
so that in the limit taking $N\to \infty$, the average obtained through this discrete sum converges towards the continuum value $\av f$. This convergence can be improved by using a Markov Chain Monte Carlo algorithm.
In such an algorithm the Dirac operators $D_j$ are generated with a probability distribution such that
\begin{align}
 P(D_j)= \frac{  e^{- S (D_j)}}{\sum_{i}^{\mathstrut} e^{-S( D_i)}} \;.
\end{align} 
This simplifies the expression for the average
\begin{align}
\av{f}_N= \frac{1}{N} \sum_{j=1}^N f(D_j)
\end{align}
and improves convergence by concentrating the sampling on regions which contribute strongly.
To generate such an ensemble of Dirac operators the Metropolis-Hastings algorithm is used \cite{hastings_monte_1970}.
In this algorithm  a proposed $D_{j+1}$ is generated from $D_j$ by a move which will be defined in the next subsection.
The proposed operator $D_{p}$ will be accepted as a new part of the chain, $D_{j+1}=D_p$, if $S(D_{p})<S(D_j)$. 
If this was the only rule to add new operators to the Markov chain the code would terminate in any sufficiently deep local minimum.
To make it possible to escape local minima, the new operator is also accepted if $\mathrm{exp}(S(D_{j})-S(D_{p}))>\mathfrak p$, with $\mathfrak p$ a uniformly distributed random number in $[0,1]$.
If $D_p$ is rejected in both tests then $D_{j+1}=D_j$.
This algorithm ensures a Markov Chain satisfying detailed balance, which ensures that the transition probability converges~\cite{newman_monte_1999}.

After a sufficient number of moves, the probability distribution for $D_j$ converges towards the desired configuration and becomes independent of the initial state $D_0$.
The states generated before this convergence are not representative of the probability distribution and can not be used to measure observables.
We checked that this burn-in process terminated by starting from different initial configurations and monitoring the convergence of the action. 

The code is implemented in C++ and all matrix algebra operations use the open source software library Eigen~\cite{eigenweb}.

\subsection{The Monte Carlo move}
To construct a Markov Chain on the space of Dirac operators $\mathcal G=\{D\}$, a move that proposes a new Dirac operator $D_p$ based on the last Dirac operator $D_j$ is needed. 
The Markov Chain property requires that the next proposed operator can only depend on the current operator $D_j$. 
The space $\mathcal G$ is a vector space, so a simple additive move
\begin{align}
D \to D + \delta D\; , 
\end{align}
with $\delta D$ a Dirac operator, will always be ergodic, and as long as  $\delta D$ does not depend on past states the Markov property is also satisfied.
As shown in section \ref{sec:dirac} the Dirac operator is defined using a choice of Hermitian matrices $H_i$ and anti-Hermitian matrices $ L_i$.
To construct $\delta D$ we define it as a Dirac operator composed from $\delta H_i,\delta L_i$.
Generate a random $n \times n$ matrix $R$ with matrix elements in the complex range $[-1-i,1+i]$ and define
\begin{align*}
	\delta H_i &=\mathfrak l\,  ( R_i + R_i^*) \\
	\delta L_i &= \mathfrak l \, ( R_i - R_i^*)
\end{align*}
where $\mathfrak l$ is a real constant that is determined at the start of each simulation.
The value of $\mathfrak l$ determines how `long' the steps in the configuration space are.
A Monte Carlo algorithm has the best thermalisation properties if the acceptance rate of proposed moves is $a_r= (\# \text{accepted moves})/(\# \text{proposed moves}) \simeq 0.5$ (where $\# \text{proposed moves}$ counts all $D_p$ generated).
At the beginning of a simulation the acceptance rate is tested and $\mathfrak l$ adjusted, larger if the acceptance rate is too large, smaller if the acceptance rate is too small, such that $a_r \simeq 0.5$ is satisfied within a tolerance of $1 \%$. The number of attempted Monte Carlo moves is called the Monte Carlo time $\tau_{\mathcal{MC}}$.

Note that the move for the $L_i$ does not preserve the condition that it is trace-free. However since the $L_i$ appear only in commutators, the trace decouples and its value does not affect the Dirac operator.

\subsection{Calculating the action}
The  expression for the Dirac operator  contains terms $\com{M}{\cdot\,}$ and $\acom{M}{\cdot\,}$ for $M \in M(n,\mathbb{C})$.
The commutators and anti-commutators require the use of the left and right actions of $M$.
These are written as matrices using the tensor product
\begin{align}
\com{M}{\cdot\,}&=  M \te \ident{n}- \ident{n} \te M^T\\
\acom{M}{\cdot\,}&= M \te \ident{n}+\ident{n} \te M^T  \;.
\end{align}
The Dirac operator $D$ can then be written as a $k n^2\times kn^2$ matrix that acts on the tensor space $V \te \C^n\te \C^n$.

The matrix operations needed in the computer code are matrix multiplication, addition and calculation of eigenvalues. The run time of these grows like $\mathcal{O}(m^2)$, $\mathcal{O}(m^2)$ and $\mathcal{O}(m^3)$ respectively for $m\times m$ matrices. Therefore it makes sense to write the action in terms of the much smaller $n\times n$ matrices $H_i$, $L_i$ to accelerate the simulations. The details of this calculation for the geometries investigated are collected in appendix \ref{sec:GeoAndAc}.

\subsection{Observables}
Given a Dirac operator $D$, the eigenvalues $\{\lambda_i\}$ can be computed. The two main observables of interest are the $j$-th eigenvalue, ordering the eigenvalues from lowest to highest
\begin{equation} f^j(D)=\lambda_j
\end{equation}
and the distribution of eigenvalues at eigenvalue $\lambda$
\begin{equation}\label{eq:density}f_\lambda(D)=\frac1{kn^2}\sum_j \delta(\lambda-\lambda_j)
\end{equation}

Since eigenvalue calculations are computationally expensive, the eigenvalues are only measured every $4n$ attempted Monte Carlo moves. This improves run time, and reduces the correlation of the measurements. The action $S$ and the acceptance rate of moves are recorded at every step to monitor the algorithm.

At later points it will become useful to measure some additional observables that are computed from the eigenvalues, for example, $\tr{D^2}$. For certain cases it has also proven instructive to examine the non-physical degrees of freedom of the matrices $H$ and $L$ via their eigenvalues.

The average of any observable can be calculated directly from the set of measurements. However to estimate the statistical error on our measurements it is necessary to take the correlation between successive states in the Markov Chain into account. The error bars shown on plots of average eigenvalues show the statistical error $\Err(\lambda_i)$ calculated as
\begin{equation}
\Err(\lambda_i)= \sqrt{\frac{2 \sigma_{\lambda_i} \tau_{A, \lambda_i}}{M}}
\end{equation}
with $\sigma_{\lambda_i}$ the variance of the eigenvalue, $\tau_{A, \lambda_i}$ the integrated autocorrelation time of the eigenvalue and $M$ the number of measurements performed \cite{newman_monte_1999}.

\begin{figure}
\subcaptionbox{Autocorrelation of $S$ for $n=5,15$}{\includegraphics[width=0.5\textwidth]{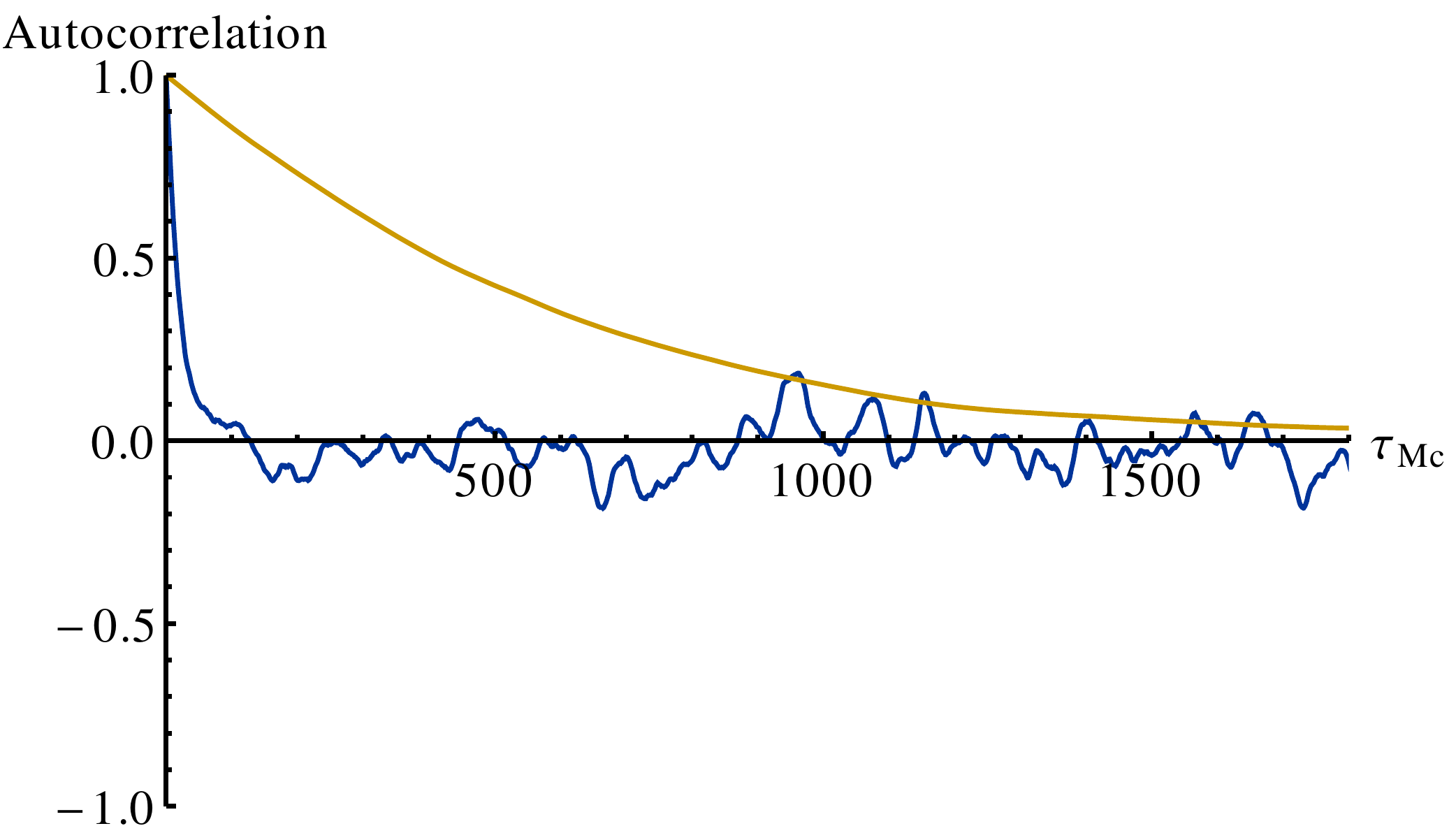}}
\subcaptionbox{Autocorrelation of $\lambda_{\mathrm{min}}$ for $n=5,15$}{\includegraphics[width=0.5\textwidth]{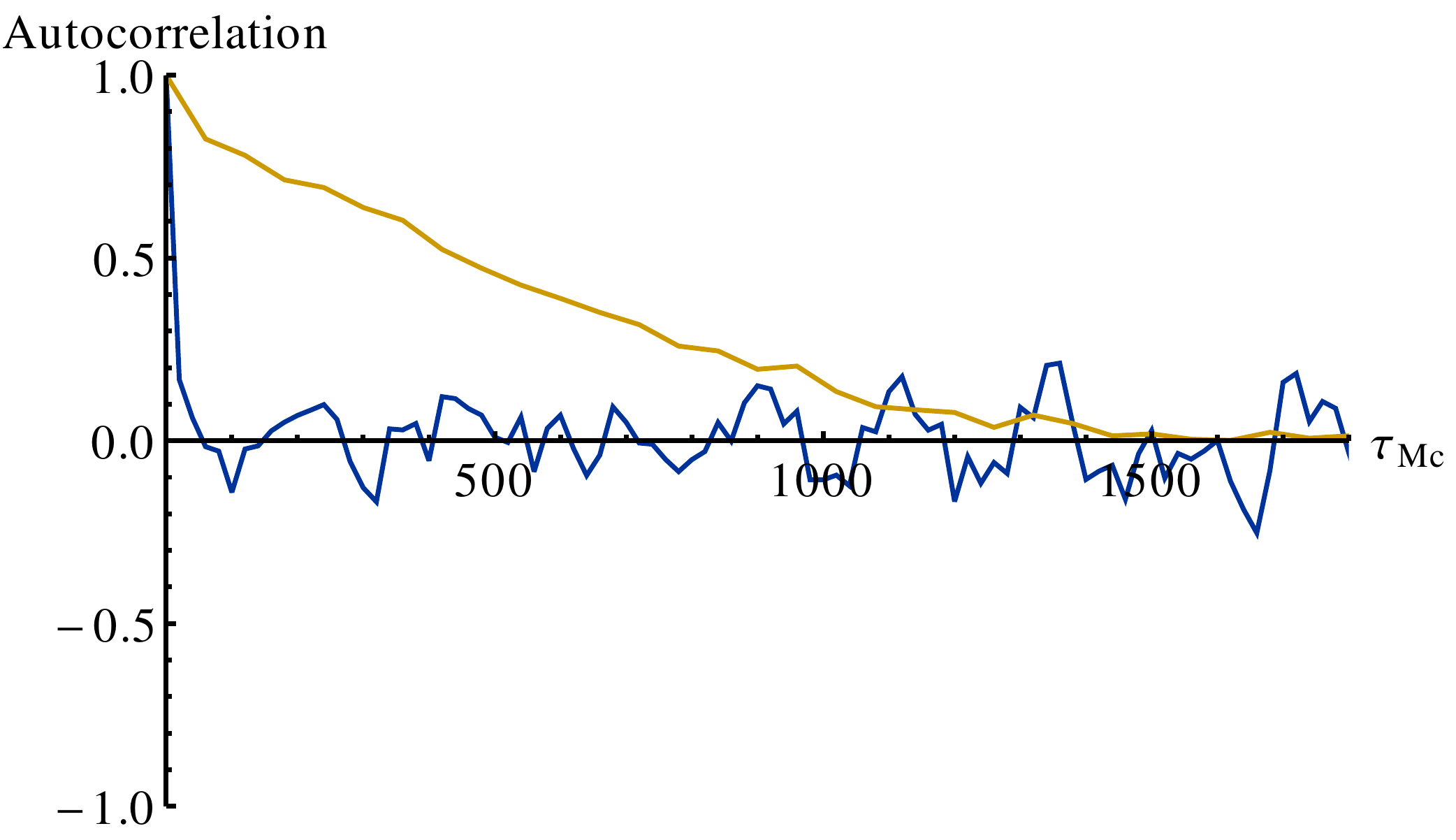}}
\caption{\label{fig:auco}Fall-off of the autocorrelation for the action and the minimum eigenvalue for a type $(1,0)$ geometry with $S=\tr{D^4}$.
The blue line is $n=5$ and the yellow line is $n=15$. The horizontal axis is Monte Carlo time.}
\end{figure}

In figure \ref{fig:auco}, autocorrelations for the simulations of a type $(1,0)$ geometry with $S=\tr{D^4}$ for size $n=5$ and $n=15$ are shown. The figures show the autocorrelation for both the action and the smallest eigenvalue of $D$. 

The autocorrelation time is determined on the data after the burn-in is completed.
In practice the burn-in phase was combined with the adjustment of $\mathfrak{l}$. Then $\tau_{MC}$ was counted from $5000$ moves after the last adjustment to $\mathfrak{l}$.

This burn-in and adjustment period takes up most of the simulations. We found that for the eigenvalue distribution and the average eigenvalues, $200$ measurements (corresponding to $4n\cdot 200$ attempted Monte Carlo moves) lead to very good results.
To determine the phase transition, $2000$ measurements were used to ensure that statistical fluctuations were not mistaken for a phase transition.

\section{Results for $D^2$ action}\label{sec:D2}
In this section the Monte Carlo simulations for the simplest possible action $S= \tr{D^2}$ are examined. The one-dimensional Clifford algebras, type $(1,0)$ and $(0,1)$ are examined first and the results understood using the standard theory of Gaussian matrix models. After this, some numerical results for the two- and three-dimensional types are shown.

\subsection{\label{sec:simp}The simplest cases: type $(1,0)$ and $(0,1)$}
The $n^2$ eigenvalues of Dirac operators \eqref{dirac10}, \eqref{dirac01} can be written in terms of the $n$ eigenvalues  $\mu_j$ of the matrix $H$ or the eigenvalues $i\mu_j$ of $L$. For the $(1,0)$ case one has eigenvalues
\begin{equation}\lambda_{jk}=\mu_j+\mu_k
\end{equation}
while for the $(0,1)$ case
\begin{equation}\lambda_{jk}=\mu_j-\mu_k
\end{equation}
This follows from the fact that eigenvectors of $D$ are of the form $u_j\otimes \overline u_k$, with $u_j$ the eigenvectors of $H$ or $L$. 

The first point is that the $(0,1)$ case has eigenvalue $0$ with multiplicity $n$ given by the terms $j=k$. This can also be seen directly from the Dirac operator:  all matrices in $M(n, \mathbb{C})$ that commute with $L$ have eigenvalue $0$, and there are always at least $n$ linearly independent such matrices. It will be seen later that a peak in the eigenvalue distribution at, or near, $0$ is a feature of some other random fuzzy spaces.

The second point is that the spectrum of the $(0,1)$ case is symmetric about the origin, as $\lambda_{jk}=-\lambda_{kj}$. This is in accordance with its signature $s=1$, which means that each Dirac operator has symmetric spectrum. The spectrum of a $(1,0)$ Dirac operator is typically not symmetric since $s=7$ in this case. This means that our simulation gives an eigenvalue distribution that is not exactly symmetric, though it will eventually converge to a symmetric distribution as the Monte Carlo time increases. 

For the $(1,0)$ case, using the simplified action \eqref{eq:S10simp} one can transform the integral over the Dirac operator into an integral over the Hermitian matrix $H$.
\begin{align}
S^{(1,0)}(D)&=\tr{D^2}=2 n \tr{H^2}+ 2 (\tr{H})^2\\
			 &= 2 n \sum_i \mu_i^2 + 2 \sum_i \sum_j \mu_i \mu_j
\end{align}

The $(0,1)$ case is similar, but one has to take into account the fact that the integration over Dirac operators is an integration over traceless matrices $L$. Using  \eqref{eq:S01simp} gives
\begin{align}
S^{(0,1)}(D)&=\tr{D^2}=-2 n \tr{L^2}\\
			 &=2 n \sum_i \mu_i^2  \;,
\end{align}

\begin{figure}[h]
\subcaptionbox{Type $(1,0)$ average eigenvalues of $H$}{\includegraphics[width=0.5\textwidth]{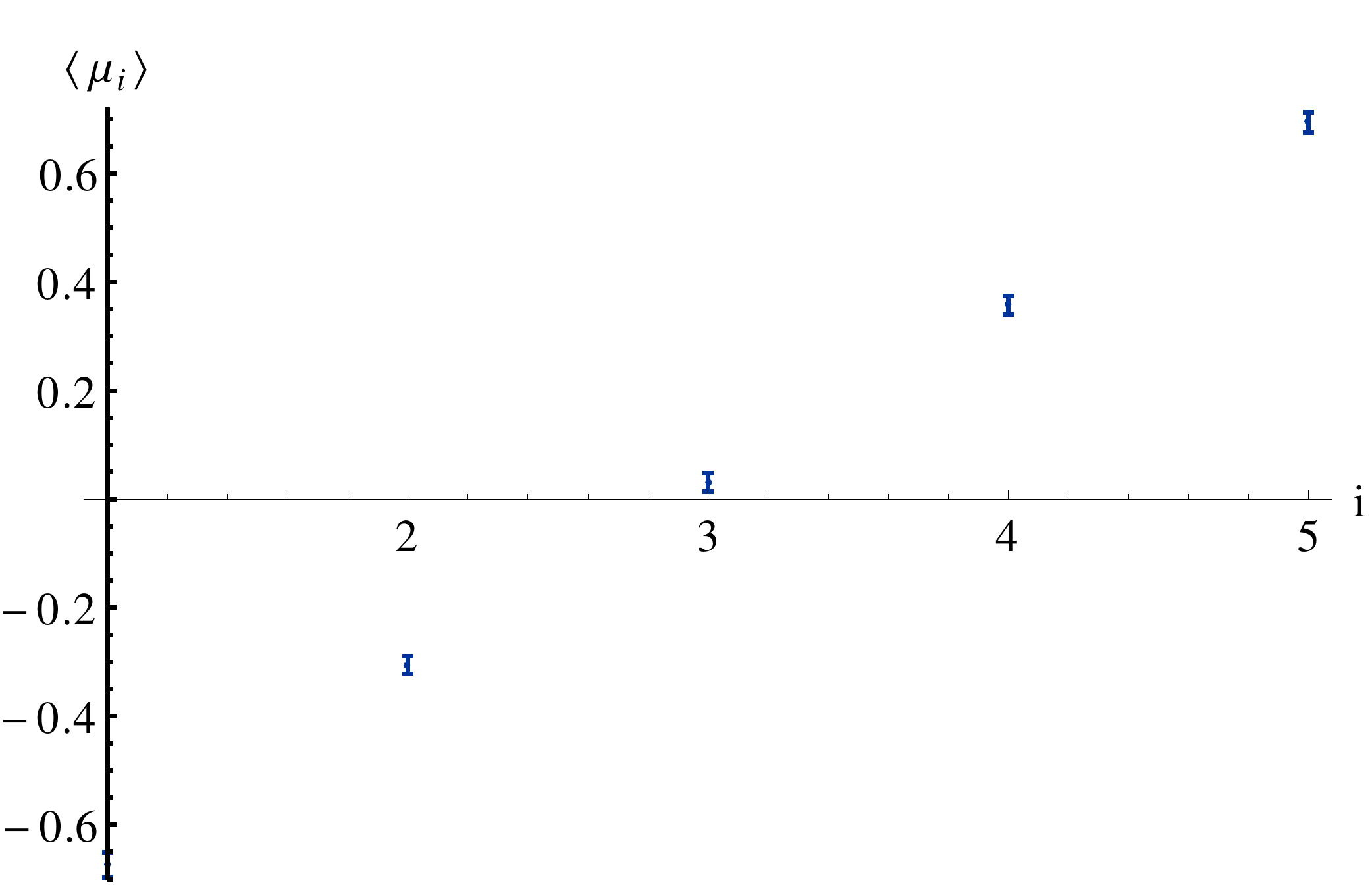}}
\subcaptionbox{Type $(0,1)$ average eigenvalues of $-iL$}{\includegraphics[width=0.5\textwidth]{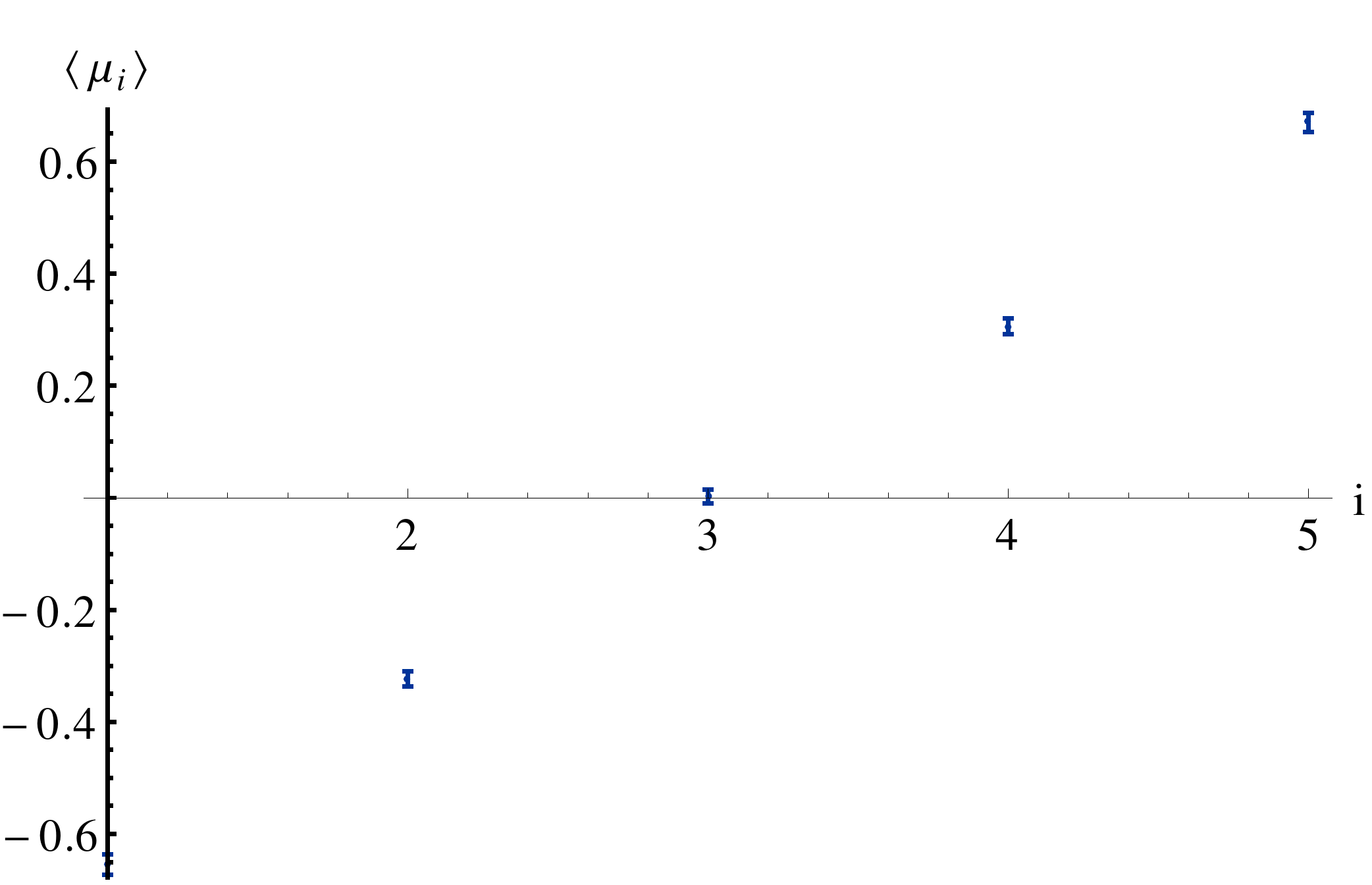}}

\subcaptionbox{Type $(1,0)$ average eigenvalues of $D$}{\includegraphics[width=0.5\textwidth]{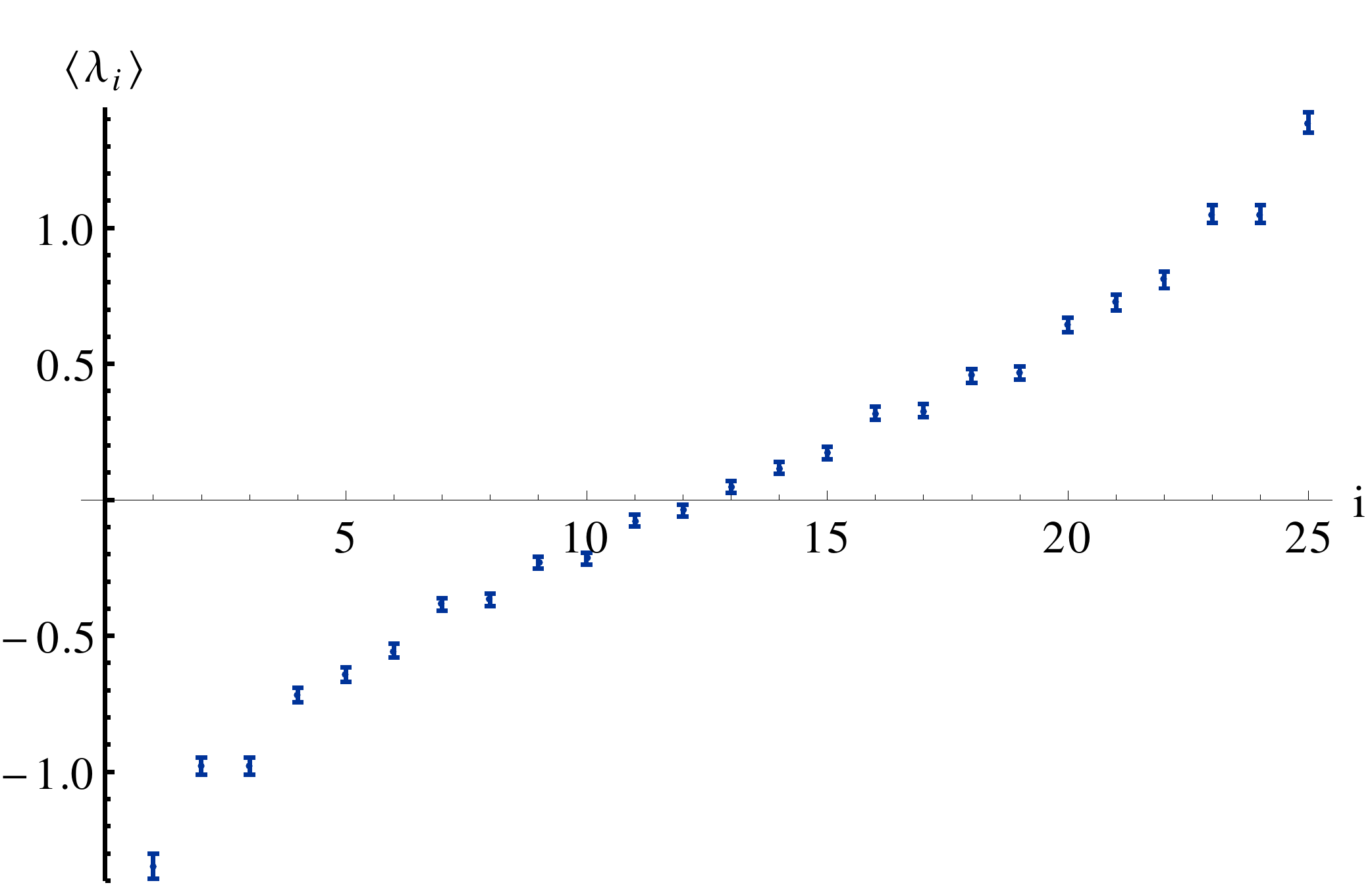}}
\subcaptionbox{Type $(0,1)$ average eigenvalues of $D$}{\includegraphics[width=0.5\textwidth]{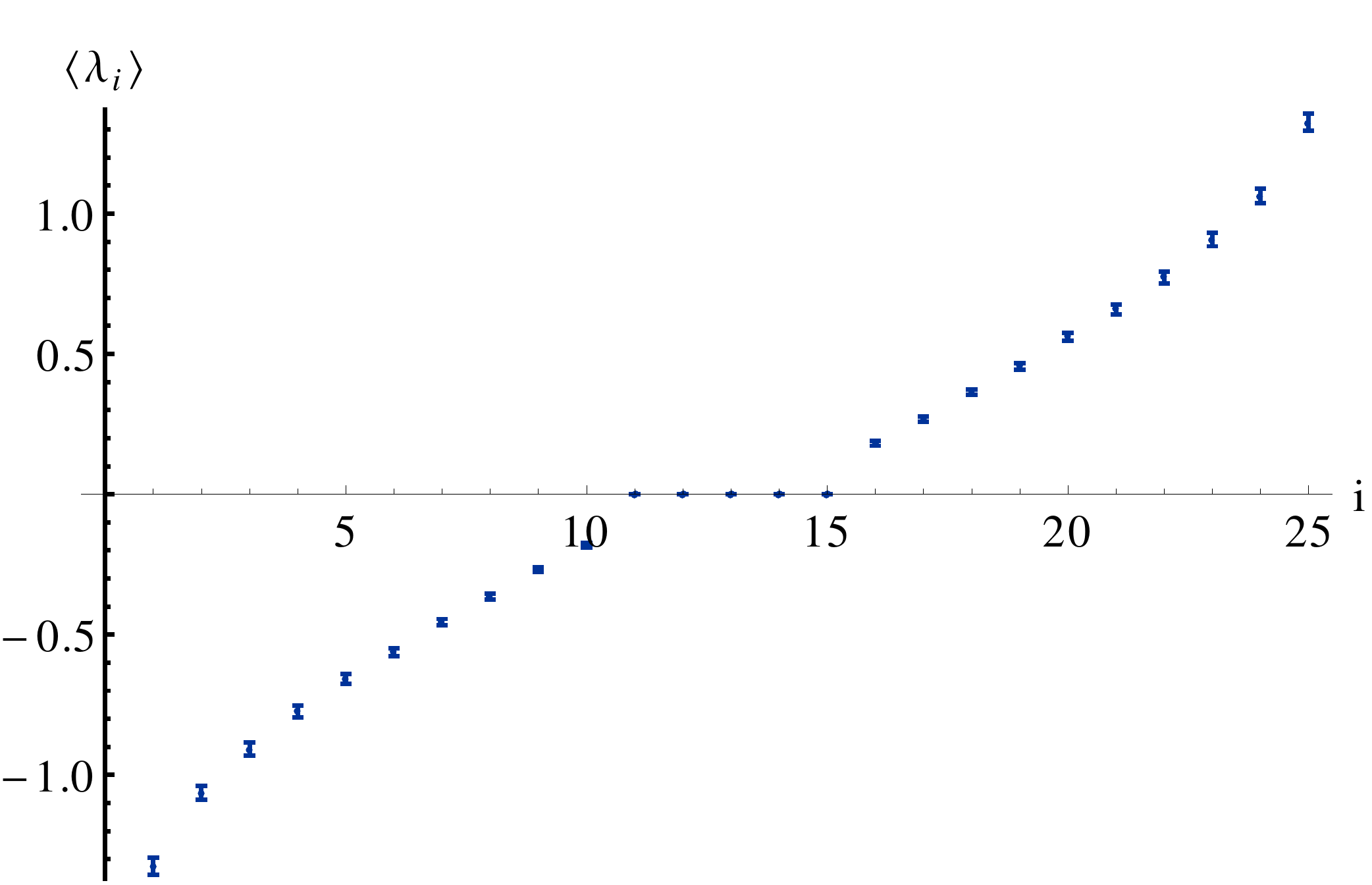}}
\caption{\label{fig:simpleCases}Average ordered eigenvalues for $H, L$ and $D$ for the cases $(1,0)$ and $(0,1)$ with $n=5$.}
\end{figure}
An example of average ordered eigenvalues generated by the Monte Carlo simulation is shown in figure \ref{fig:simpleCases}.

These random matrix models are close to the Gaussian Hermitian matrix model \cite{mehta_density_1960,chadha_method_1981}, 
which  has the similar action
\begin{equation}\widetilde S(M)=2n \tr{M^2}=2 n \sum_j \mu_j^2  \;,
\end{equation}
with integration over all Hermitian matrices.

A standard technique in random matrix models is to calculate the joint probability density for the eigenvalues $\mu_j$. The formula is \cite{mehta_random_2004}
\begin{align}
P(\mu_1,\dots,\mu_n)=C \exp\left(-\widetilde S(\mu_1,\ldots,\mu_n) \right) \prod_{j<k} (\mu_j-\mu_k)^2\;.
\end{align}
The terms with the differences of eigenvalues result from the Jacobian for the change of variables from the matrix elements to the eigenvalues. Since this term is small when two eigenvalues are close, this results in the phenomenon of the repulsion of eigenvalues.

The matrix $M$ can be split into traceless and trace parts, and these are statistically independent. It follows that expectation values in the $(0,1)$ model can be calculated as expectation values of observables in the Gaussian Hermitian matrix model that are independent of the trace of $M$. This is done by writing $M=-iL$. The action in the $(1,0)$ model transforms directly to the Gaussian Hermitian matrix model by rescaling the trace by a factor $\sqrt2$. The transformation is
\begin{equation}M=H+\frac1n(\sqrt2-1)\tr H.
\end{equation}
A standard result (the Wigner semicircle law \cite{wigner_distribution_1958}) is that the analogue of the eigenvalue distribution  \eqref{eq:density} for the Gaussian Hermitian matrix model converges as $n\to\infty$ to the density of states
\begin{align}\label{eq:semicirc}
\sigma(\mu)=\lim_{n\to\infty} <f_\mu(M)> &= \begin{cases}
\frac2{\pi A} \sqrt{ A -\mu^2} & \text{for } -\sqrt A\le\mu\le\sqrt A \\
0	& \text{ everywhere else}
\end{cases}
\end{align}
with $A=1$.



\begin{figure}
\subcaptionbox{Type $(1,0)$ $n=2$}{\includegraphics[width=0.5\textwidth]{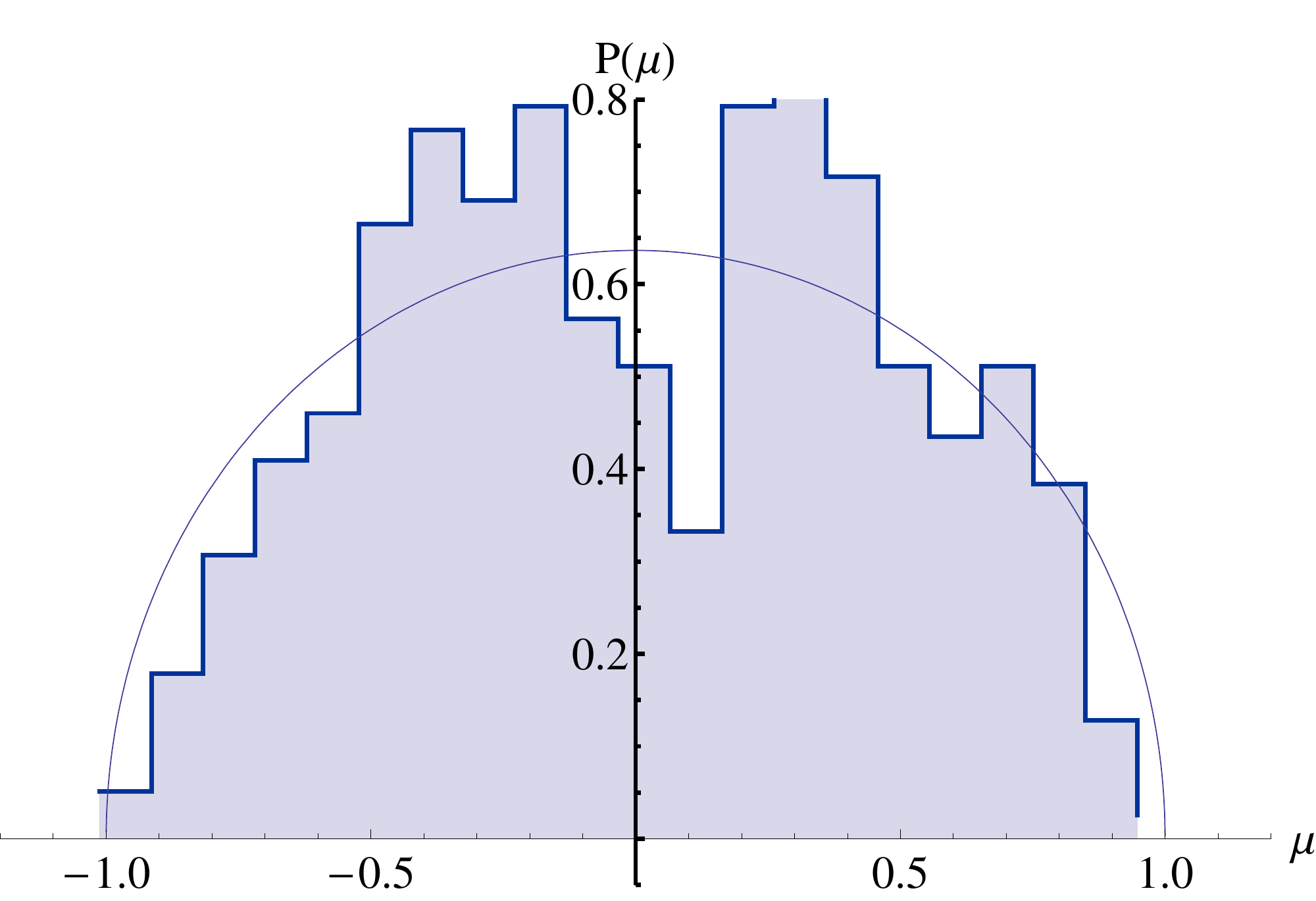}}
\subcaptionbox{Type $(0,1)$ $n=2$}{\includegraphics[width=0.5\textwidth]{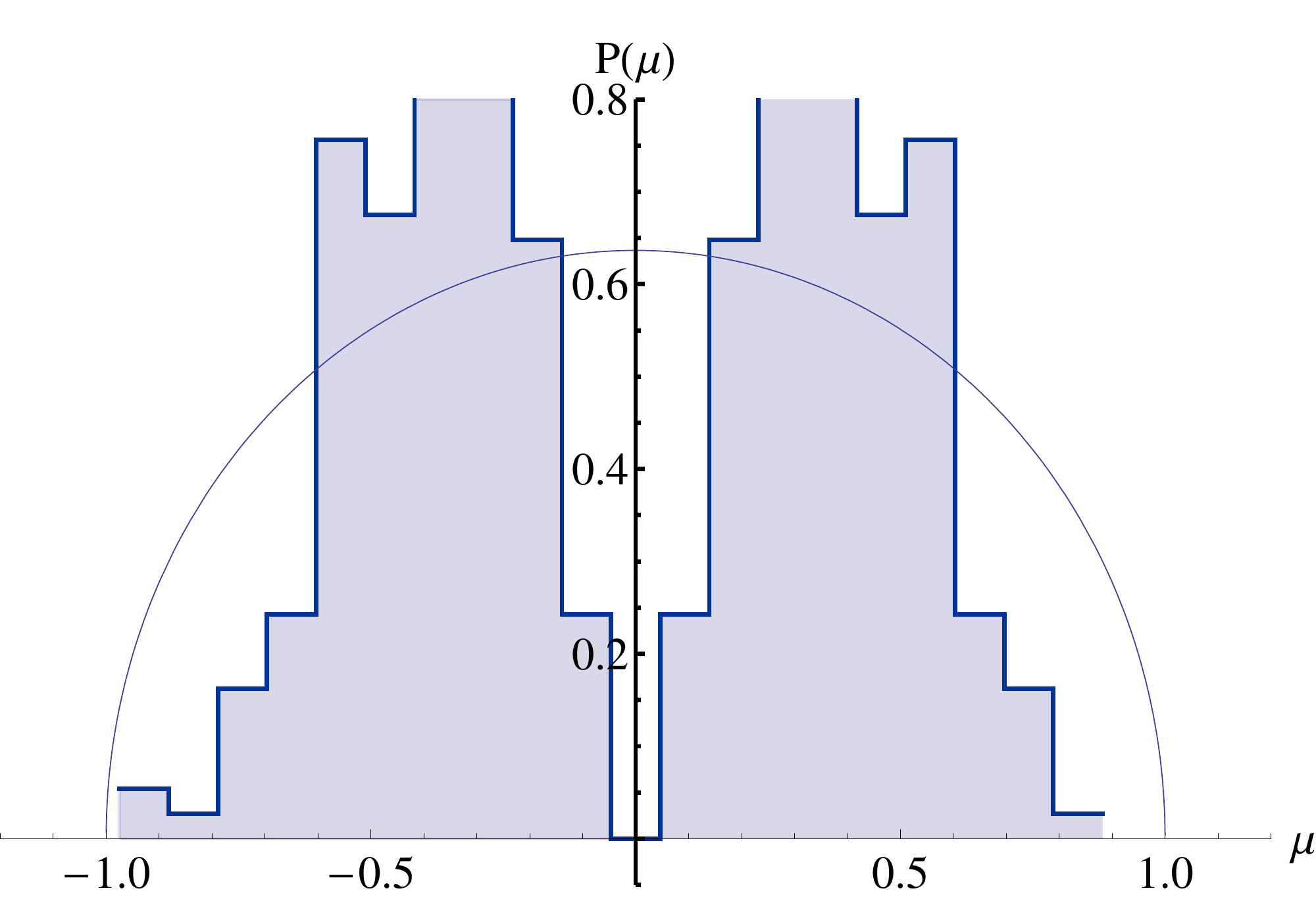}}

\subcaptionbox{Type $(1,0)$ $n=5$}{\includegraphics[width=0.5\textwidth]{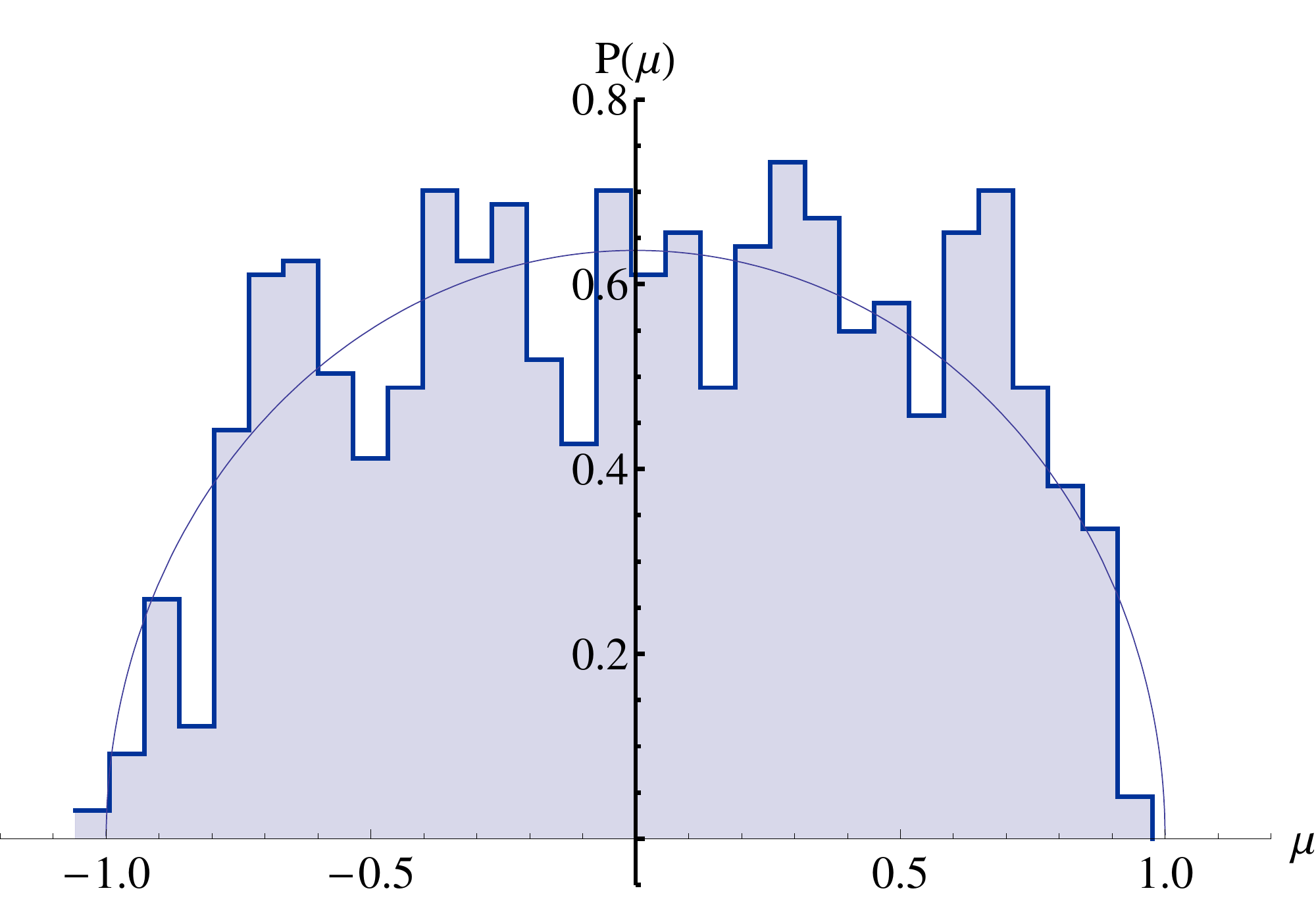}}
\subcaptionbox{Type $(0,1)$ $n=5$}{\includegraphics[width=0.5\textwidth]{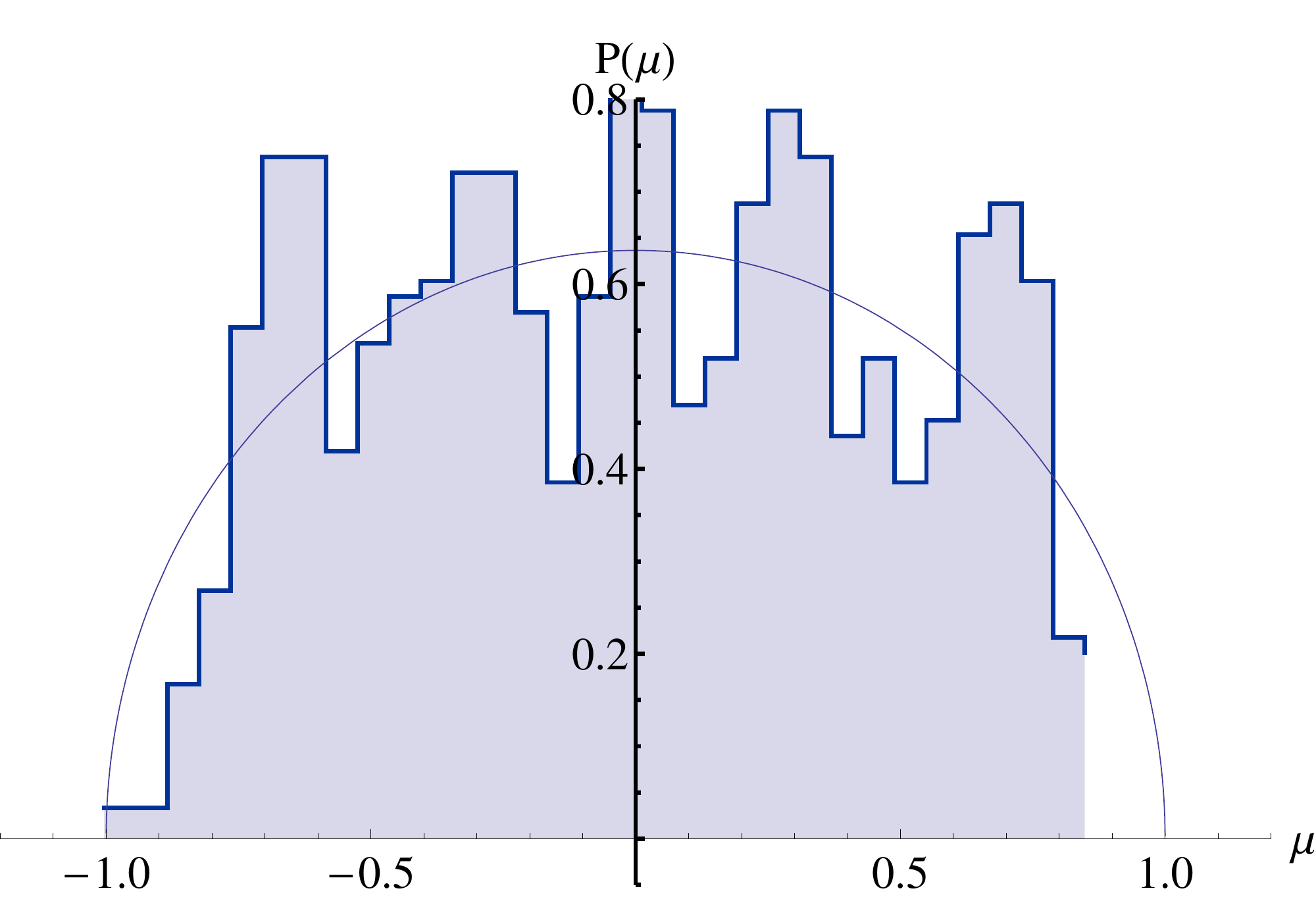}}

\subcaptionbox{Type $(1,0)$ $n=15$}{\includegraphics[width=0.5\textwidth]{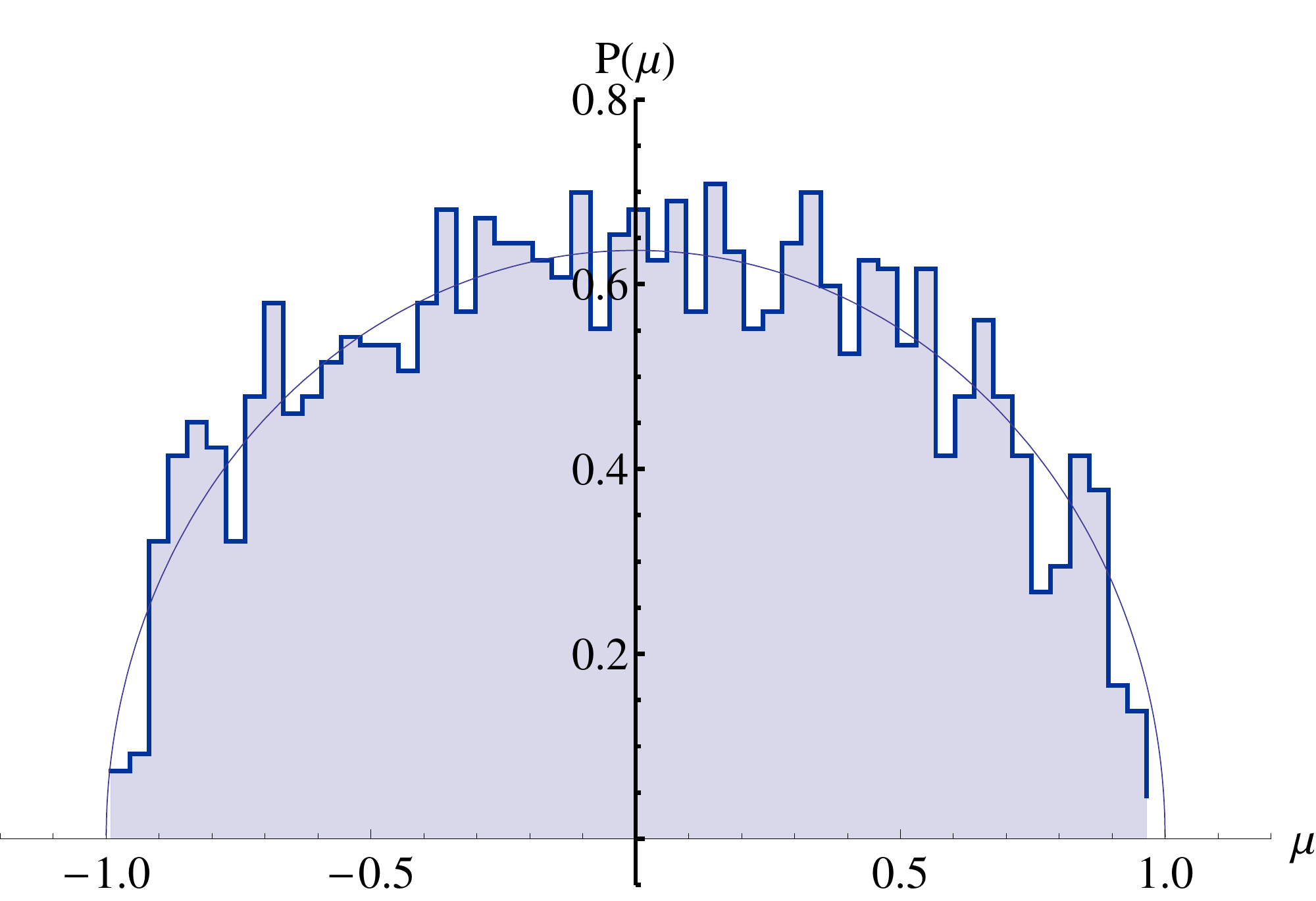}}
\subcaptionbox{Type $(0,1)$ $n=15$}{\includegraphics[width=0.5\textwidth]{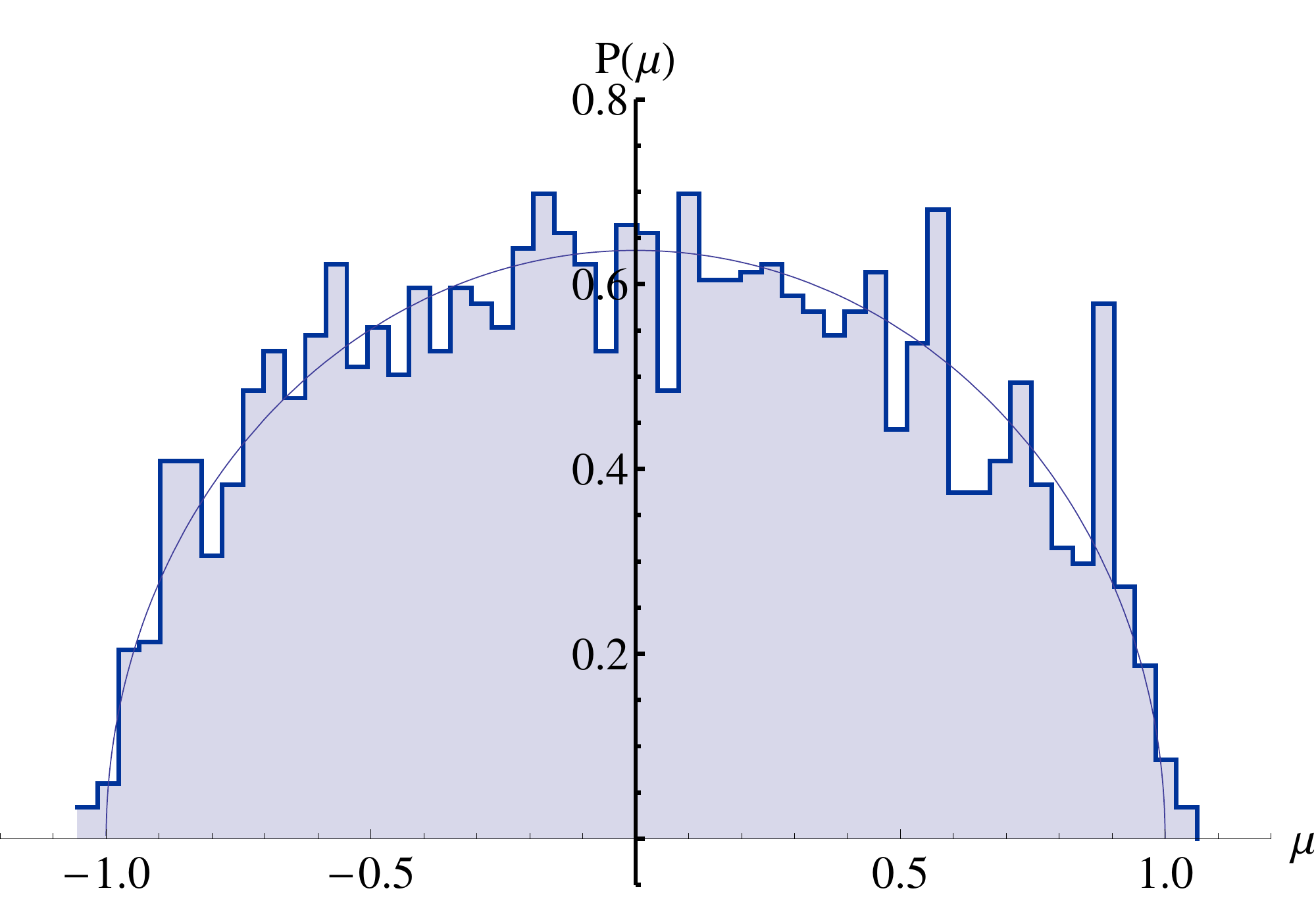}}
\caption{\label{fig:semiC} The semicircle law is compared with the density of states for $H$ or $L$.}
\end{figure}

In our simulations using actions $S^{(1,0)}$ and $S^{(0,1)}$ we find that the semicircle law is also a good approximation for the eigenvalues of $H$ and $L$. It is already well-satisfied for $n=5$ and improves at higher $n$, as shown in figure \ref{fig:semiC}. The reason for this is that in the Gaussian Hermitian matrix model, the variable $\frac1n\tr M $ is normally-distributed with variance $1/(4n^2)$, and so adjusting the eigenvalues with a fixed multiple of this makes no difference to the density of states in the limit $n\to\infty$.

Another standard result from random matrix theory is that the correlation between different fixed eigenvalues of $M$ vanishes as $n\to\infty$. Thus for large $n$, the joint probability distribution away from the diagonal $\mu_1=\mu_2 $ is simply the product of the density of states \cite{Pastur}. Therefore, for large $n$, one can calculate the eigenvalue distribution of the Dirac operator from the semicircle law as a convolution, with a correction for the behaviour of the correlations on the diagonal. 

\begin{figure}
\subcaptionbox{$(1,0)$ $n=5$}{\includegraphics[width=0.5\textwidth]{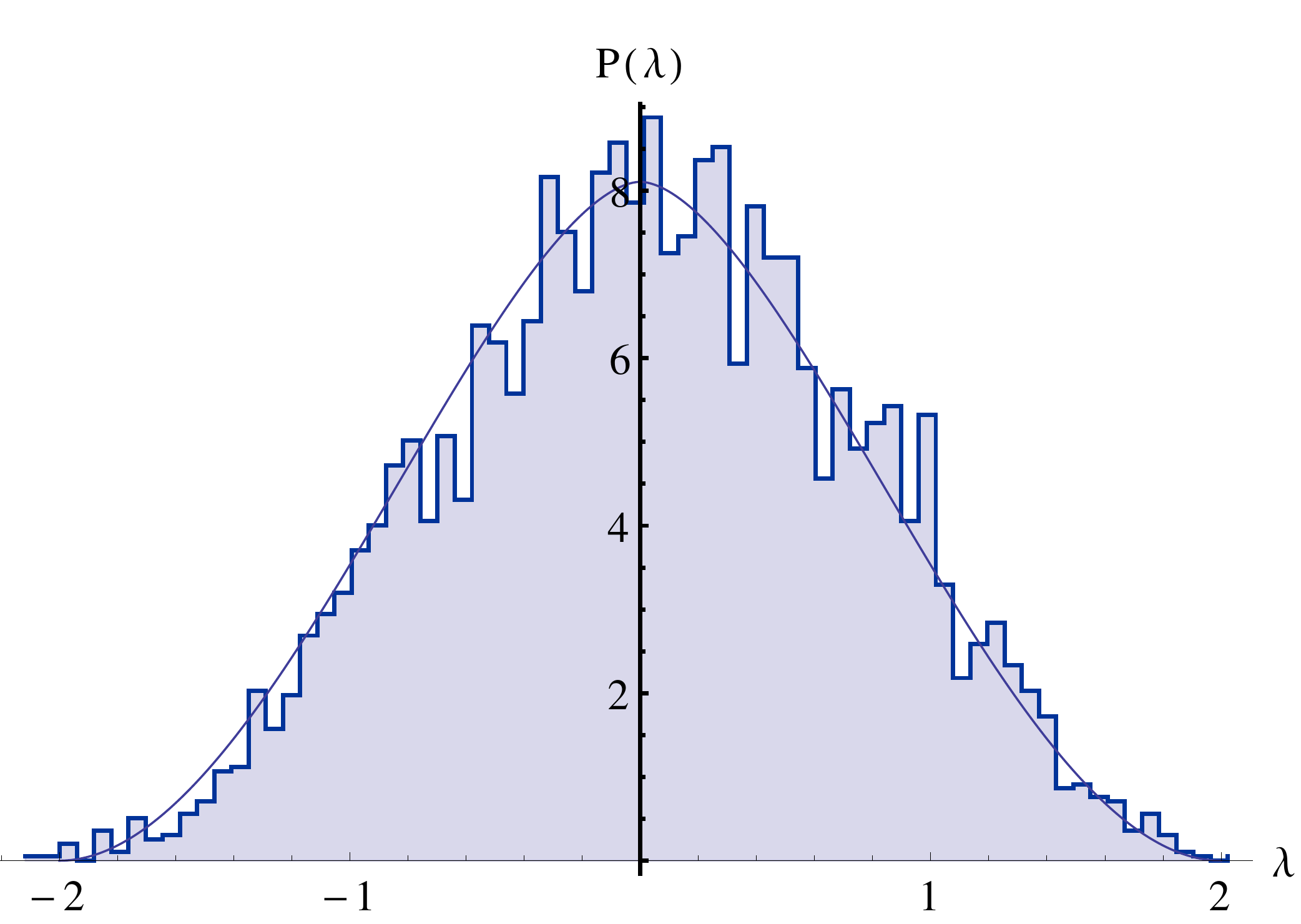}}
\subcaptionbox{$(0,1)$ $n=5$}{\includegraphics[width=0.5\textwidth]{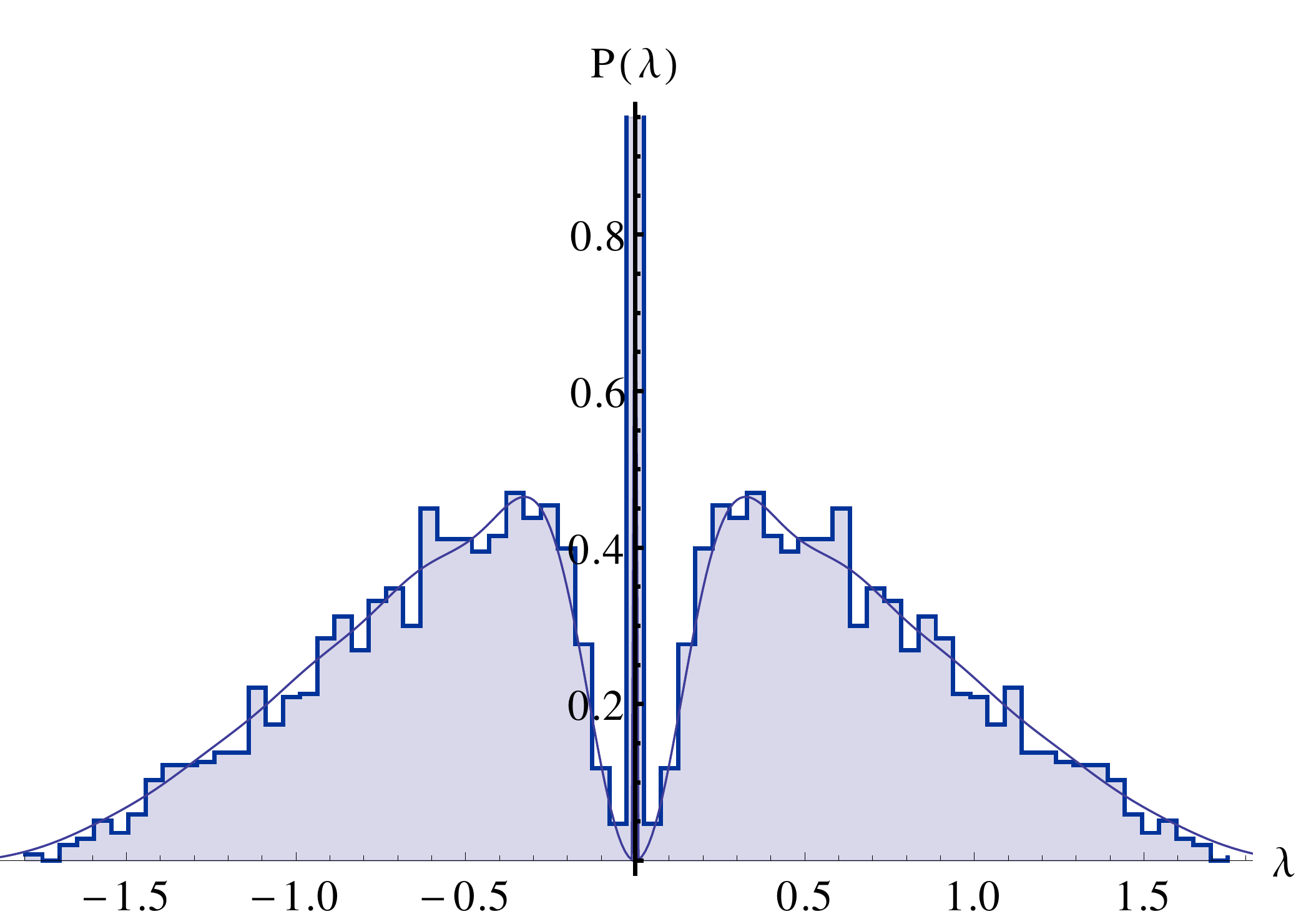}}
\caption{\label{fig:folded} The eigenvalue density for the Dirac operator compared with the convolution of two semicircle functions $\sigma_D$, with correction applied in the $(0,1)$ case. }
\end{figure}

This is shown as follows. Let $f(\lambda)$ be an observable for a random fuzzy space, with $\lambda=\mu_1\pm\mu_2$. Then assuming a product probability distribution, one has
\begin{equation}\av f=\int \sigma(\mu_1)\sigma(\mu_2)f(\mu_1\pm\mu_2) \md\mu_1\md\mu_2=\int\sigma_D(\lambda)f(\lambda)\md\lambda
\end{equation}
  with density of states for the Dirac operator the convolution
\begin{align}
\sigma_D(\lambda)=\int \sigma(\lambda\mp\mu) \sigma(\mu) \md \mu \;,
\end{align}
which is an elliptic integral.
This integral is the same for type $(1,0)$ and $(0,1)$, since $\sigma(\mu)=\sigma(-\mu)$. The Monte Carlo simulation of the eigenvalue density at finite $n$ is shown  in figure \ref{fig:folded}. The continuous line is the curve $\sigma_D(\lambda)$ for the $(1,0)$ case but a significant correction term is added to $\sigma_D$ for the $(0,1)$ case.

The correction to the product probability density gives a contribution only near the diagonal $\mu_1=\mu_2$. The approximate form is an additional contribution to $\av f$ of \cite{Pastur}
\begin{equation} -\int \frac{\sin^2\bigl(\pi n(\mu_1-\mu_2)\sigma((\mu_1+\mu_2)/2)\bigr)}{\pi^2 n^2(\mu_1-\mu_2)^2}f(\mu_1\pm\mu_2)\; \md\mu_1\md\mu_2
\end{equation}
 For the $(0,1)$ case $(-)$, this formula contributes significantly near $\lambda=0$, accounting for the gap at the origin in figure \ref{fig:folded}(b) with a width that scales as $1/n$.

\subsection{Higher types}\label{sec:higher}
\begin{figure}
\subcaptionbox{Type $(2,0)$}{	\includegraphics[width=0.5\textwidth]{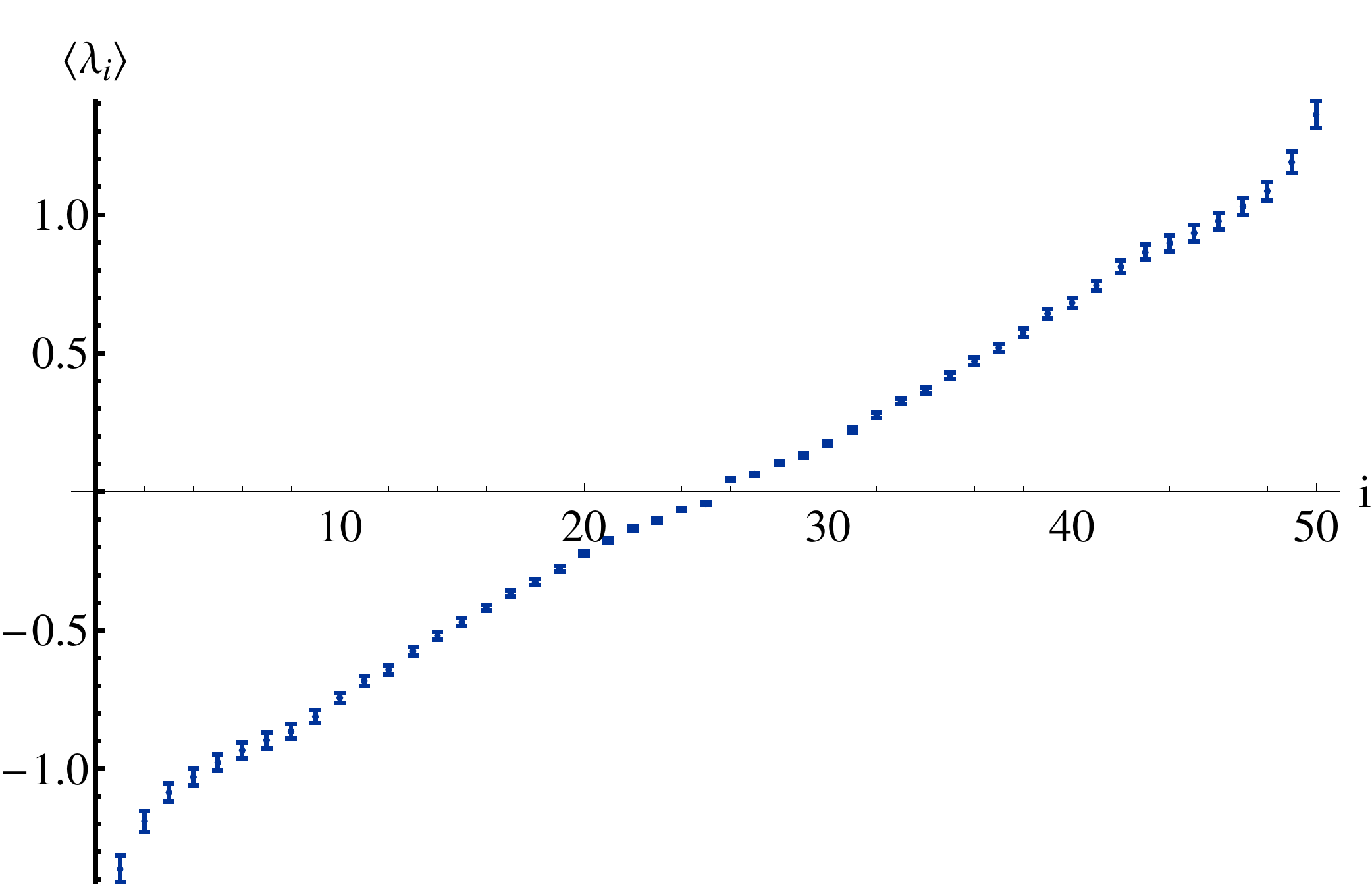}}
\subcaptionbox{Type $(2,0)$}{	\includegraphics[width=0.5\textwidth]{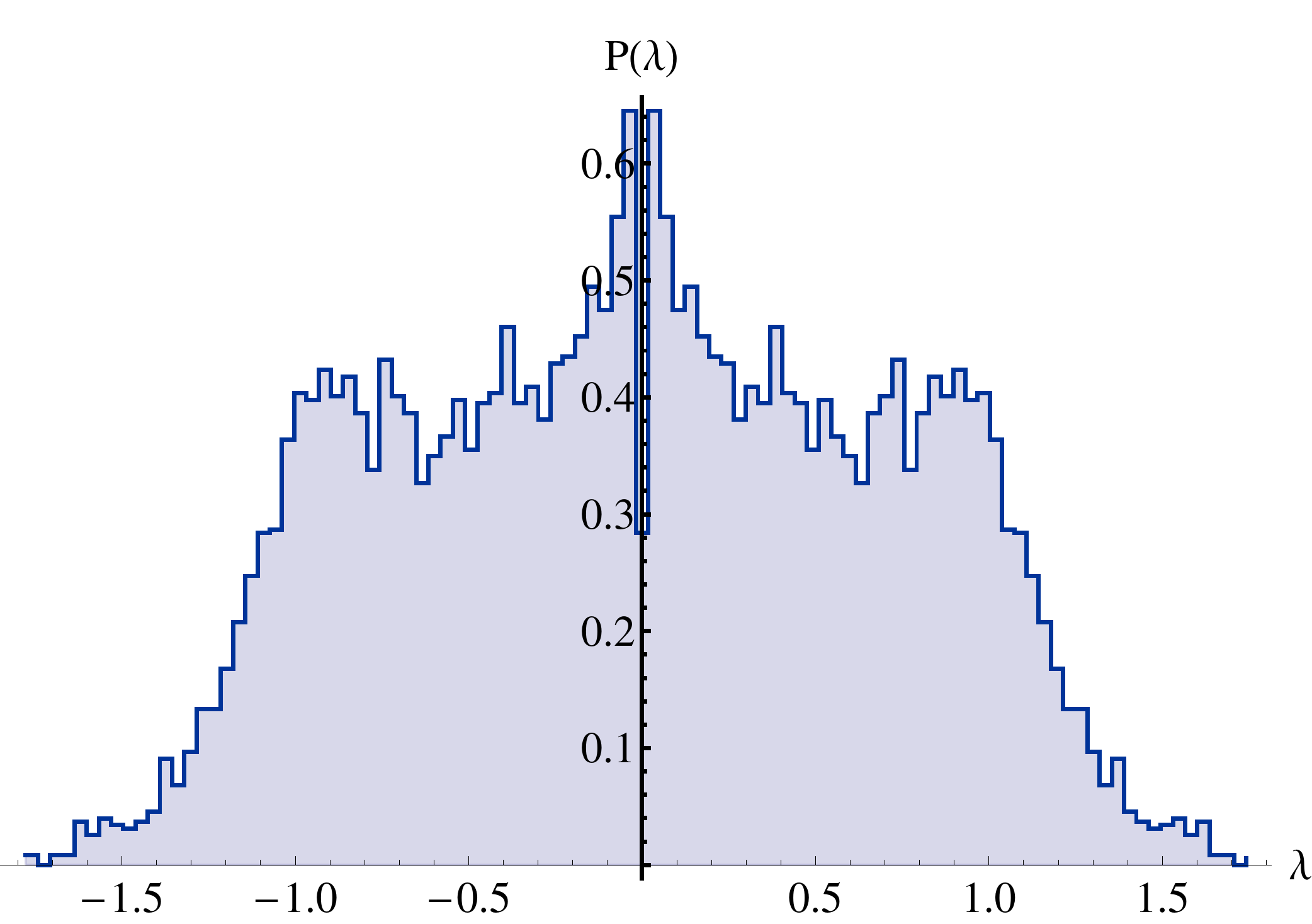}}
	
\subcaptionbox{Type $(1,1)$}{	\includegraphics[width=0.5\textwidth]{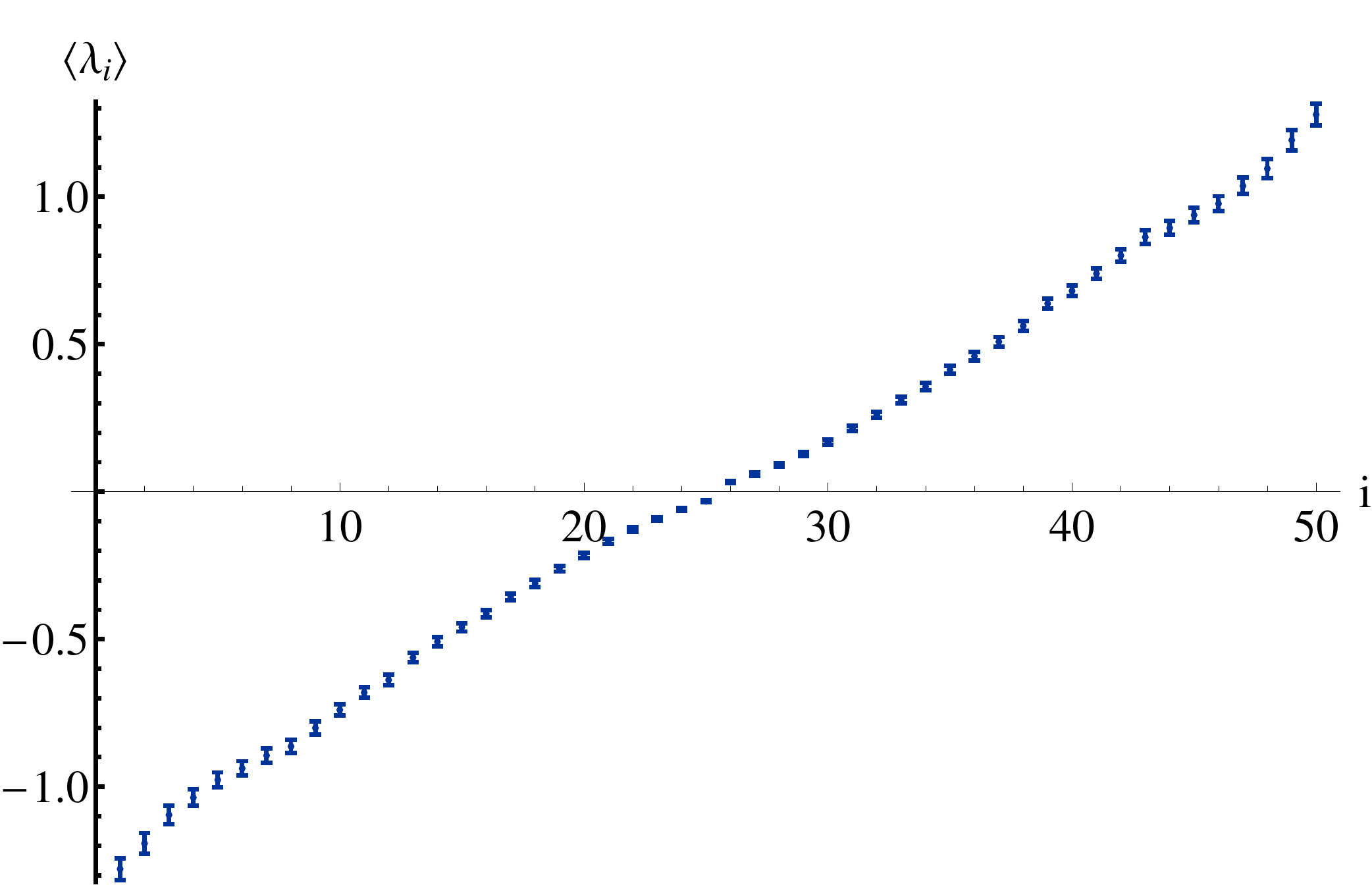}}
\subcaptionbox{Type $(1,1)$}{	\includegraphics[width=0.5\textwidth]{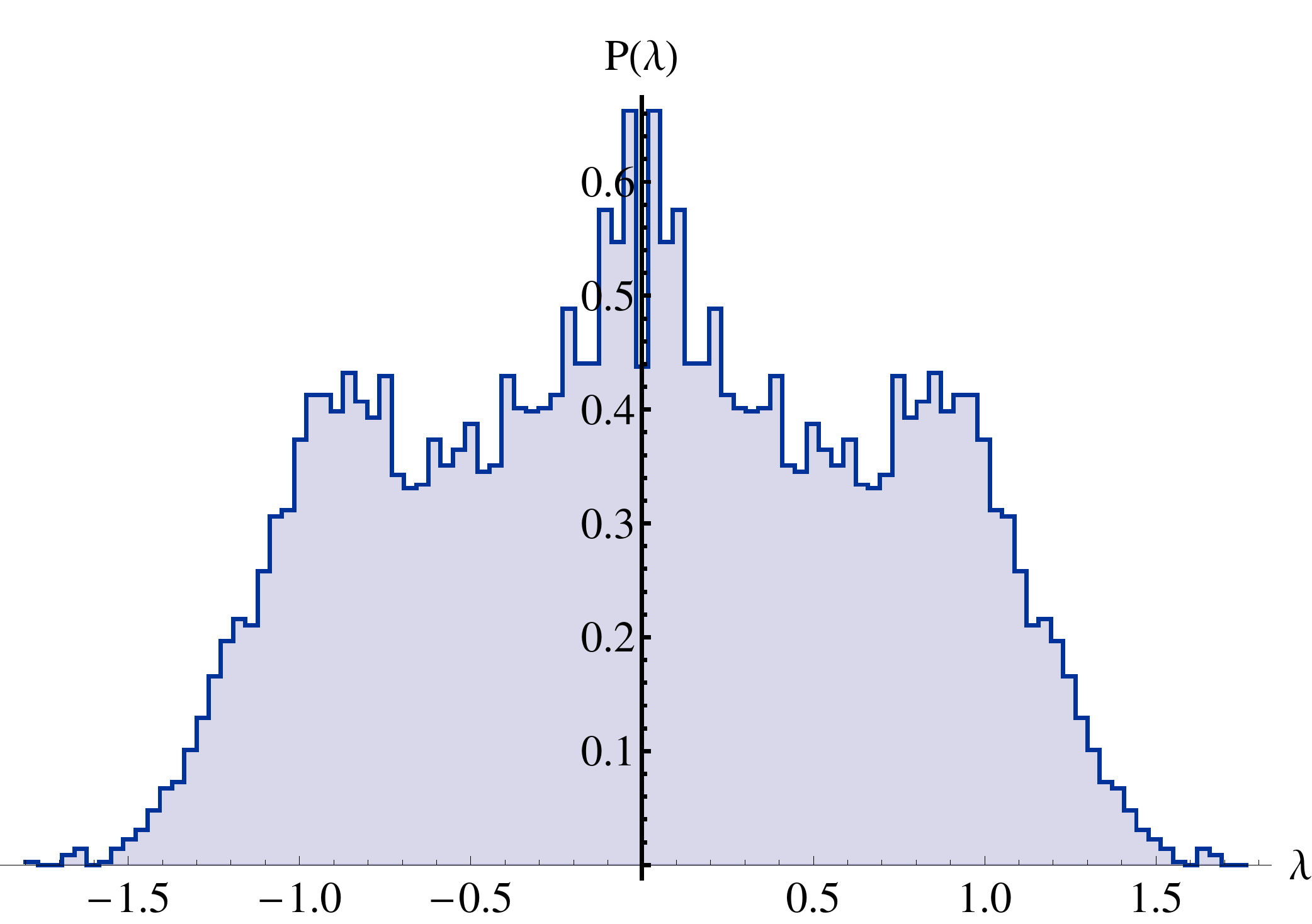}}
	
\subcaptionbox{Type $(0,2)$}{	\includegraphics[width=0.5\textwidth]{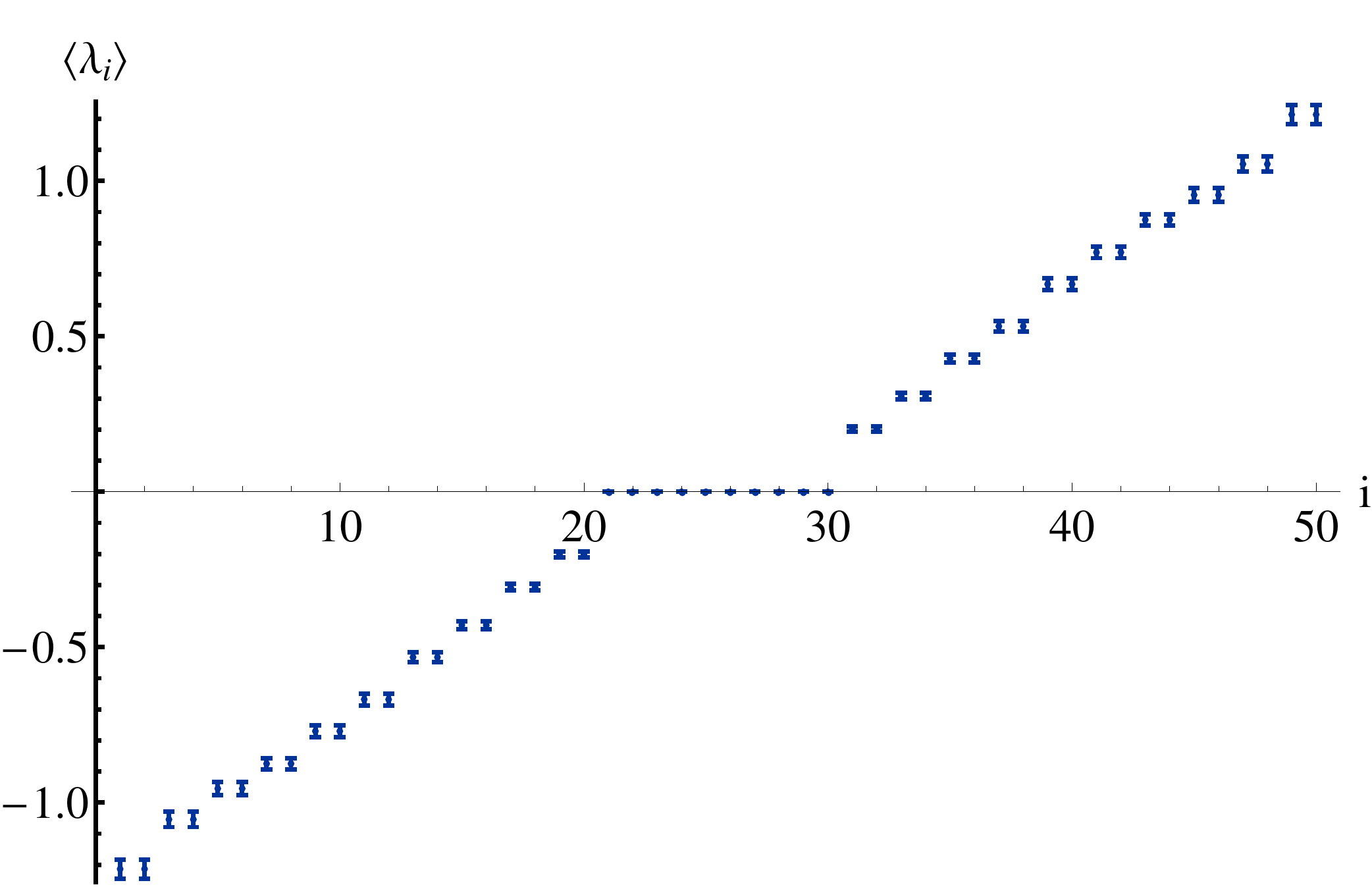}}
\subcaptionbox{Type $(0,2)$}{	\includegraphics[width=0.5\textwidth]{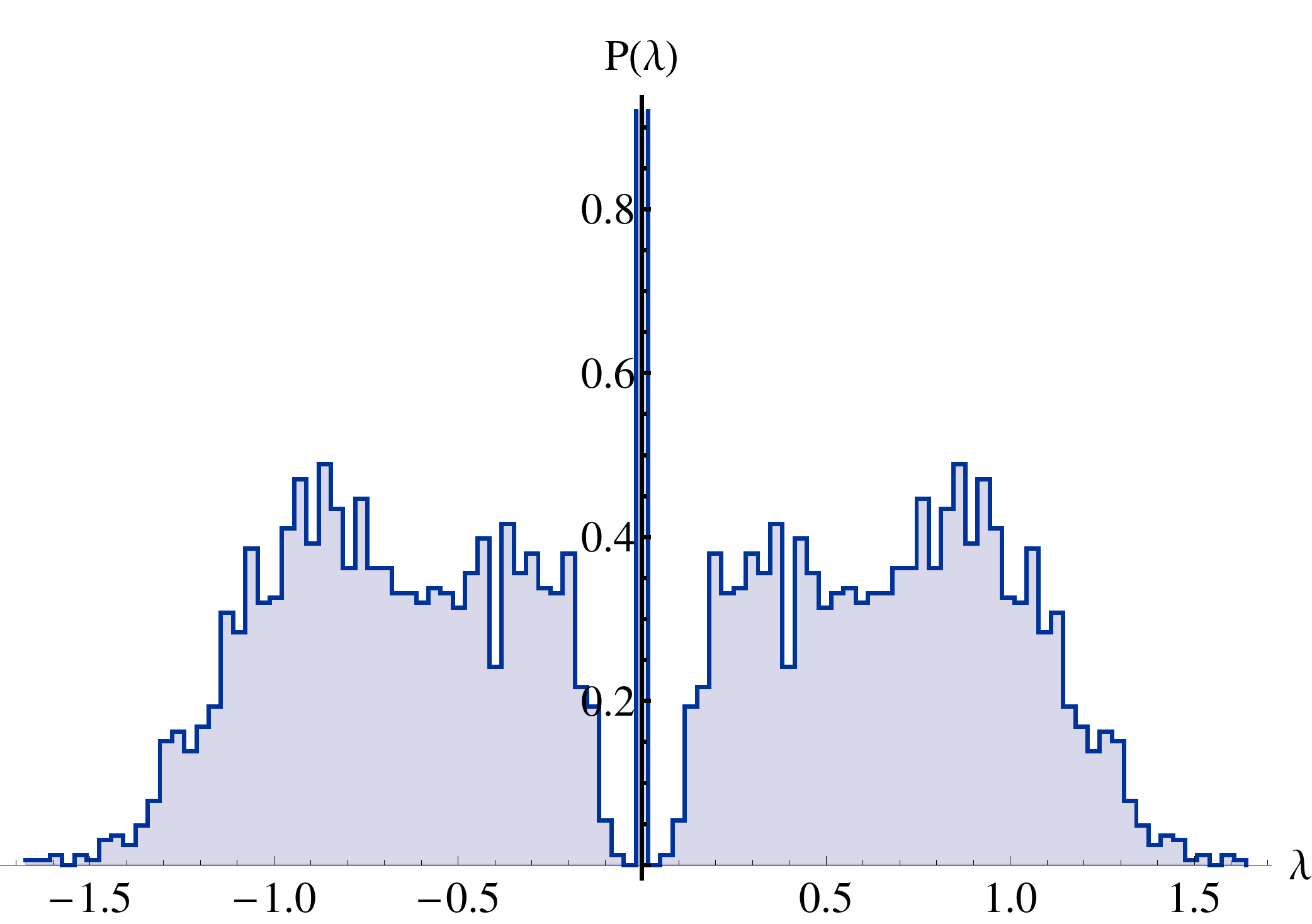}}

\caption{\label{fig:chiralspec5}The average eigenvalues, and the histograms of the eigenvalue distribution for the different types of two-dimensional Clifford algebra. The action is $S(D)=\tr{D^2}$ and the matrix size $n=5$.}
\end{figure}

\begin{figure}
\subcaptionbox{Type $(2,0)$}{	\includegraphics[width=0.5\textwidth]{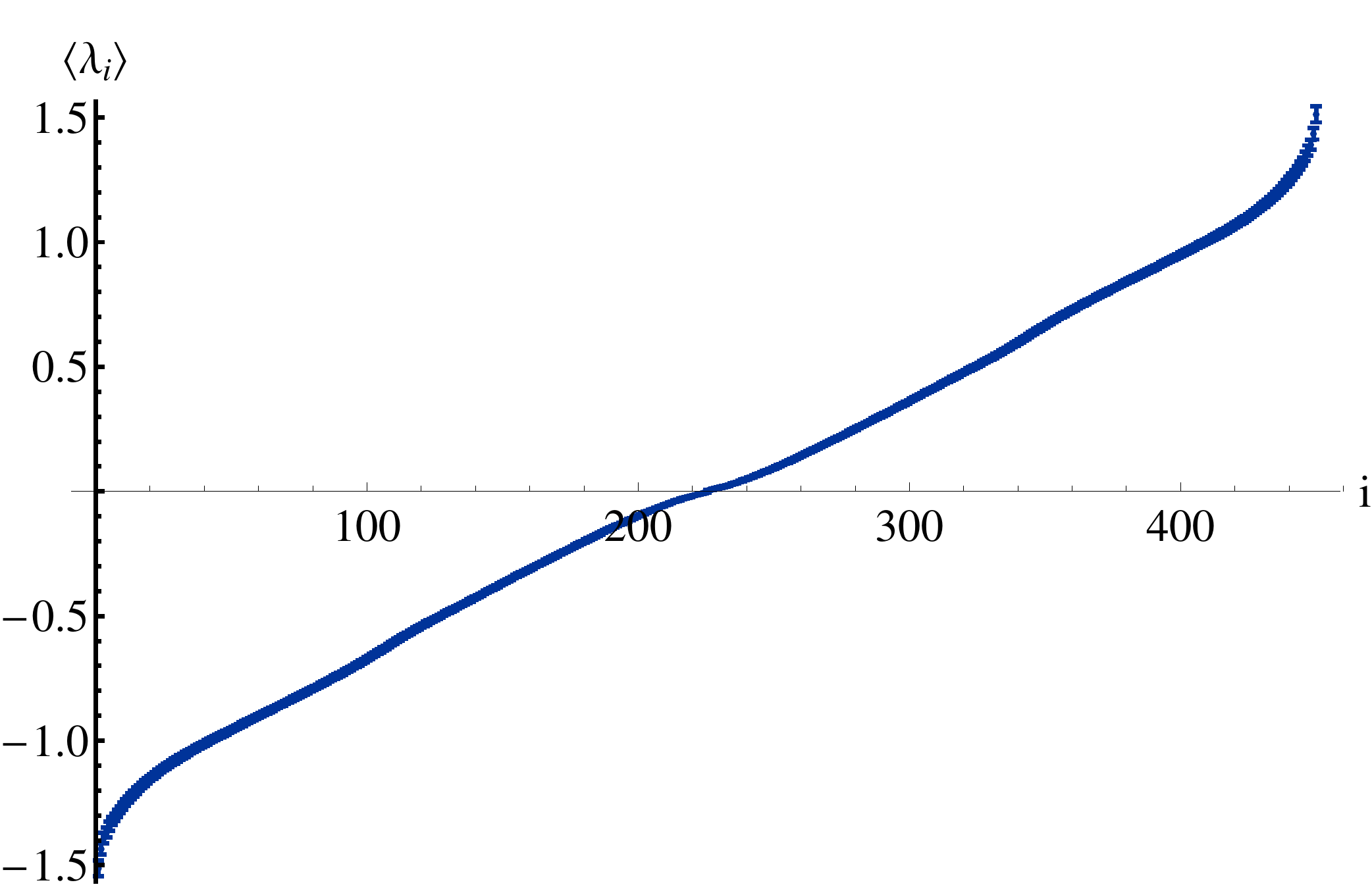}}
\subcaptionbox{Type $(2,0)$}{	\includegraphics[width=0.5\textwidth]{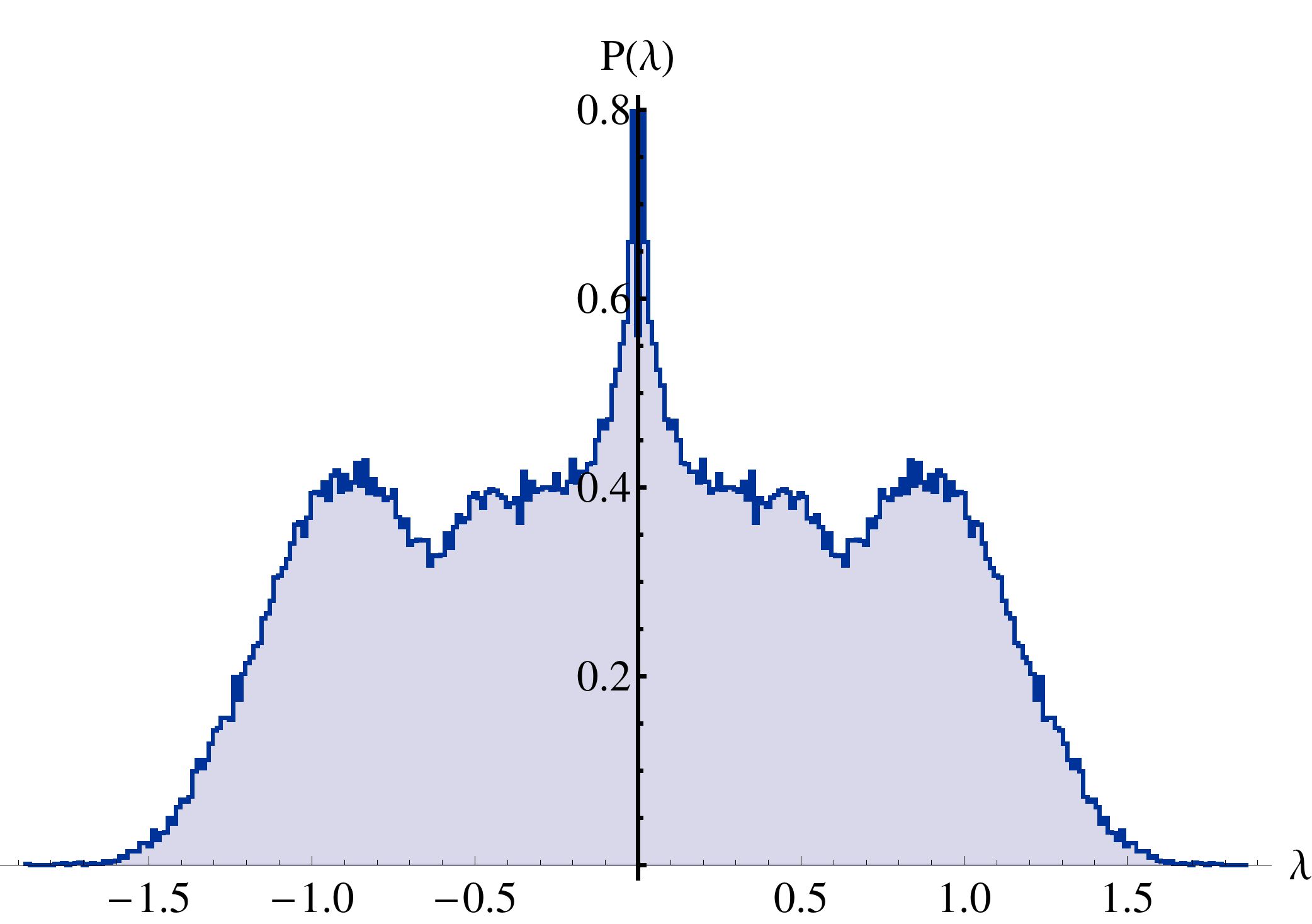}}
	
\subcaptionbox{Type $(1,1)$}{	\includegraphics[width=0.5\textwidth]{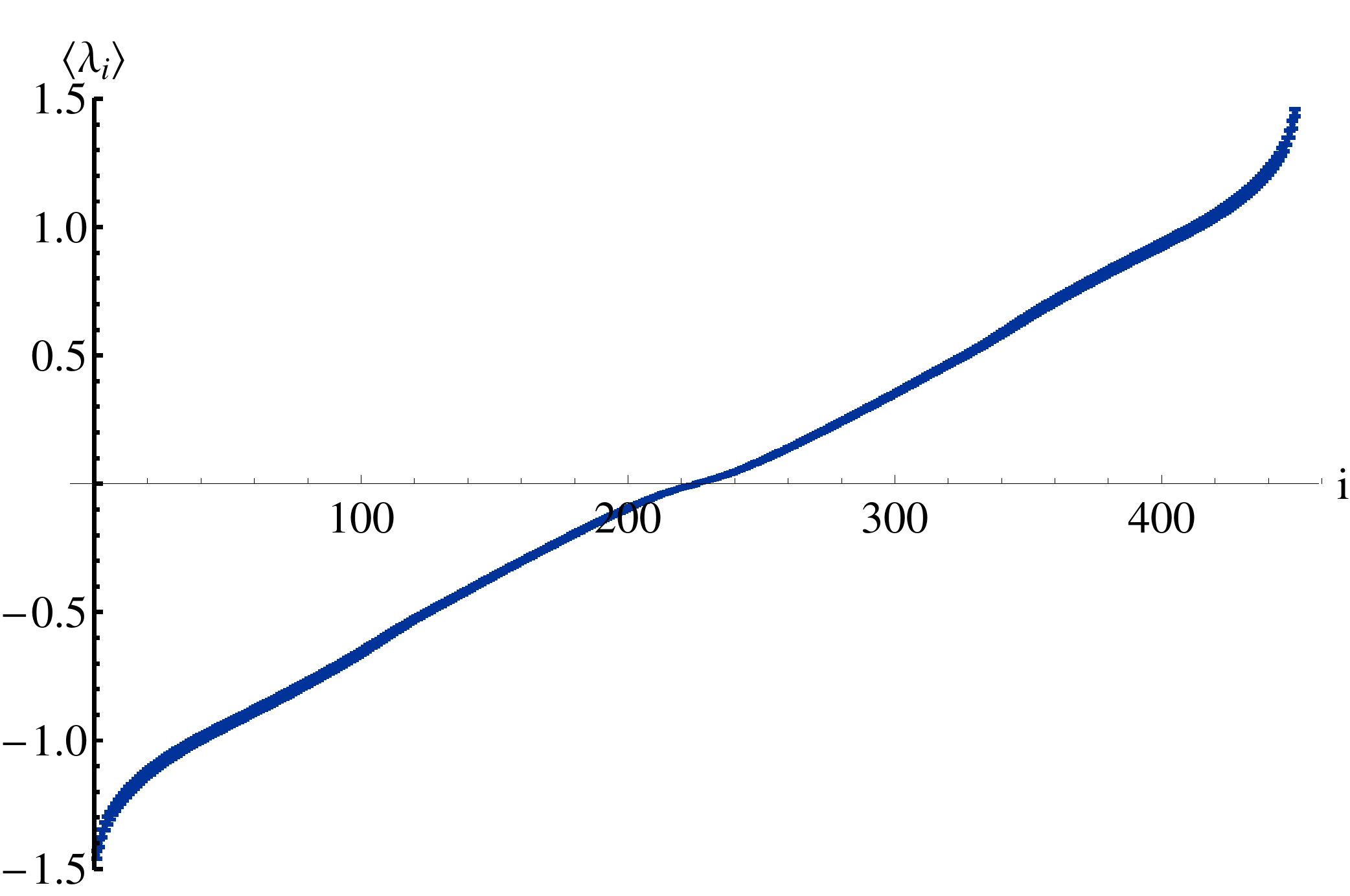}}
\subcaptionbox{Type $(1,1)$}{	\includegraphics[width=0.5\textwidth]{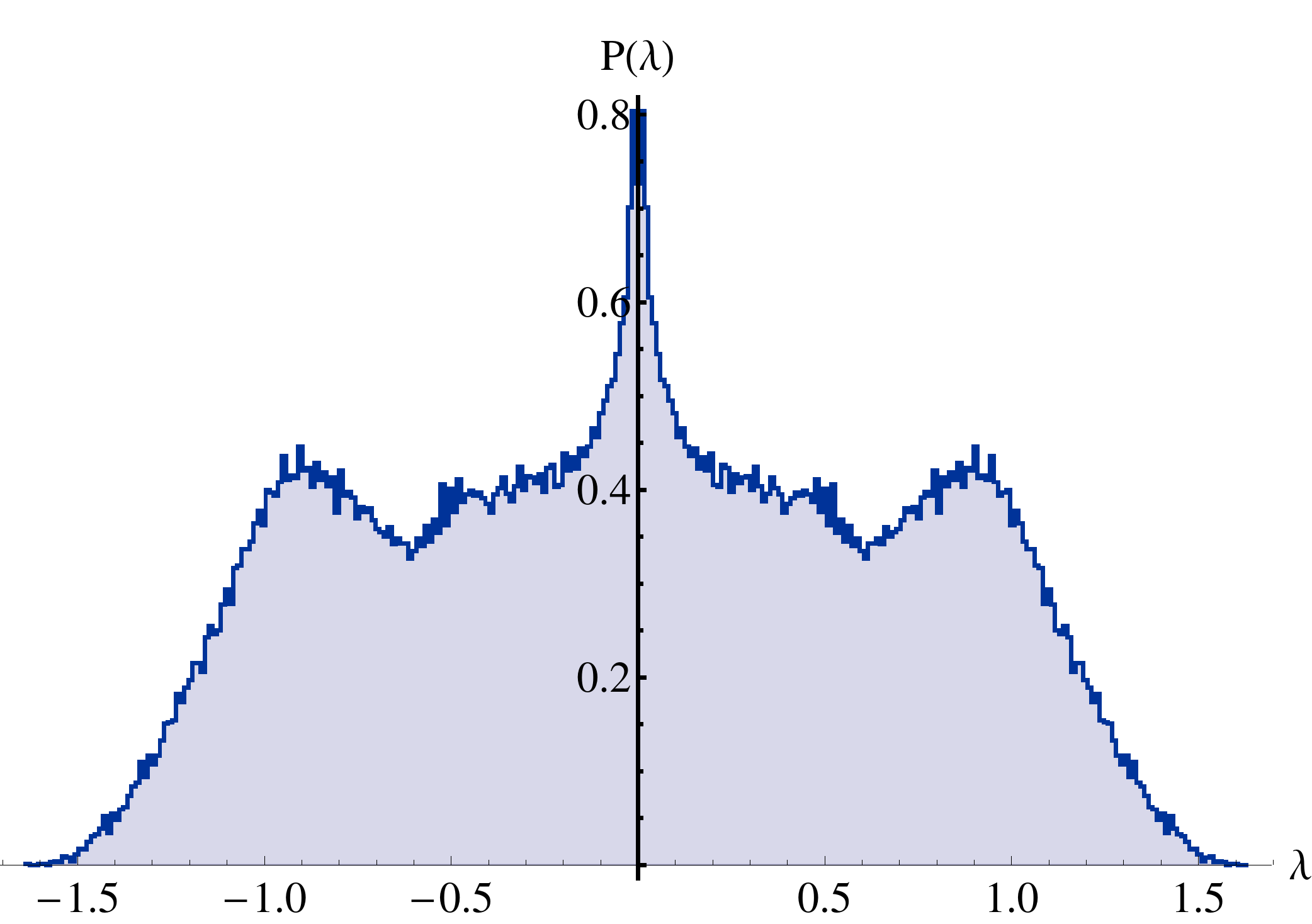}}
	
\subcaptionbox{Type $(0,2)$}{	\includegraphics[width=0.5\textwidth]{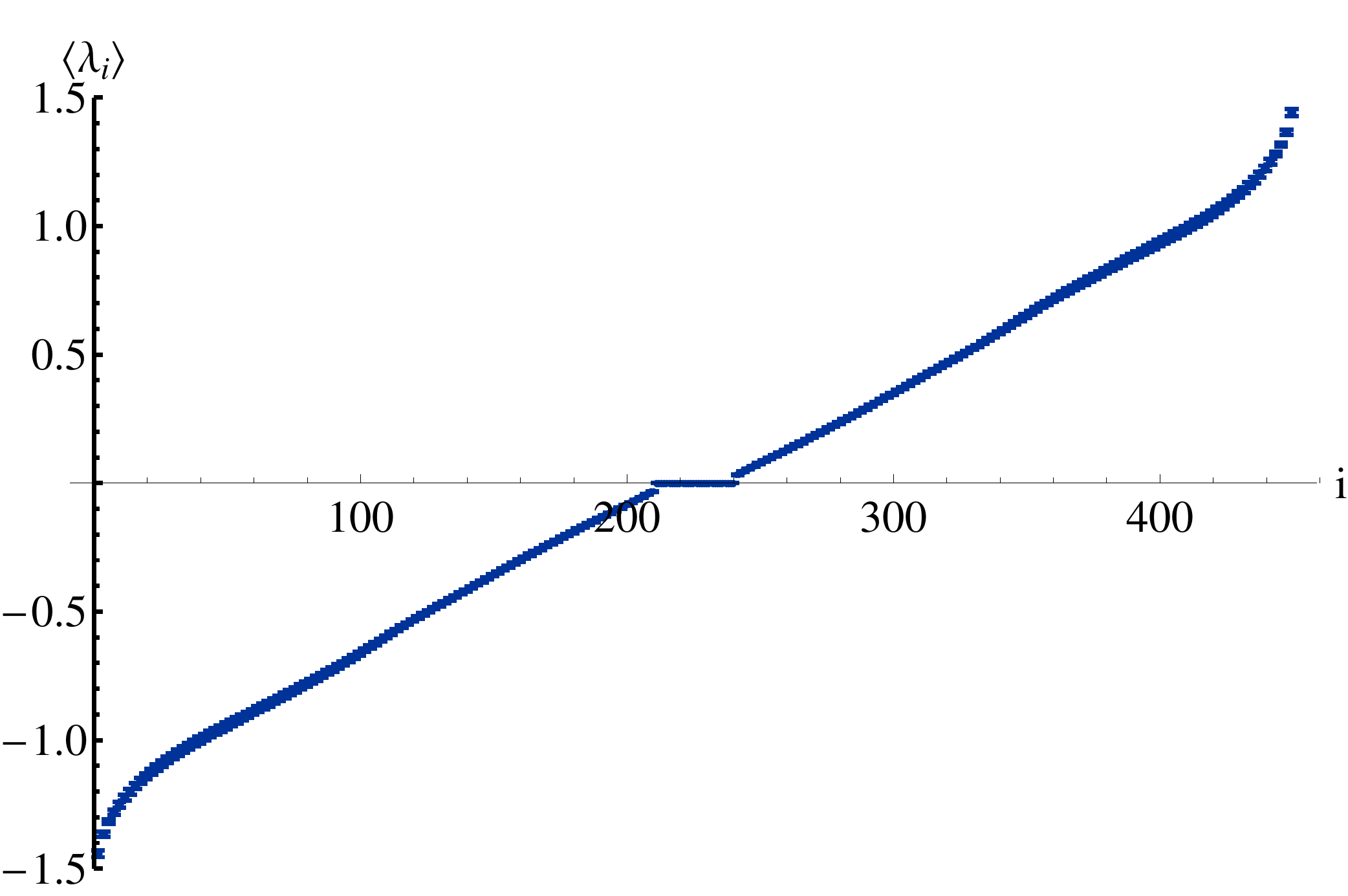}}
\subcaptionbox{Type $(0,2)$}{	\includegraphics[width=0.5\textwidth]{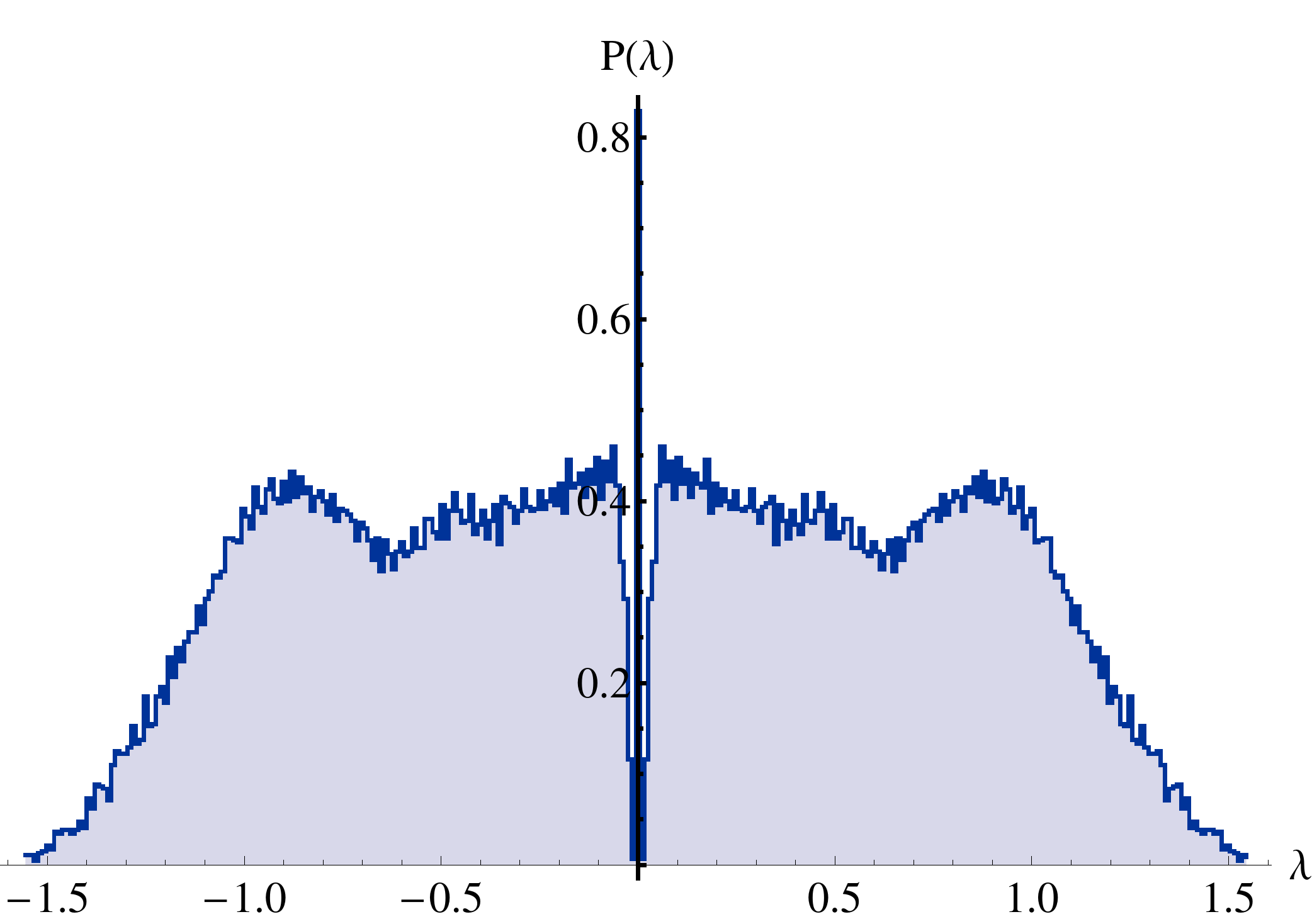}}

\caption{\label{fig:chiralspec15}The average eigenvalues, and the histograms of the eigenvalue distribution for the different types of two-dimensional Clifford algebra. The action is $S(D)=\tr{D^2}$ and the matrix size $n=15$.}
\end{figure}

While the spectra for geometries with one-dimensional Clifford algebra are easy to understand, those with a two-dimensional Clifford algebra are less straightforward. The average eigenvalues and the eigenvalue distributions are shown in figure  \ref{fig:chiralspec5} for the case $n=5$ and in figure \ref{fig:chiralspec15} for the larger matrices $n=15$. The individual eigenvalues are more easily seen in  figure  \ref{fig:chiralspec5}. All three types are symmetric about the origin and the third one exhibits eigenvalue doubling, all in accordance with the properties for $s=6$, $0$ and $2$
derived in section (\ref{sec:dirac}).

The action $\tr D^2$ is $2$ times the sum of quadratic actions for each matrix $L_i$ or $H_i$, these quadratic actions being exactly the $(0,1)$ and $(1,0)$ actions previously analysed. In particular, these matrices are statistically independent.   The eigenvalues of the $H_i$, $L_i$ are still approximated well by the semicircle law \eqref{eq:semicirc} with $A=1/2$. However, the main difference in analysing the two-dimensional cases is that the eigenvalues of the Dirac operator are not simply related to the eigenvalues of $L_i$, $H_i$.


For the case $(0,2)$ the multiplicity of eigenvalue $0$ is at least $2n$, as shown  by examining the Dirac operator 
\begin{align}
D^{(0,2)}	&=\g{1}\te \com{L_1}{ \cdot }+\g{2}\te \com{L_2}{ \cdot } \\ 
			&=\g{1}\left(\com{L_1}{ \cdot }-\g{1} \g{2}\te \com{L_2}{ \cdot }\right)\\
			&= \g{1} \begin{pmatrix}
					\com{L_1+i L_2}{ \cdot} & 0 \\
						0 &\com{L_1-i L_2}{ \cdot}  \\
						\end{pmatrix}\;,
\end{align}
using a basis so that $\g{1}\g{2}=\mathrm{diag}(i,-i)$.
The Dirac operator acts on the space $\C^2\te M(n,\mathbb{C})$. All $v\te m$ in this space for which 
\begin{align}
&\begin{pmatrix}
		\com{L_1+ i L_2}{\cdot} & 0 \\
		0 &\com{L_1- iL_2}{\cdot}  \\
\end{pmatrix}
\begin{pmatrix} 
		v_1 m \\ 
		v_2 m\\
\end{pmatrix}\\
=&
\begin{pmatrix}
	v_1	\com{L_1+i L_2}{m}  \\
		 v_2 \com{L_1-i L_2}{m}  \\
\end{pmatrix}
=
0
\end{align}						
have  eigenvalue $0$. 
Picking a basis on $v$ one can choose $v_1=1, v_2=0$ and $v_1=0, v_2=1$.
There will then be $n$ linearly independent matrices $m$ that commute with $L_1+i L_2$ and $n$ that commute with $L_1-i L_2$, hence $2n$ eigenvalues equal to $0$ for the $(0,2)$ type geometry. 
Just as in the $(0,1)$ case, there is a gap in the eigenvalue spectrum around the spike at $0$. This shows there is  eigenvalue repulsion for this Dirac operator also, though we do not have a theoretical understanding of this phenomenon.

\begin{figure}
\subcaptionbox{Average eigenvalues}{\includegraphics[width=0.5\textwidth]{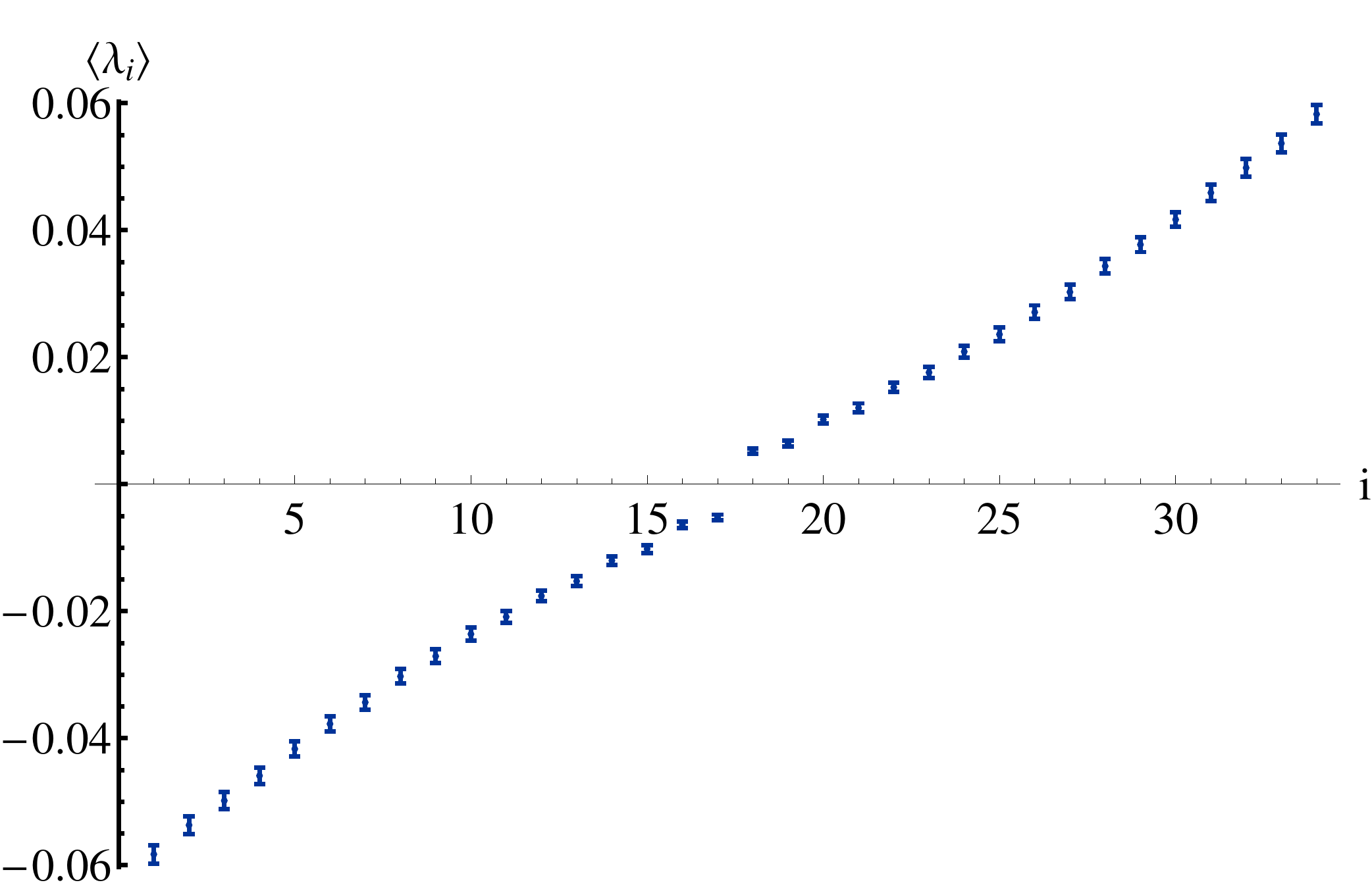}}
\subcaptionbox{Eigenvalue distribution}{\includegraphics[width=0.5\textwidth]{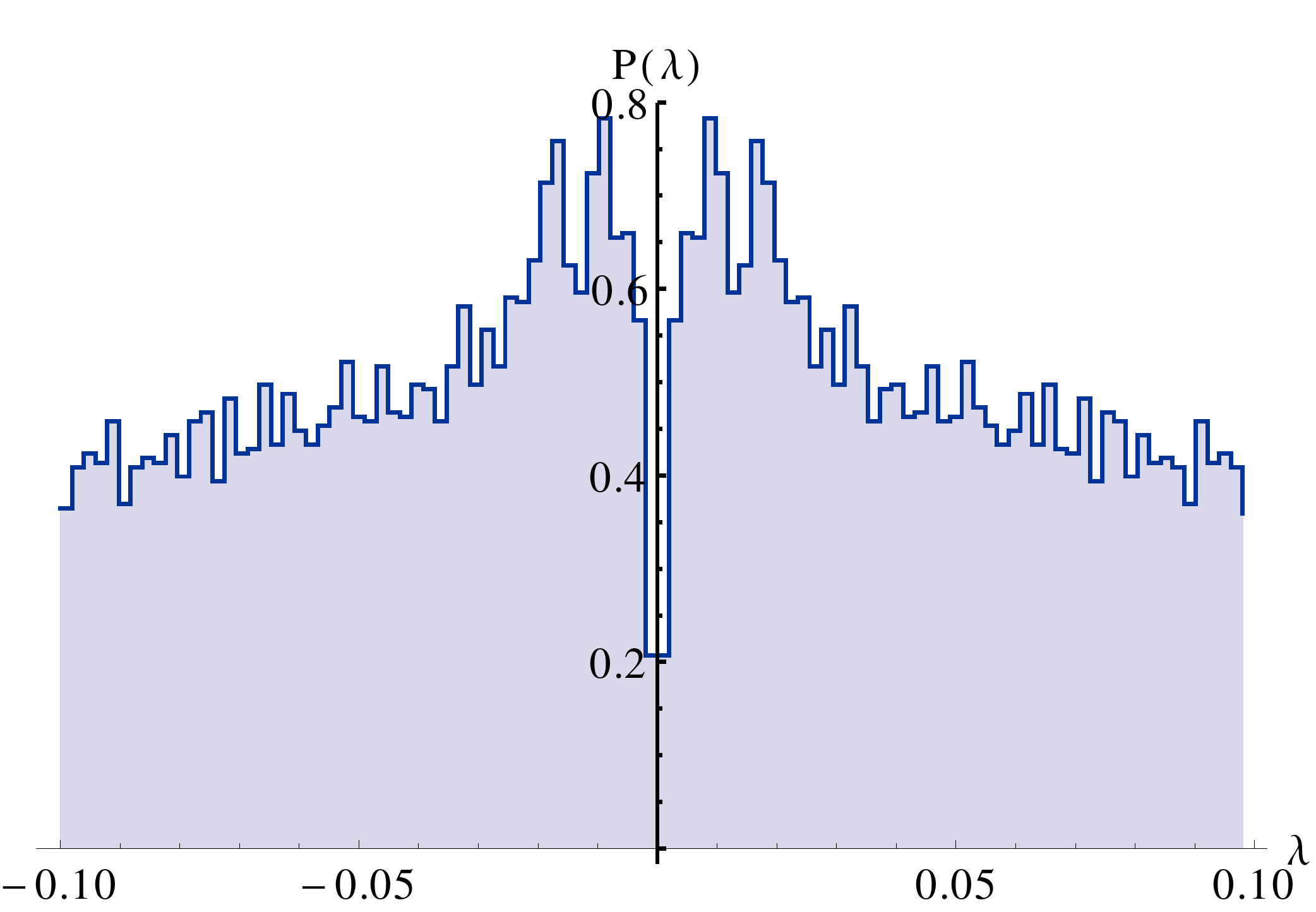}}
\caption{\label{fig:zoom20}Zooming in to a region near eigenvalue $0$. Type $(2,0)$, $n=15$. }
\end{figure}

The types $(2,0)$ and $(1,1)$ also have a feature at the origin. The density of eigenvalues is sharply lower in a narrow dip at the origin and there is a somewhat wider upward spike around this. This is shown for the $(2,0)$ case in figure \ref{fig:zoom20}, which zooms in on a region around eigenvalue $0$.
The gap in the middle is further evidence of eigenvalue repulsion, this time between eigenvalue $\lambda$ and the opposite eigenvalue $-\lambda$ that is required by the symmetry of the spectrum of $D$ about $0$.


\begin{figure}
\subcaptionbox{Type $(2,0)$ $n=5$}{\includegraphics[width=0.3\textwidth]{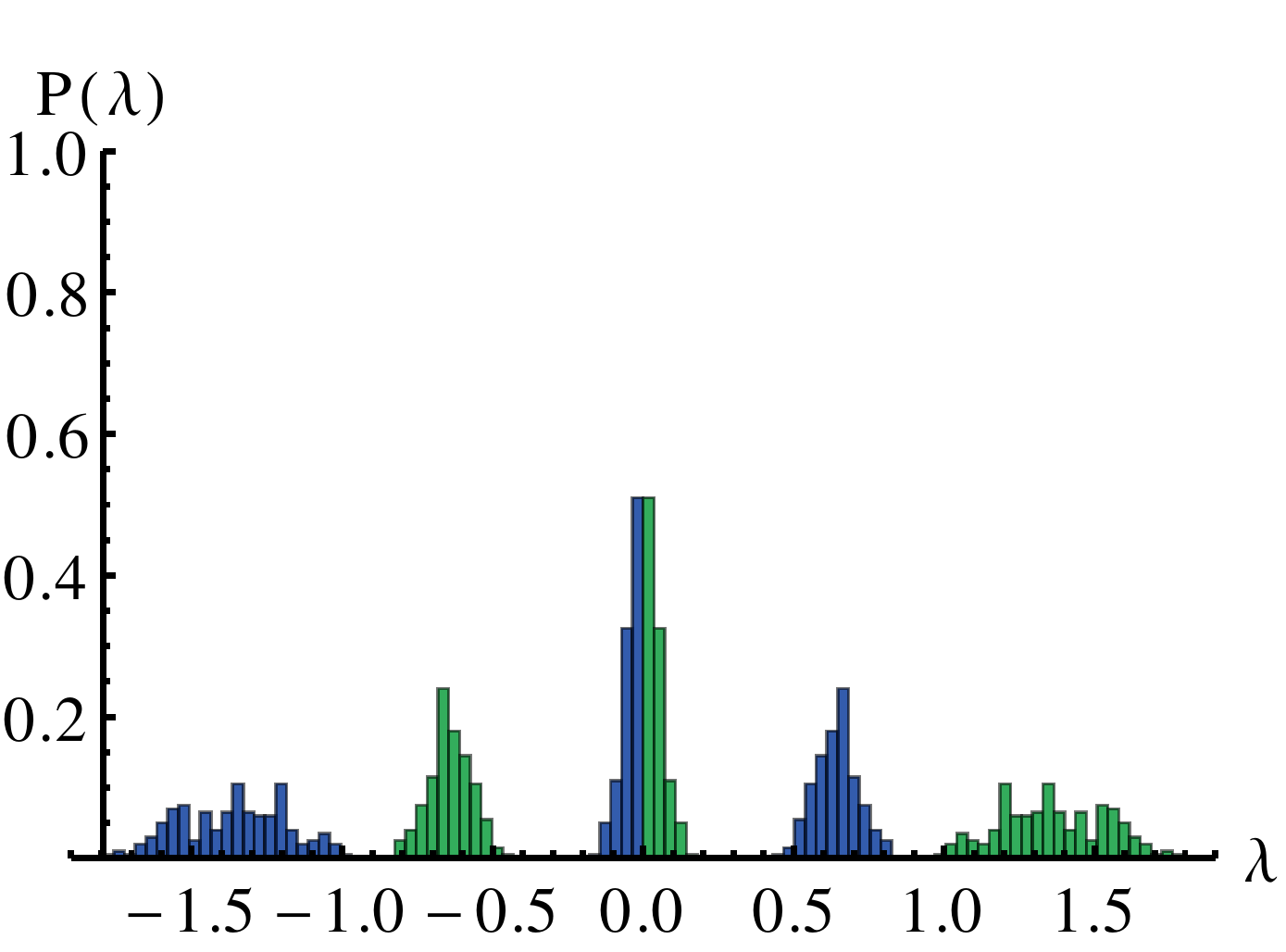}}
\subcaptionbox{Type $(2,0)$ $n=10$}{\includegraphics[width=0.3\textwidth]{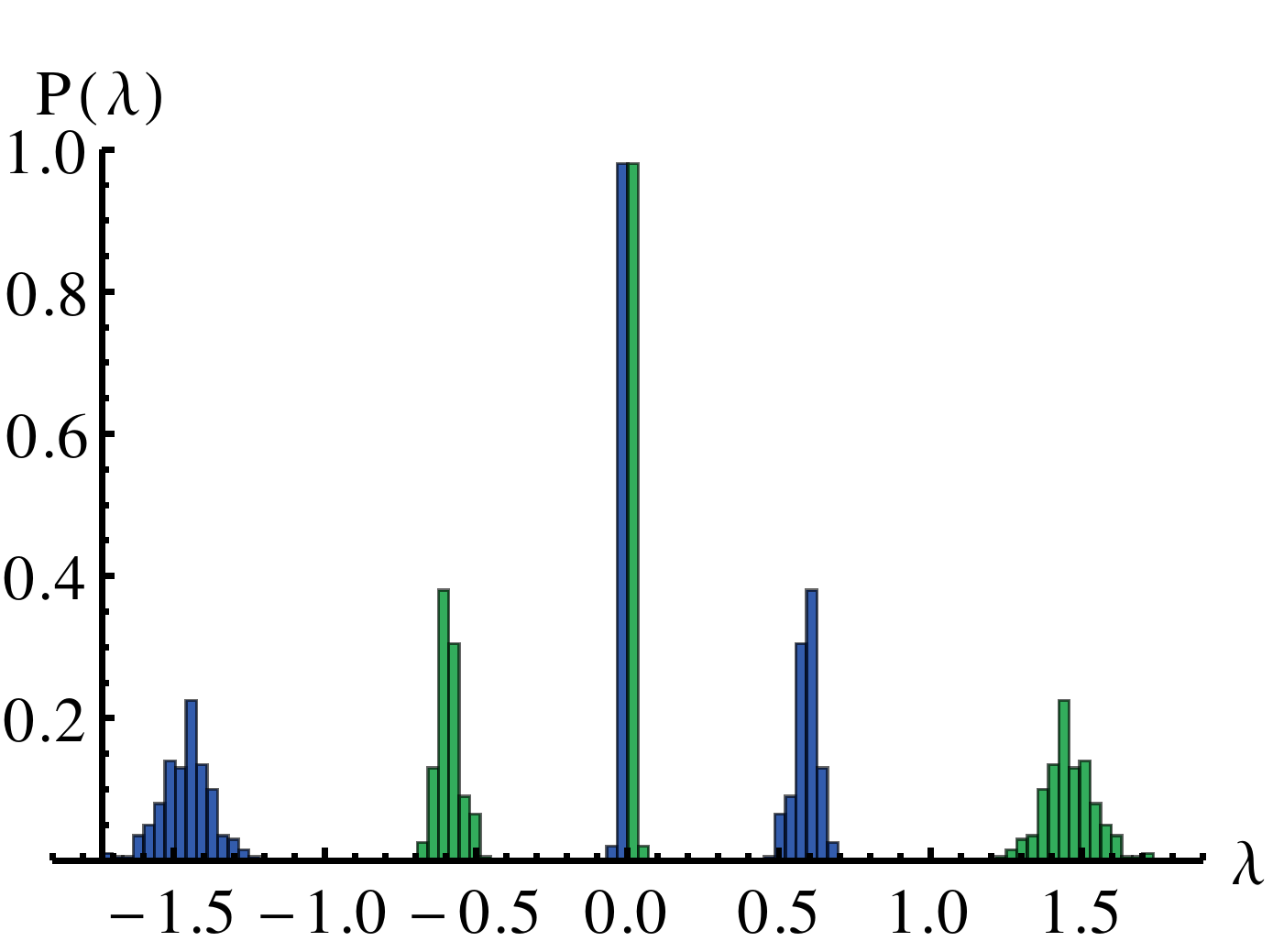}}
\subcaptionbox{Type $(2,0)$ $n=15$}{\includegraphics[width=0.3\textwidth]{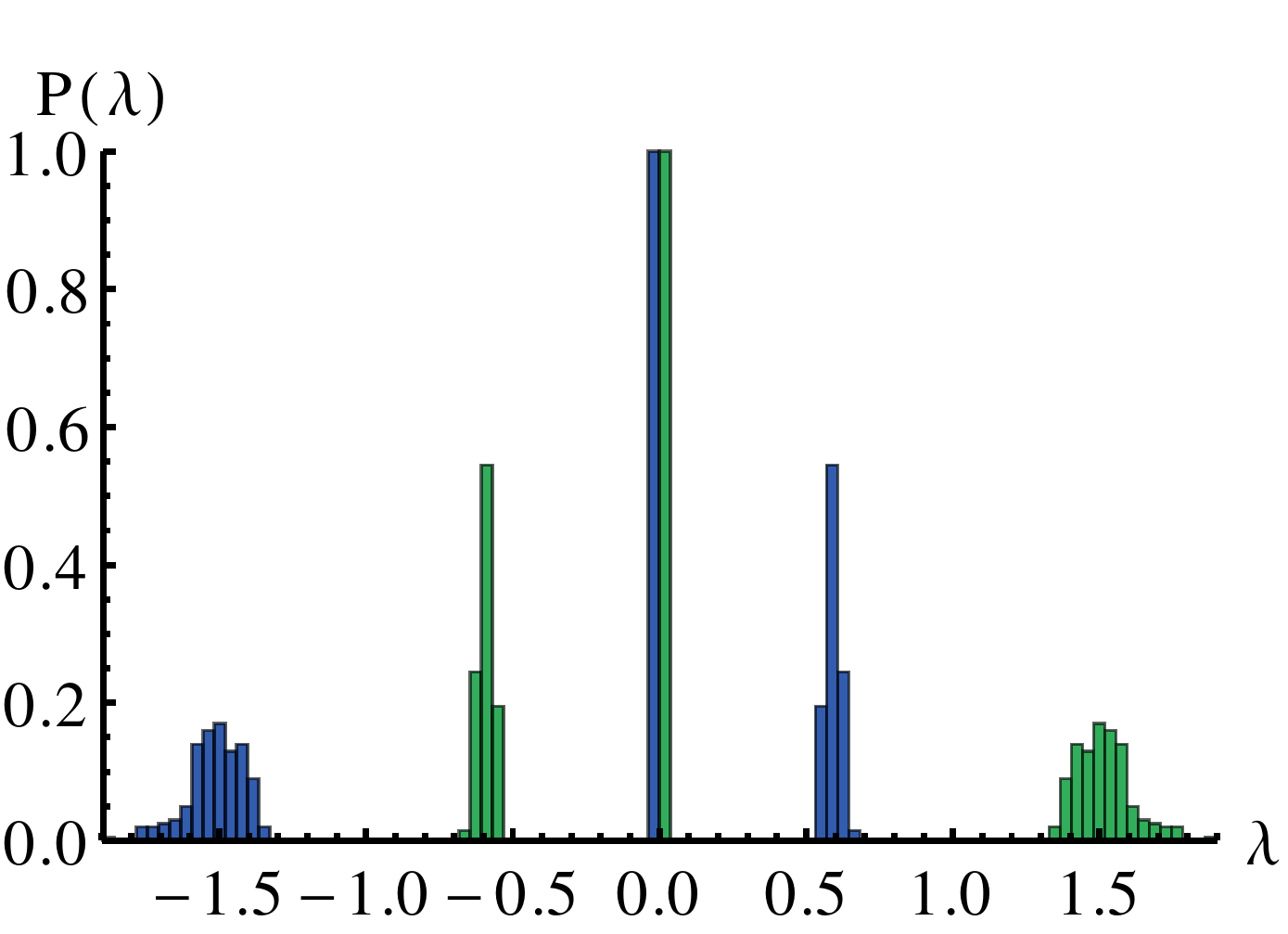}}

\subcaptionbox{Type $(0,2)$ $n=5$}{\includegraphics[width=0.3\textwidth]{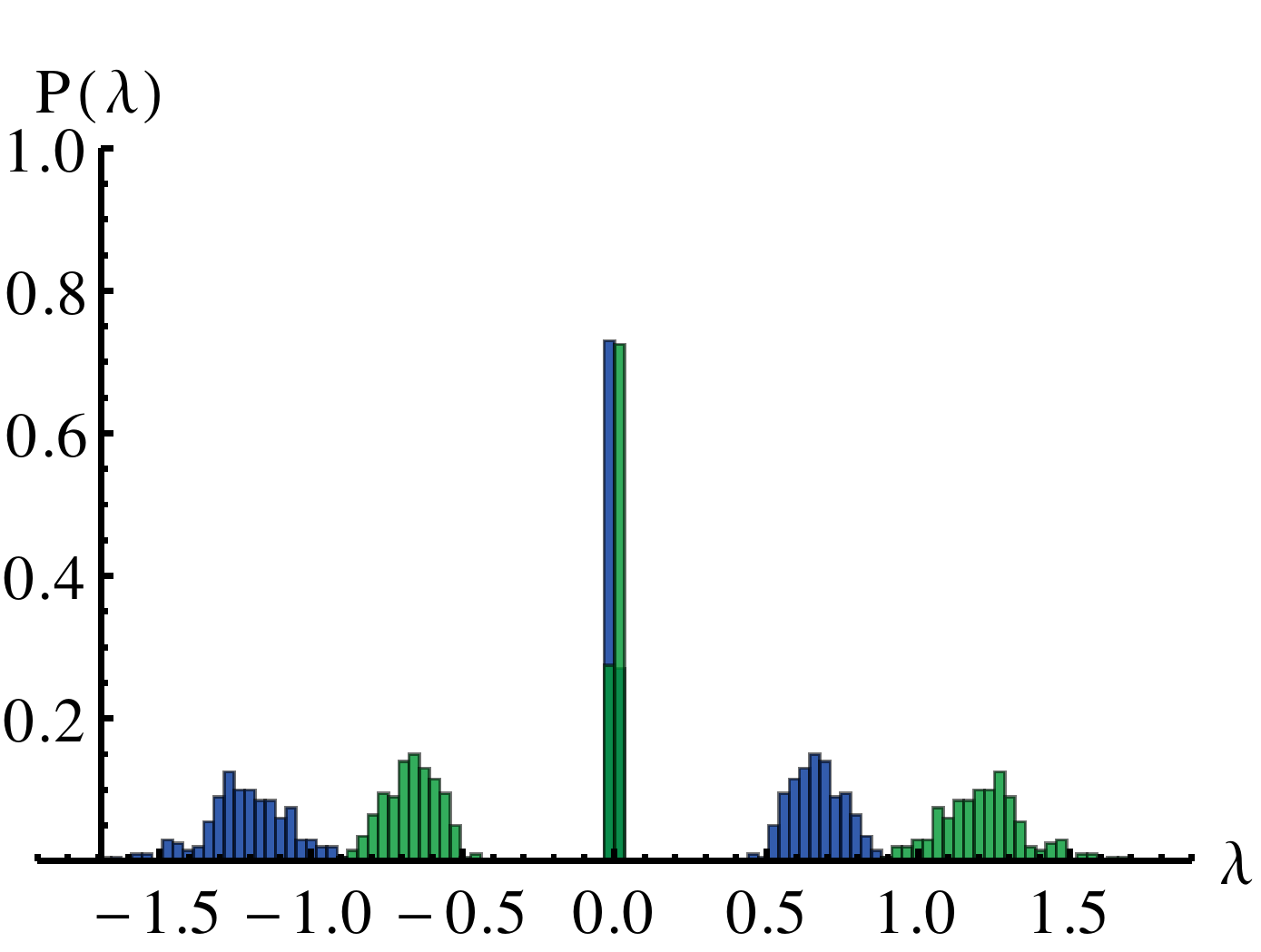}}
\subcaptionbox{Type $(0,2)$ $n=10$}{\includegraphics[width=0.3\textwidth]{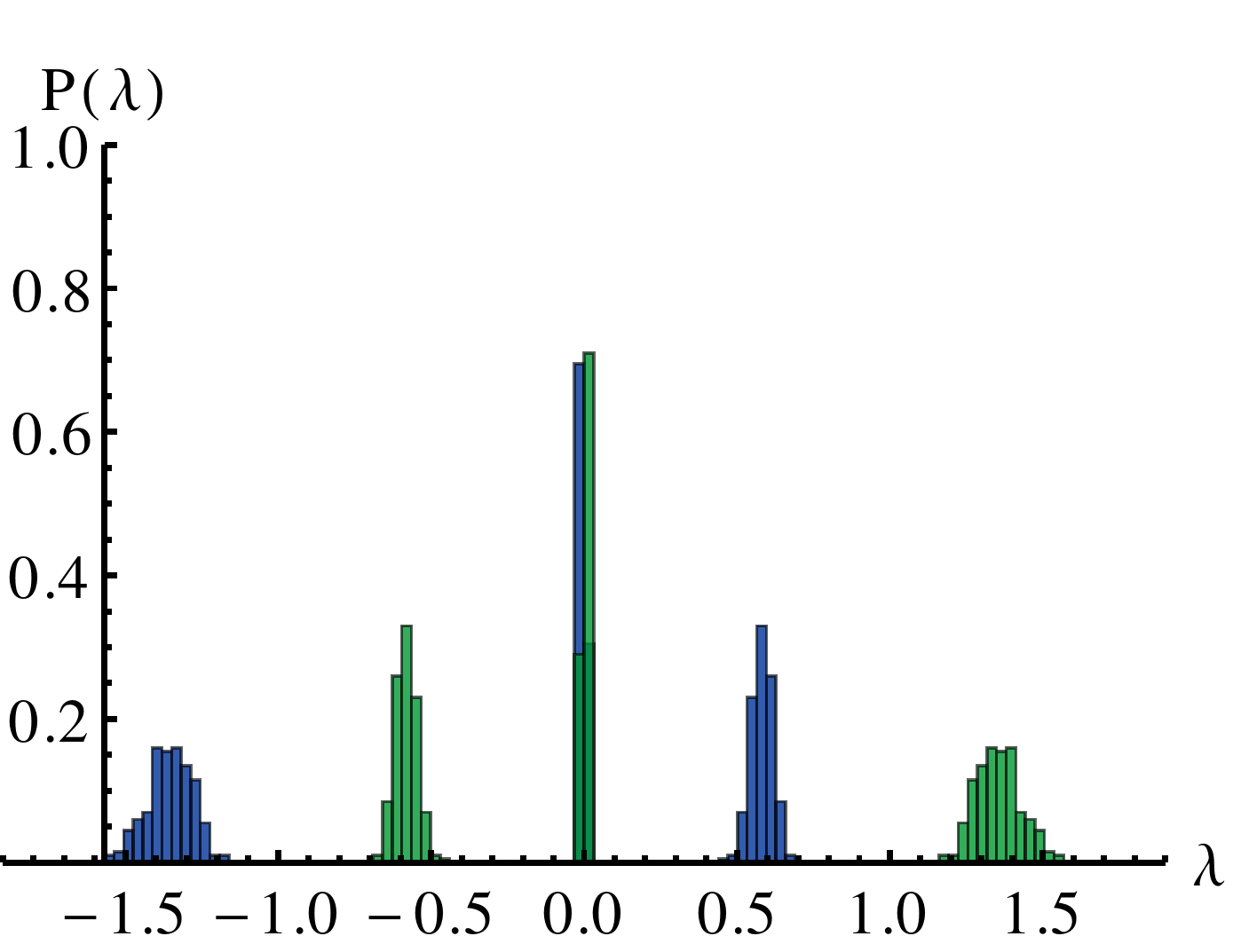}}
\subcaptionbox{Type $(0,2)$ $n=15$}{\includegraphics[width=0.3\textwidth]{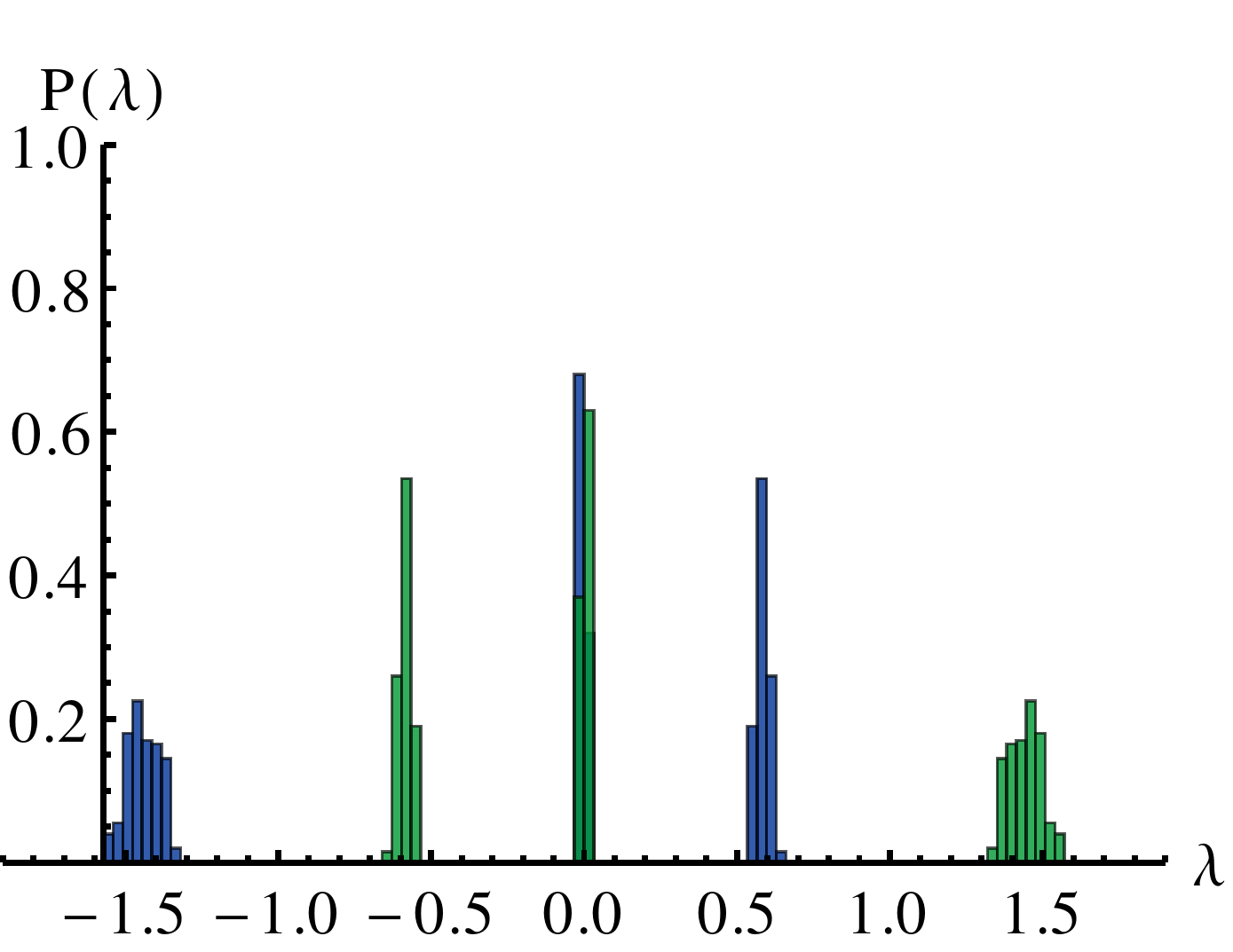}}
\caption{\label{fig:SingleEV}The distribution of single eigenvalues at different sizes.}
\end{figure}

The numerical results also indicate that the range of the eigenvalues remains unchanged under the change of matrix size, and the distribution becomes smoother, appearing to converge to a smooth limiting distribution in the same way as for random matrices. Another similar feature is that for larger $n$ the fluctuation of each individual eigenvalue becomes smaller.
This can be seen in figure \ref{fig:SingleEV}.
The leftmost bump in each plot is the smallest eigenvalue while the rightmost bump is the largest, the eigenvalues in between were chosen to be symmetric, and include the central most eigenvalues.

\begin{figure}
\subcaptionbox{Type $(0,3)$}{\includegraphics[width=0.5\textwidth]{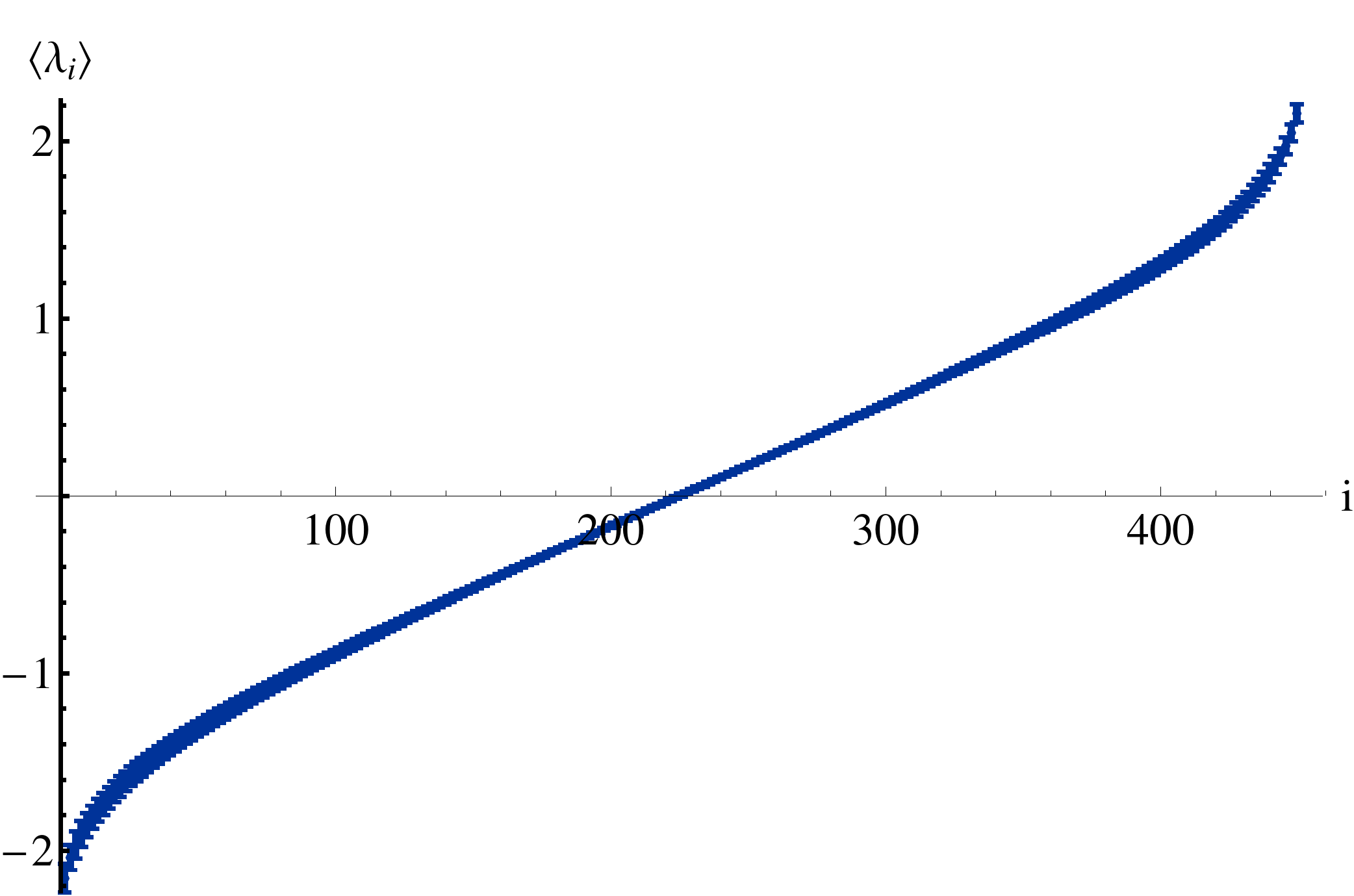}	}
\subcaptionbox{Type $(0,3)$}{	\includegraphics[width=0.5\textwidth]{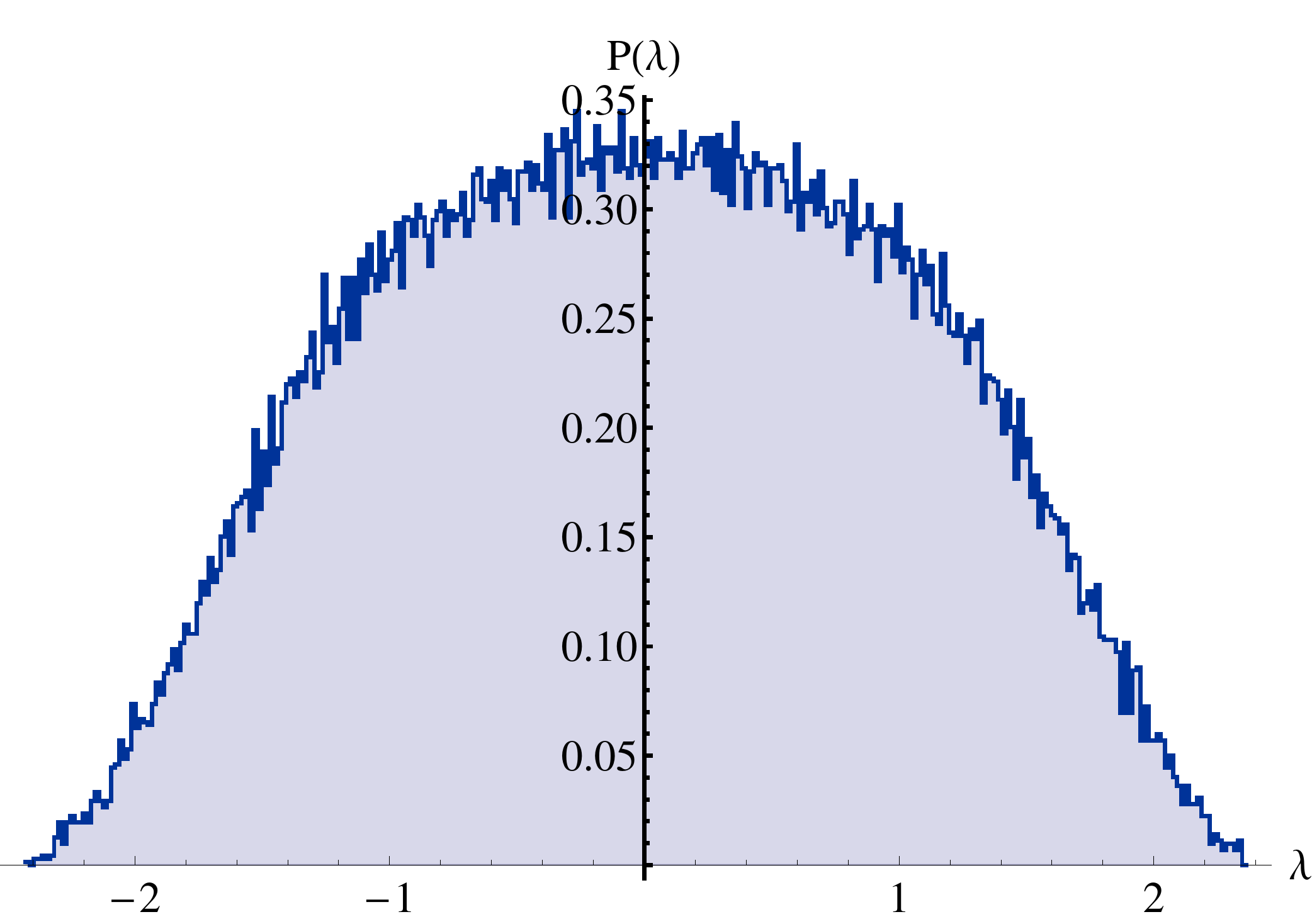}	}
\caption{\label{fig:evdoublespec} The average eigenvalues and the eigenvalue distribution for type $(0,3)$. The action is $S=\tr{D^2}$  and the matrix size $n=15$.}
\end{figure}

The eigenvalue distribution for the type $(0,3)$ case is plotted in figure \ref{fig:evdoublespec}. This appears to be smooth at the origin, similar to the $(1,0)$ case. The common property of these cases is that the Dirac operators do not have a symmetric spectrum. Thus a small eigenvalue $\lambda$ does not have to be close to any other eigenvalue. There is nothing special about the origin, and in particular, the eigenvalue repulsion hypothesis does not lead to any special behaviour here.

\section{Results for actions with $D^4$ term}\label{sec:D4}
The $\tr{D^4}$ term in the action leads to interactions between the $L_i, H_i$ that compose the Dirac operator.
An extreme case of this is for type $(0,3)$, in which a four point interaction of all four matrices $H$, $L_1$, $L_2$ and $L_3$ is present. These terms make it harder to understand the system analytically, however for the simulations they are no obstacle.

\begin{figure}
\subcaptionbox{$(1,0)$ $n=15$}{\includegraphics[width=0.5\textwidth]{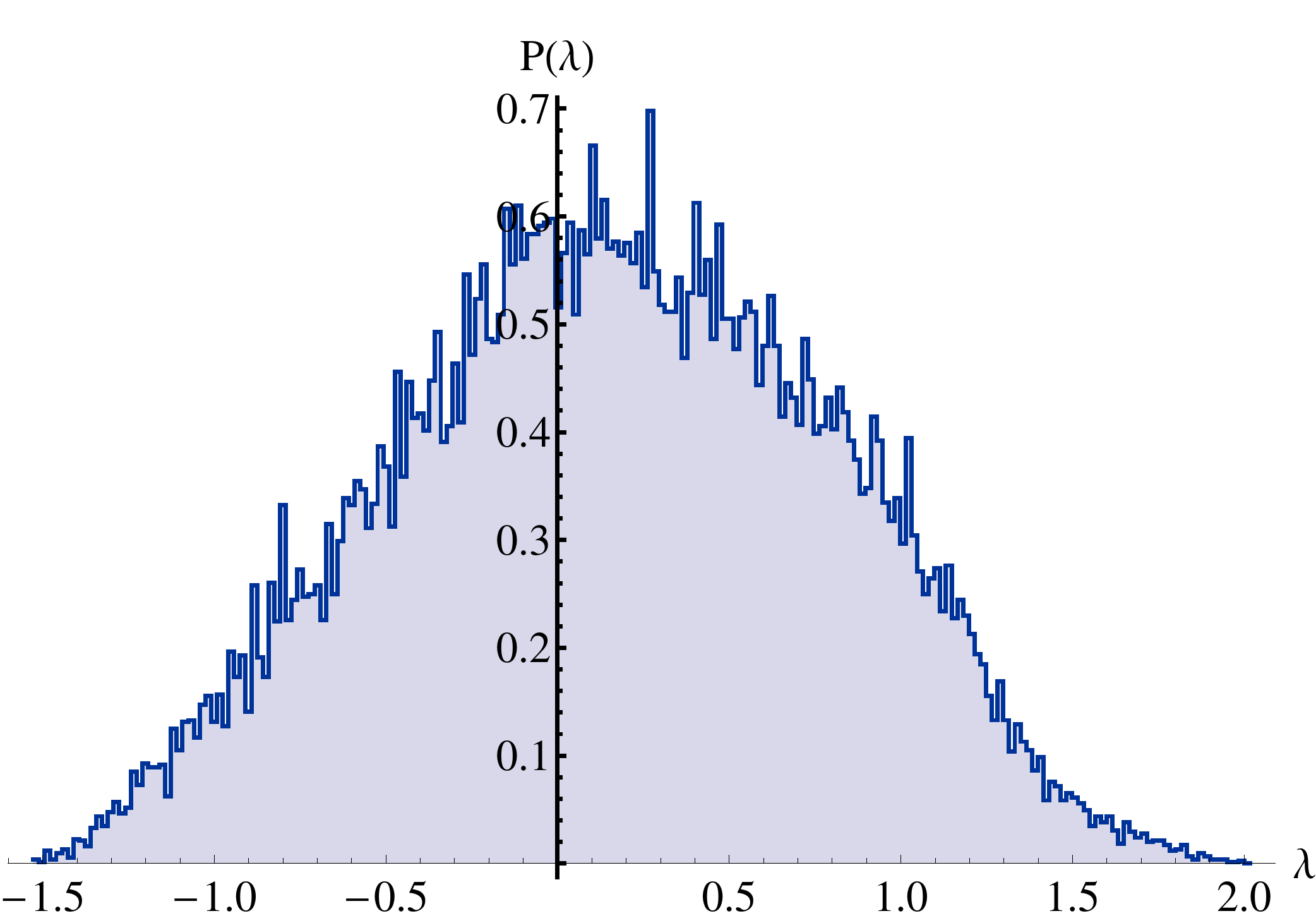}}
\subcaptionbox{$(0,1)$ $n=15$}{\includegraphics[width=0.5\textwidth]{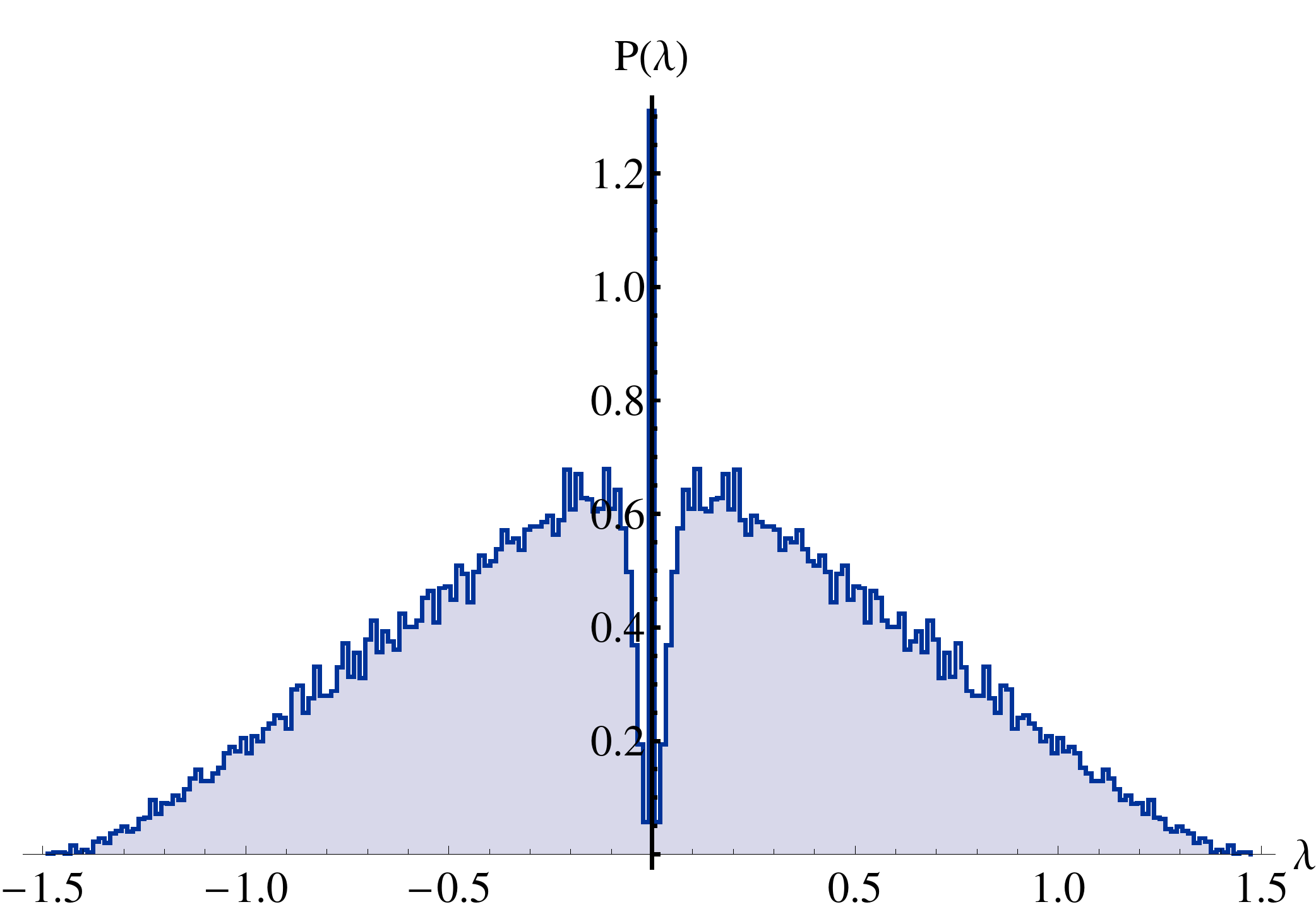}}

\subcaptionbox{$(2,0)$ $n=15$}{\includegraphics[width=0.5\textwidth]{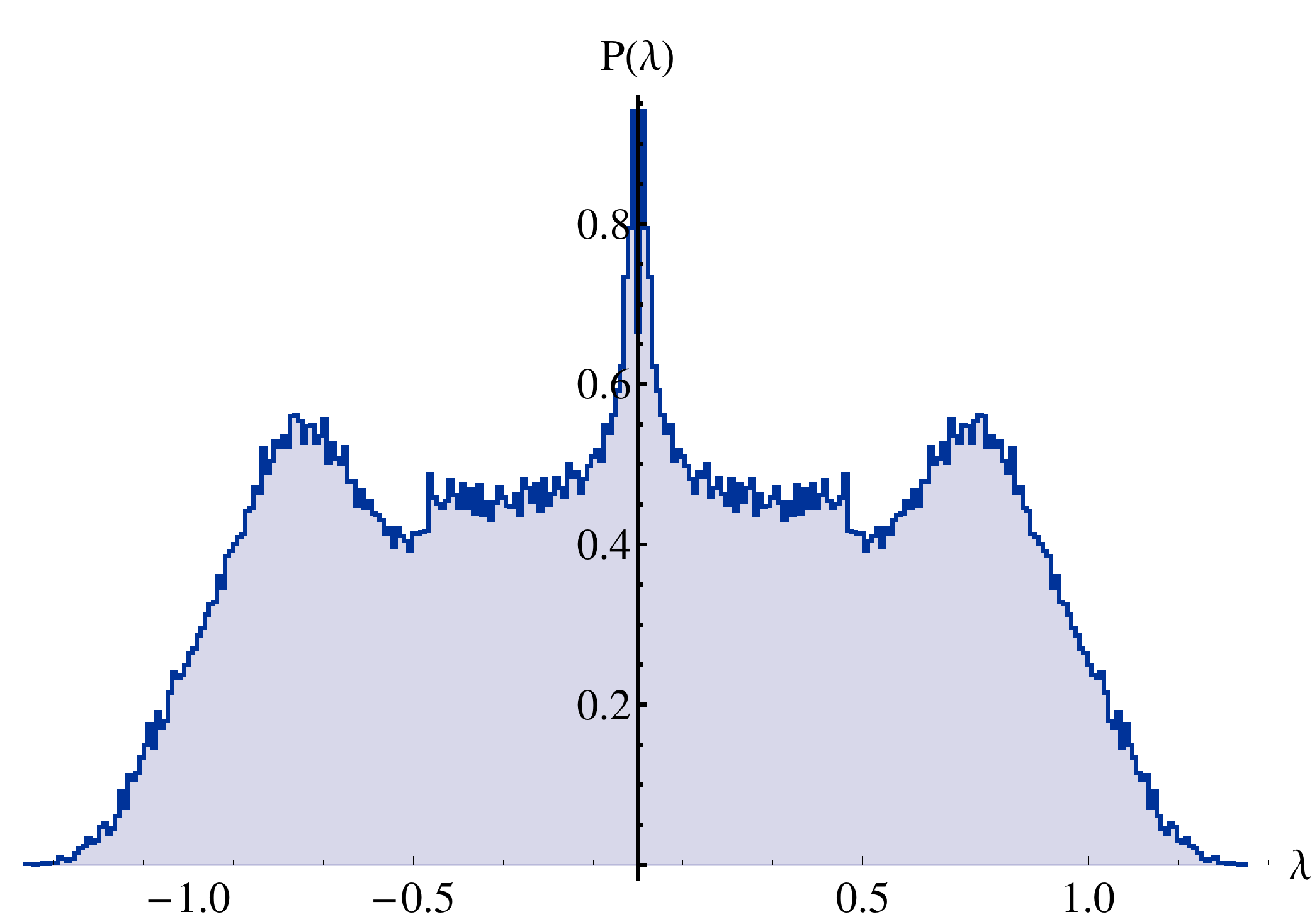}}
\subcaptionbox{$(1,1)$ $n=15$}{\includegraphics[width=0.5\textwidth]{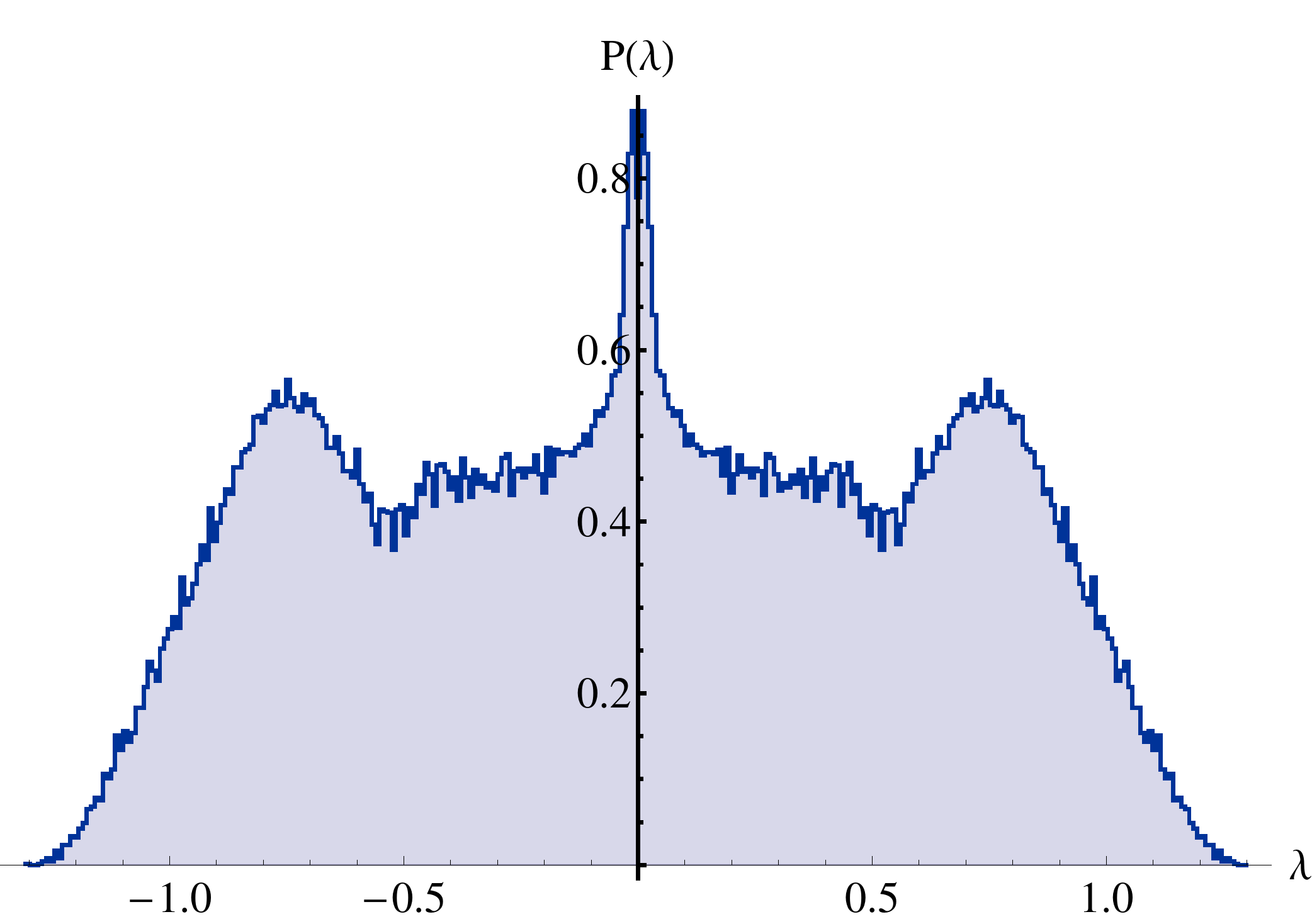}}

\subcaptionbox{$(0,2)$ $n=15$}{\includegraphics[width=0.5\textwidth]{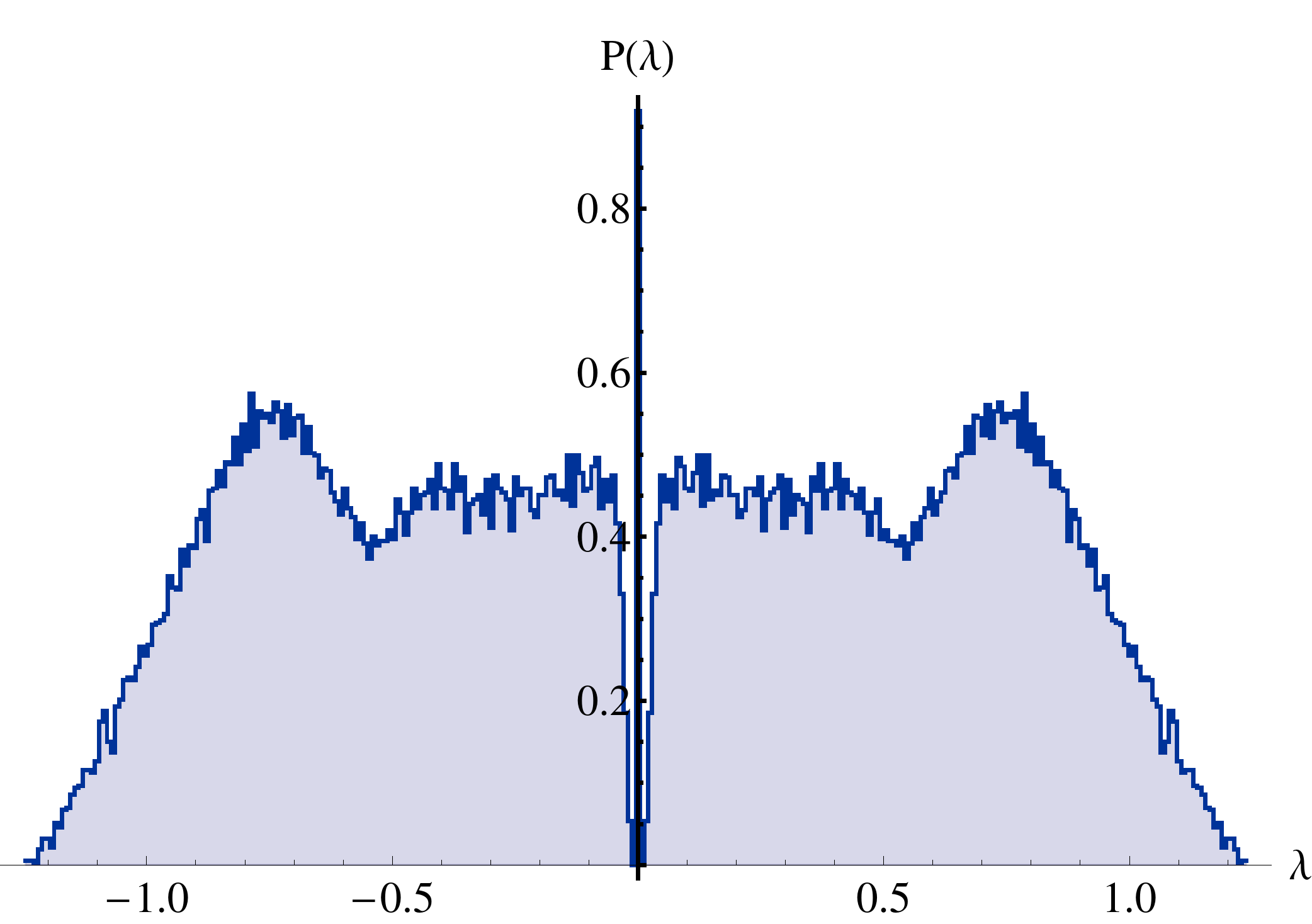}}
\subcaptionbox{$(0,3)$ $n=15$}{\includegraphics[width=0.5\textwidth]{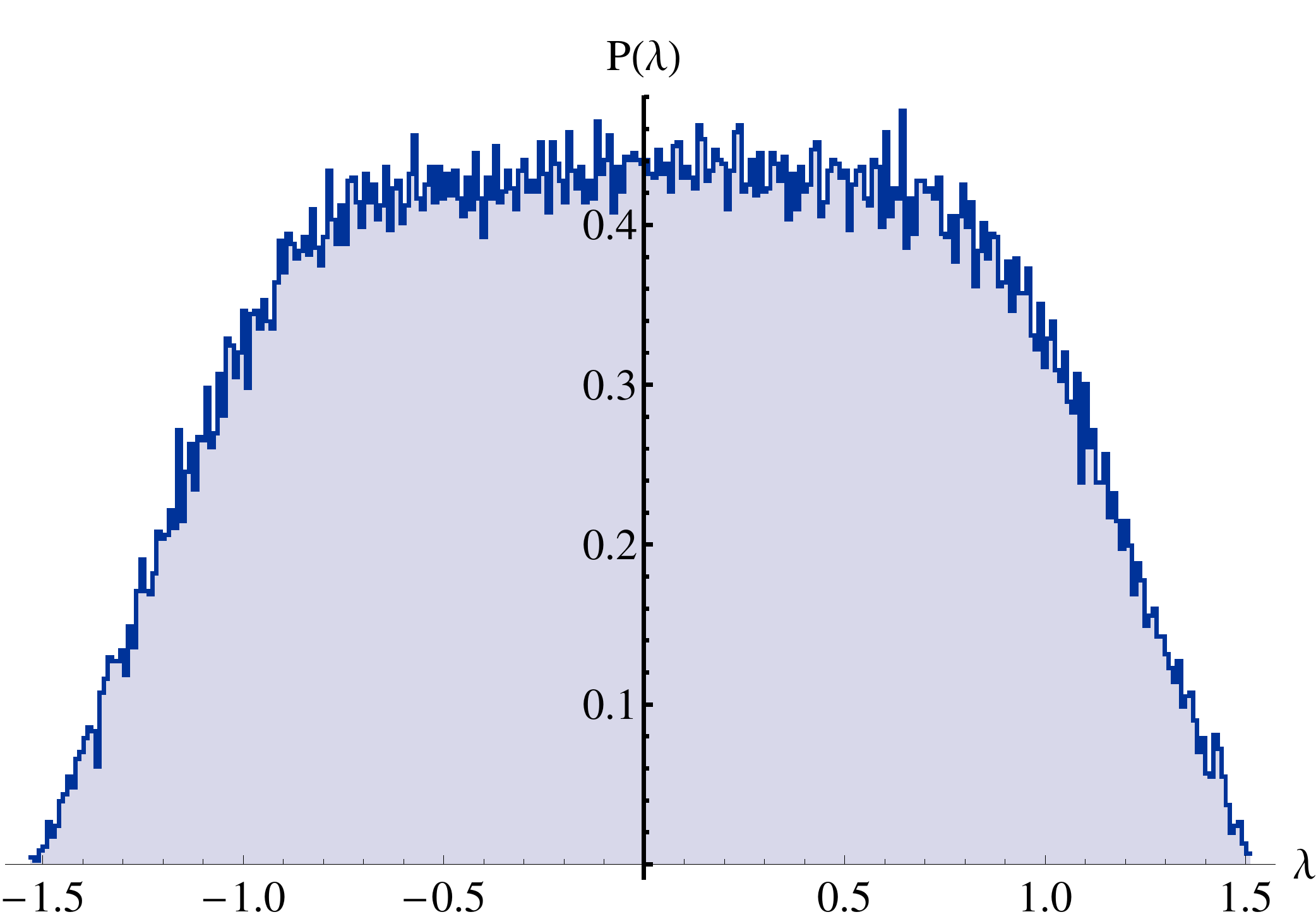}}
\caption{\label{fig:TrD4} The eigenvalue distribution for the action $S=\tr{D^4}$.}
\end{figure}

The simple action $S(D)=\tr D^4$ leads to behavior very similar to that for the action $\tr{D^2}$.
This is shown in figure \ref{fig:TrD4}. Some characteristics, like the shoulders for type $(1,1)$ and $(2,0)$ are more pronounced, but the overall shape is quite similar.

\begin{figure}
\centering{\includegraphics[width=0.5\textwidth]{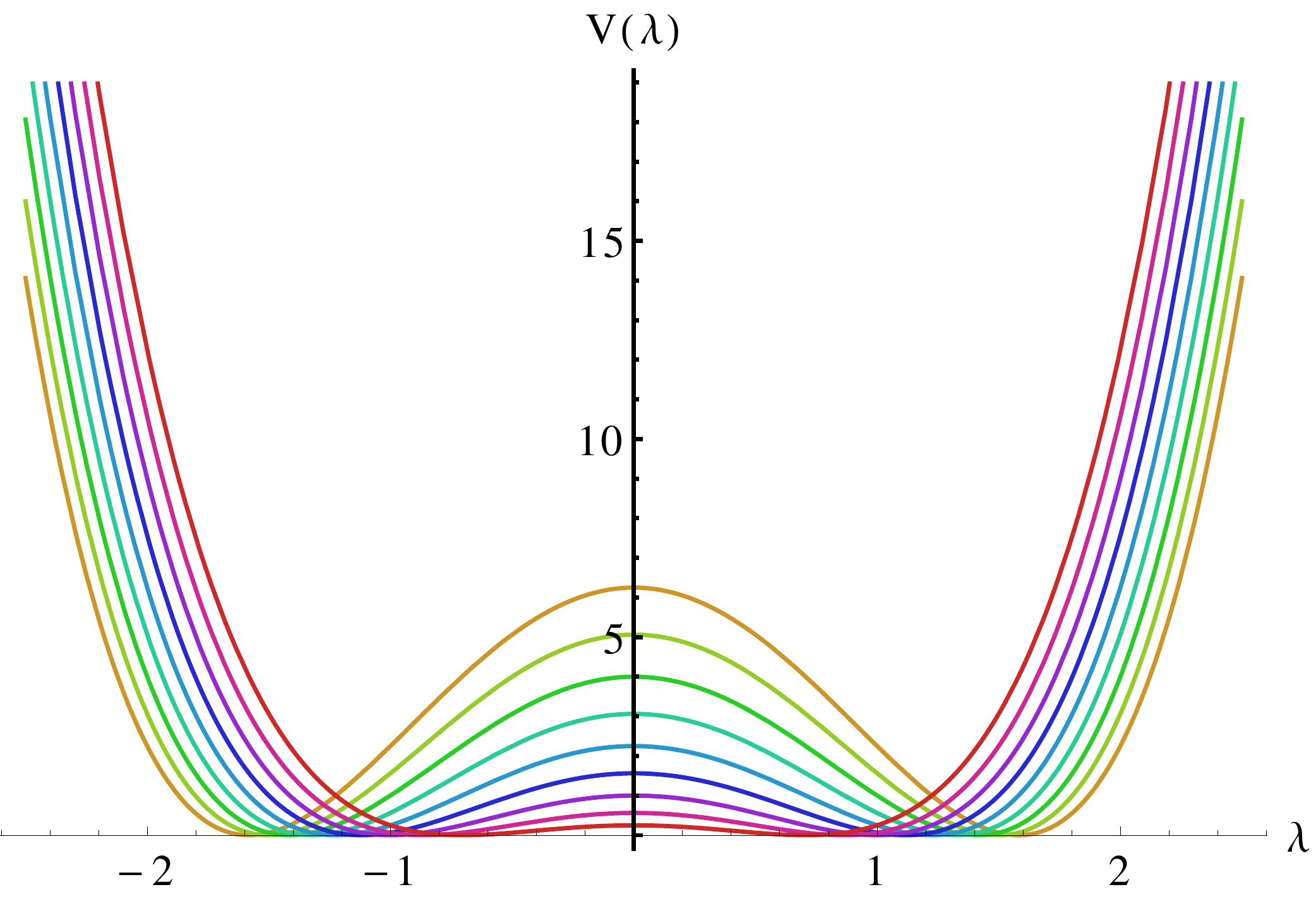}}
\caption{\label{fig:potentials} The potential $V=\lambda^4+g_2\lambda^2$ for $g_2=-1$,  $-1.5$,  $-2$,  $-2.5$,  $-3$,  $-3.5$,  $-4$,  $-4.5$,  $-5$. The lines are coloured from red ($g_2=-1$) through to yellow ($g_2=-5$). }
\end{figure}

\begin{figure}
\subcaptionbox{Type $(1,0)$}{\includegraphics[width=0.5\textwidth]{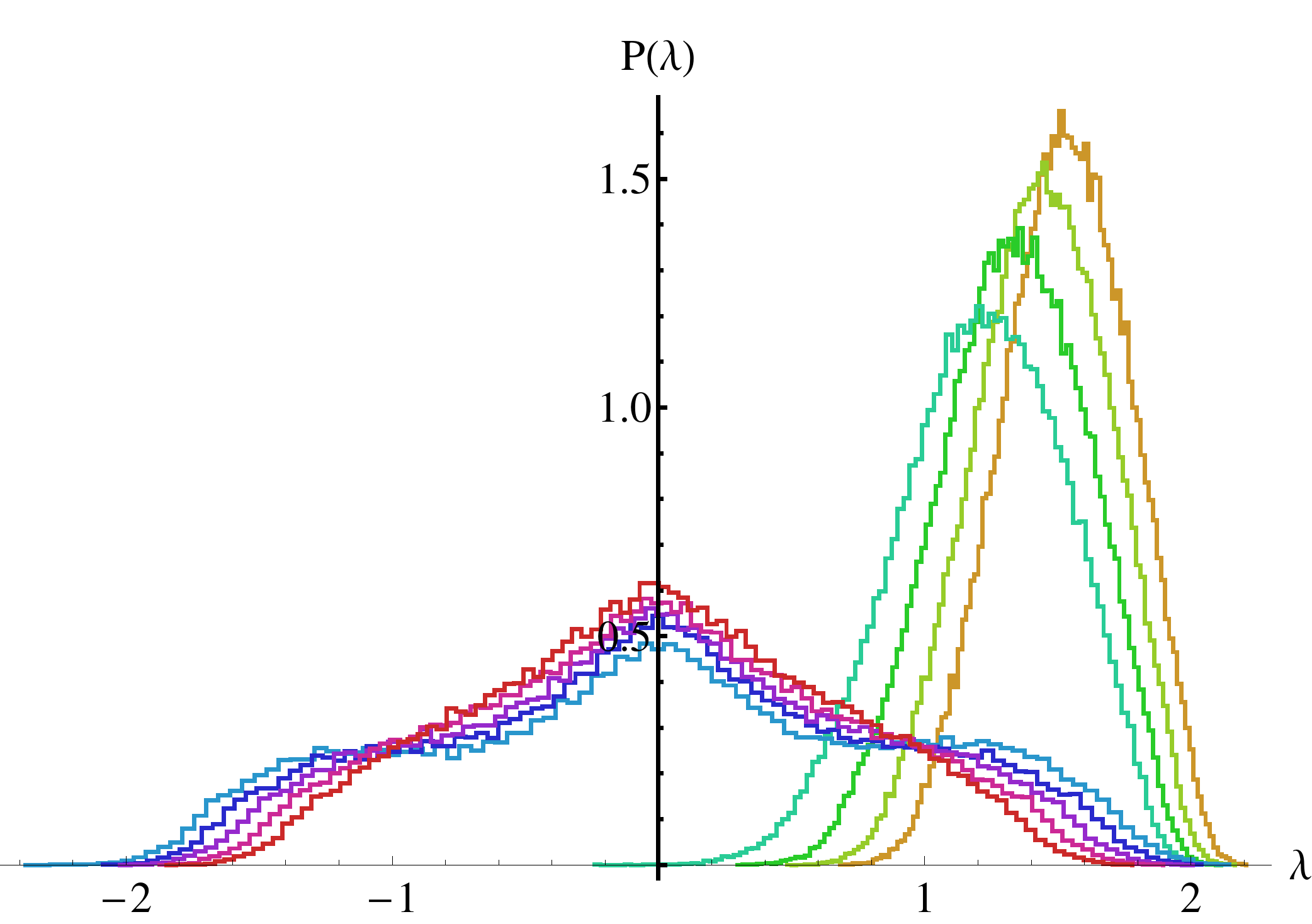}}
\subcaptionbox{Type $(0,1)$}{\includegraphics[width=0.5\textwidth]{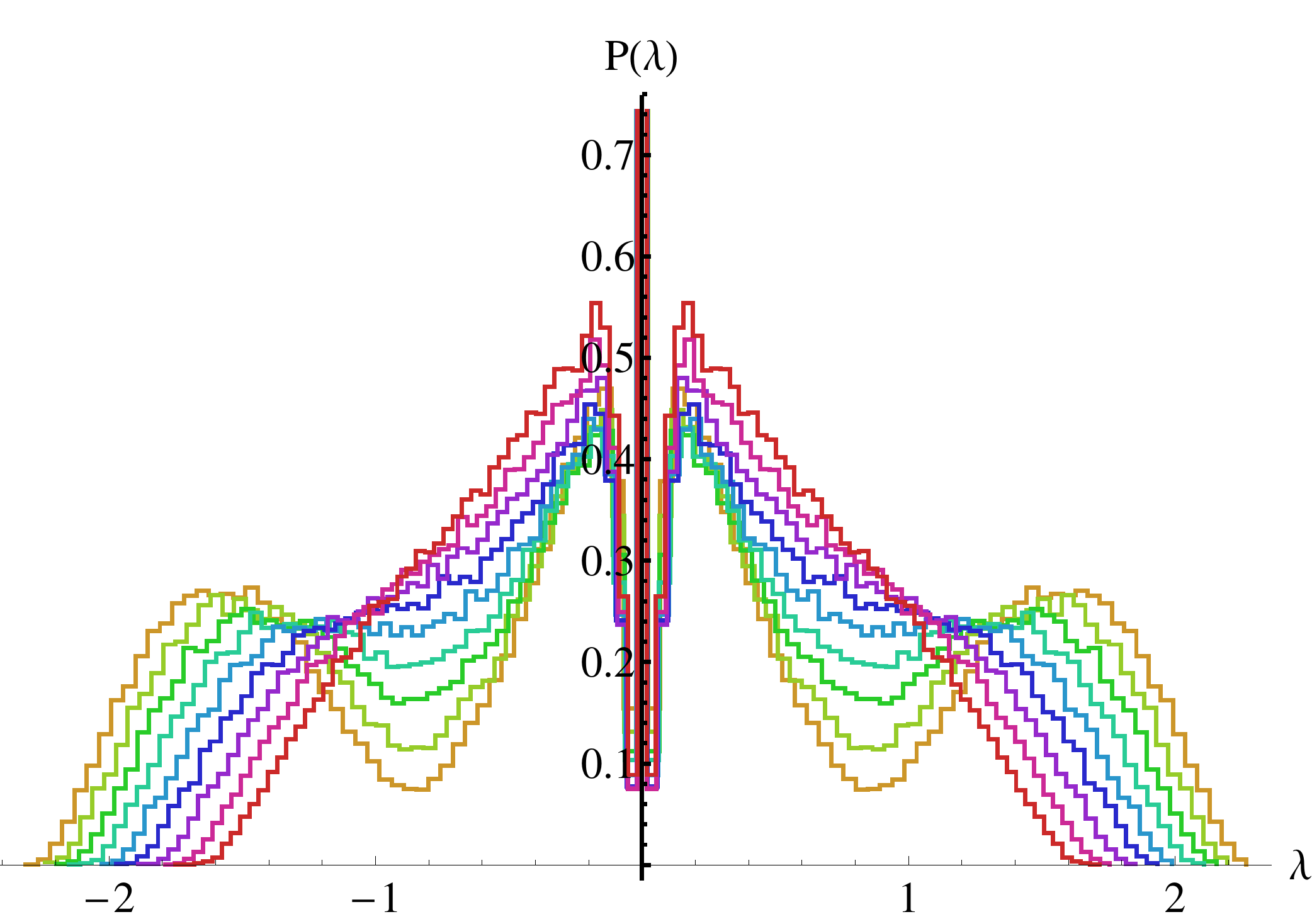}}

\subcaptionbox{Type $(2,0)$}{\includegraphics[width=0.5\textwidth]{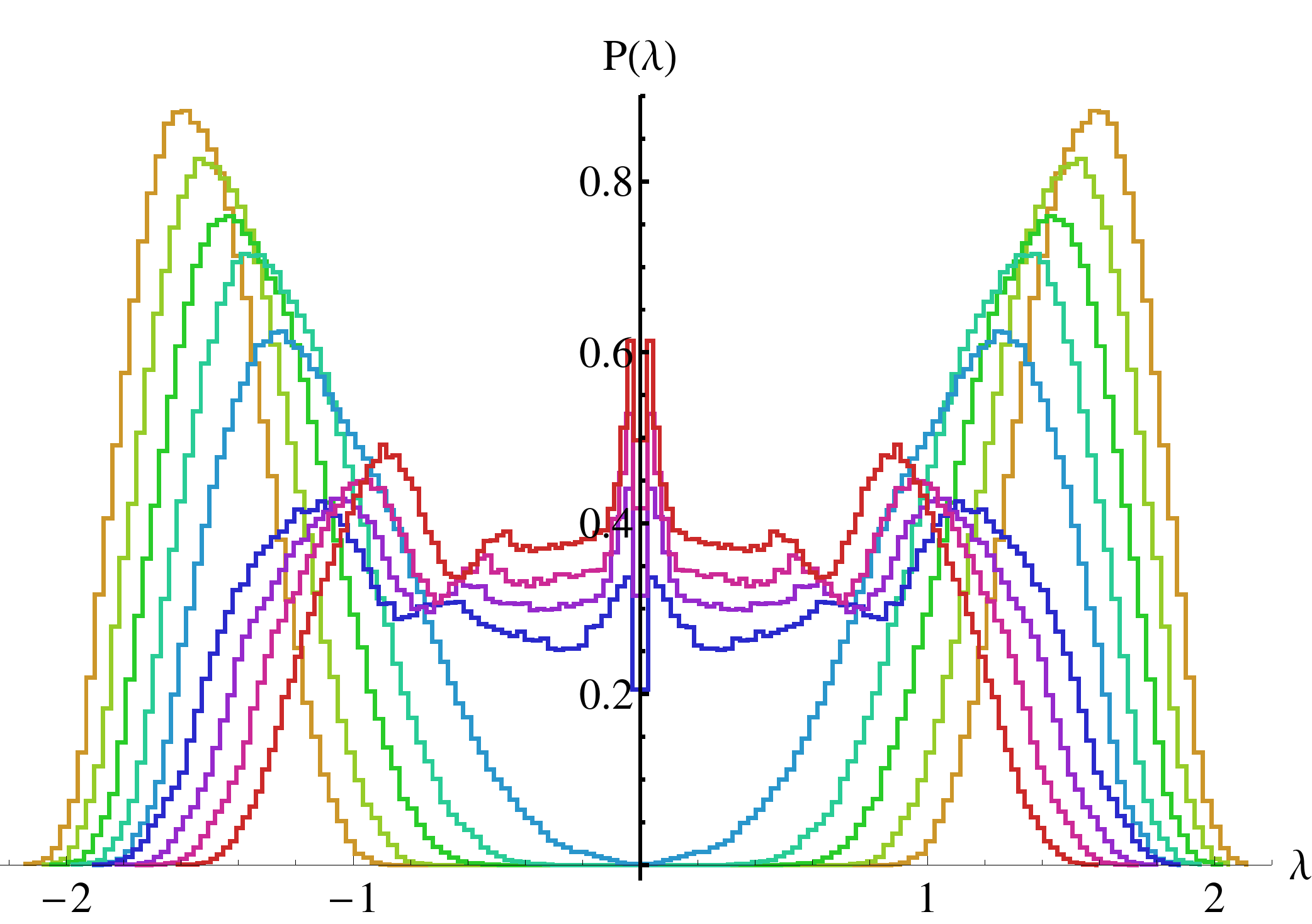}}
\subcaptionbox{Type $(1,1)$}{\includegraphics[width=0.5\textwidth]{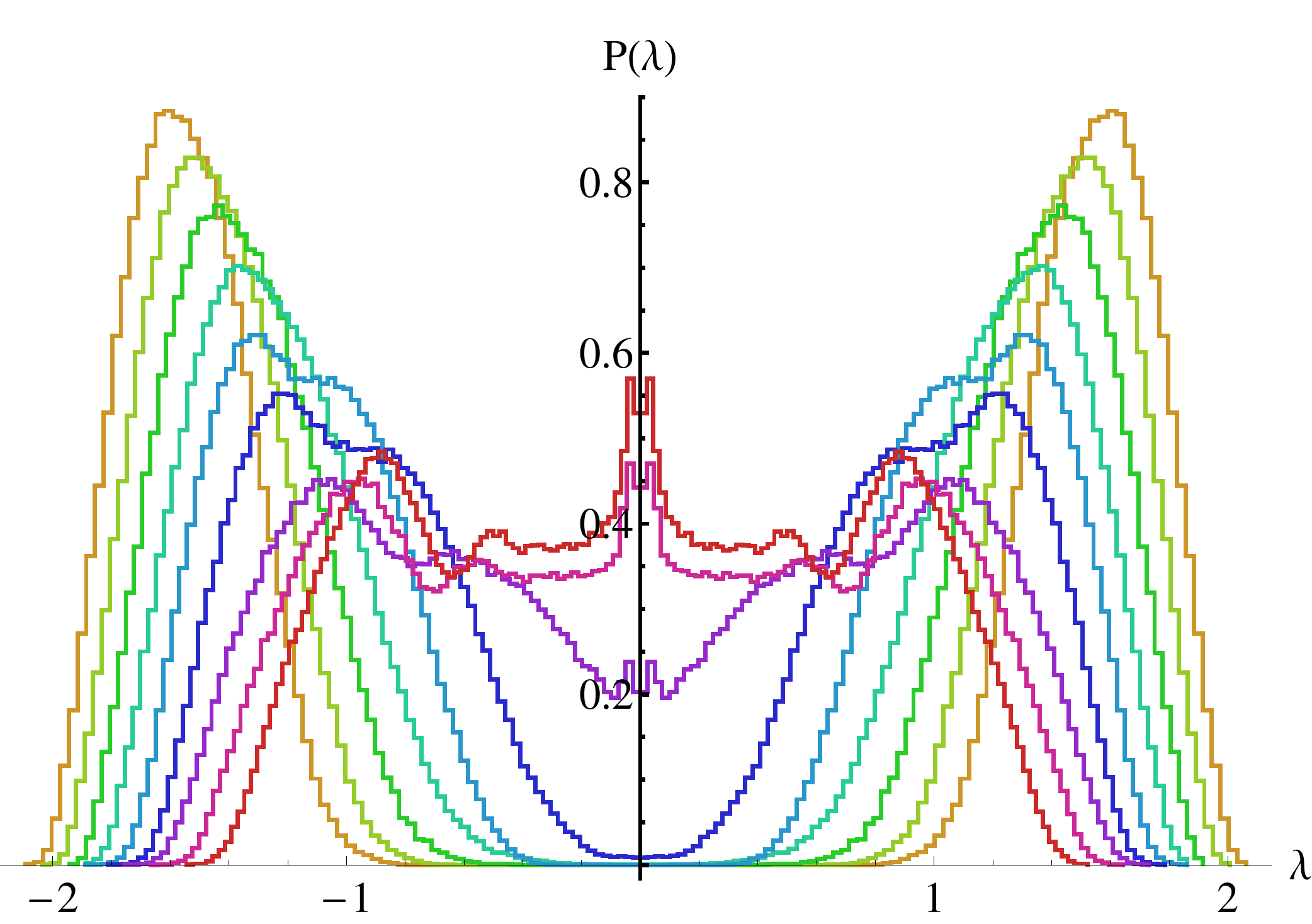}}

\subcaptionbox{Type $(0,2)$}{\includegraphics[width=0.5\textwidth]{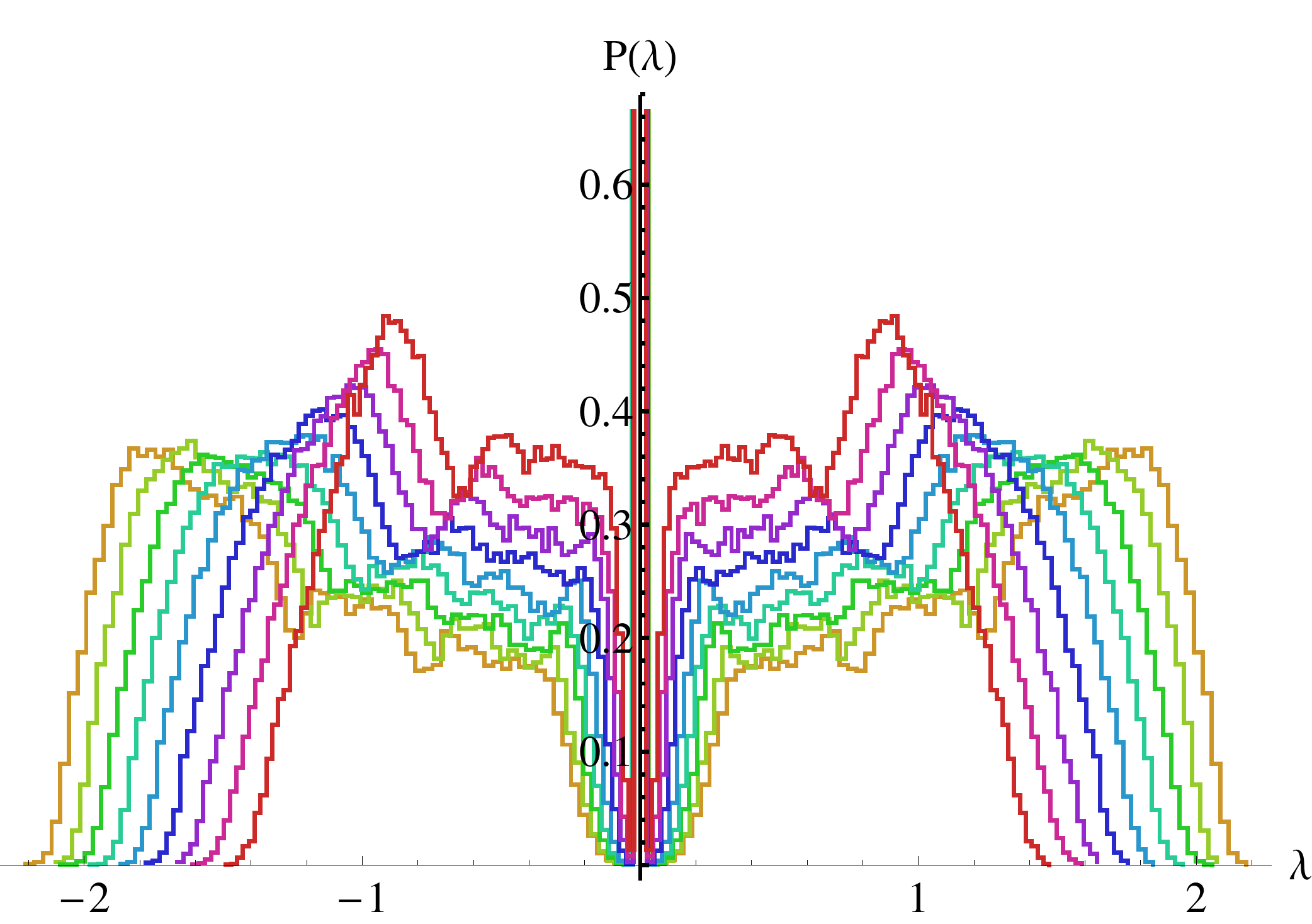}}
\subcaptionbox{\label{fig:03minima}Type $(0,3)$}{\includegraphics[width=0.5\textwidth]{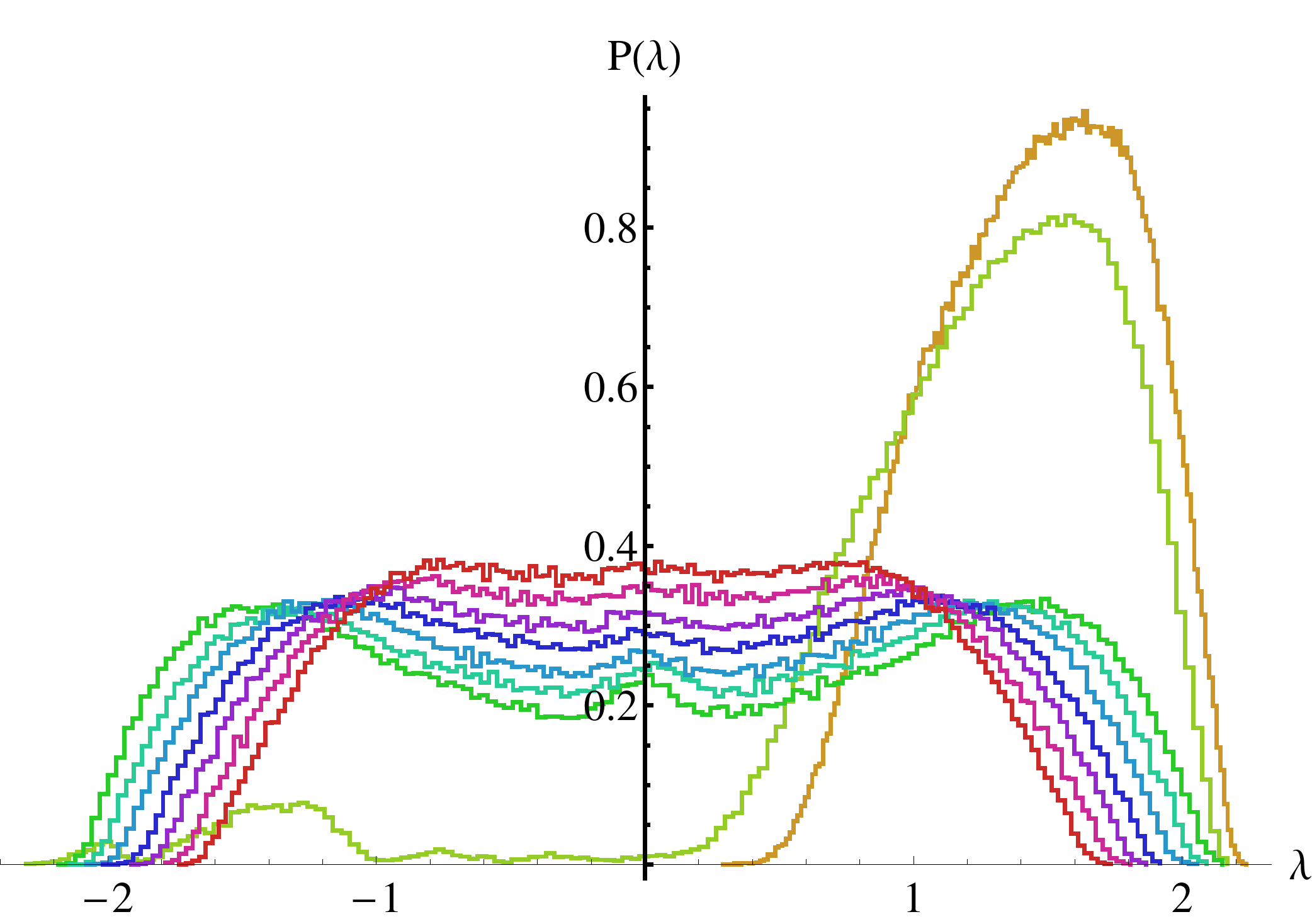}}
\caption{\label{fig:minima}The eigenvalues of $S= \Tr{D^4 +g_2 D^2}$ for $n=10$ and $g_2=-1$,  $-1.5$,  $-2$,  $-2.5$,  $-3$,  $-3.5$,  $-4$,  $-4.5$,  $-5$. The lines are coloured from red ($g_2=-1$) through to yellow ($g_2=-5$).}
\end{figure}

Combining the two terms together gives the action
\begin{equation}S= \Tr{g_2  D^2 +  D^4}.
\end{equation}
For positive values of $g_2$ the behaviour of the numerical simulations is somewhere between the $\tr{D^2}$ case and the $\tr D^4$ and does not show qualitatively new features. However when $g_2$ is negative this is a symmetry-breaking potential with two minima, shown in figure \ref{fig:potentials}.
The question of how the eigenvalues behave in this case is interesting and a variety of behaviours is exhibited depending on the type of the gamma matrices. This is shown in figure \ref{fig:minima}.
The different types are described here for values of $g_2$ decreasing from $0$.
\begin{description} \item[$(1,0)$] Two peaks start to form at around $g_2=-3$ then grow and separate sharply between $g_2=-3$ and $g_2=-3.5$, leaving the centre of the distribution empty. Since the Dirac operator is not symmetric, the Monte Carlo simulation can and does settle in just one of the peaks, though one expects that the Markov Chain would eventually explore both peaks equally given a long enough run. 
\item[$(0,1)$] Two peaks form at around $g_2=-3$ and grow slowly and steadily. The central part of the distribution remains. One can understand this from the fact that the eigenvalues of $L$ settle into two peaks,
 and since it is traceless, the favoured configuration has the same number of eigenvalues in each peak. The differences between eigenvalues of $L$ in the same minimum remains small, giving the central peak in the distribution for $D$.
The $n$ eigenvalues exactly $0$ are also still present.
\item[$(2,0)$] Two peaks develop at small $g_2$ and grow until the central part of the distribution vanishes suddenly between $g_2=-2.5$ and $-3$. 
\item[$(1,1)$]  Two peaks develop at small $g_2$ and grow until the central part of the distribution vanishes suddenly between $g_2=-2$ and $-2.5$. 
\item[$(0,2)$]  This is the most mysterious case. Two slight peaks develop but the eigenvalues do not separate into two peaks for the whole of the range of $g_2$ tested. Instead some further substructure to the eigenvalue distribution develops.
\item[$(0,3)$] This is similar to the $(0,1)$ case, with the sharp change occurring between $g_2=-4$ and $g_2=-4.5$. In figure \ref{fig:03minima} the Markov Chain for $g_2=-4.5$ has to a certain degree explored both minima.
\end{description}

\begin{figure}
\subcaptionbox{Type $(1,0)$}{\includegraphics[width=0.45\textwidth]{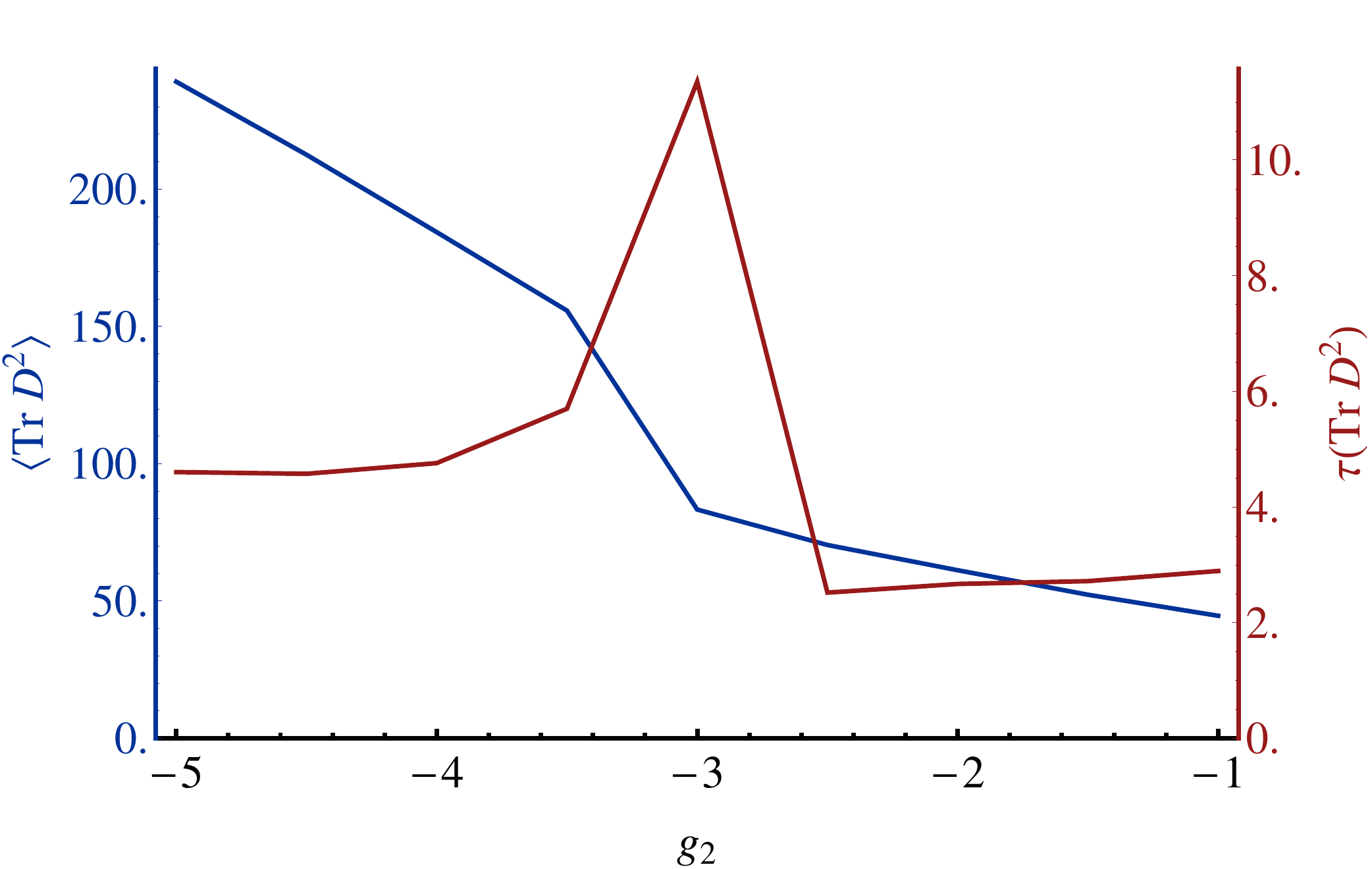}}
\subcaptionbox{Type $(0,1)$}{\includegraphics[width=0.45\textwidth]{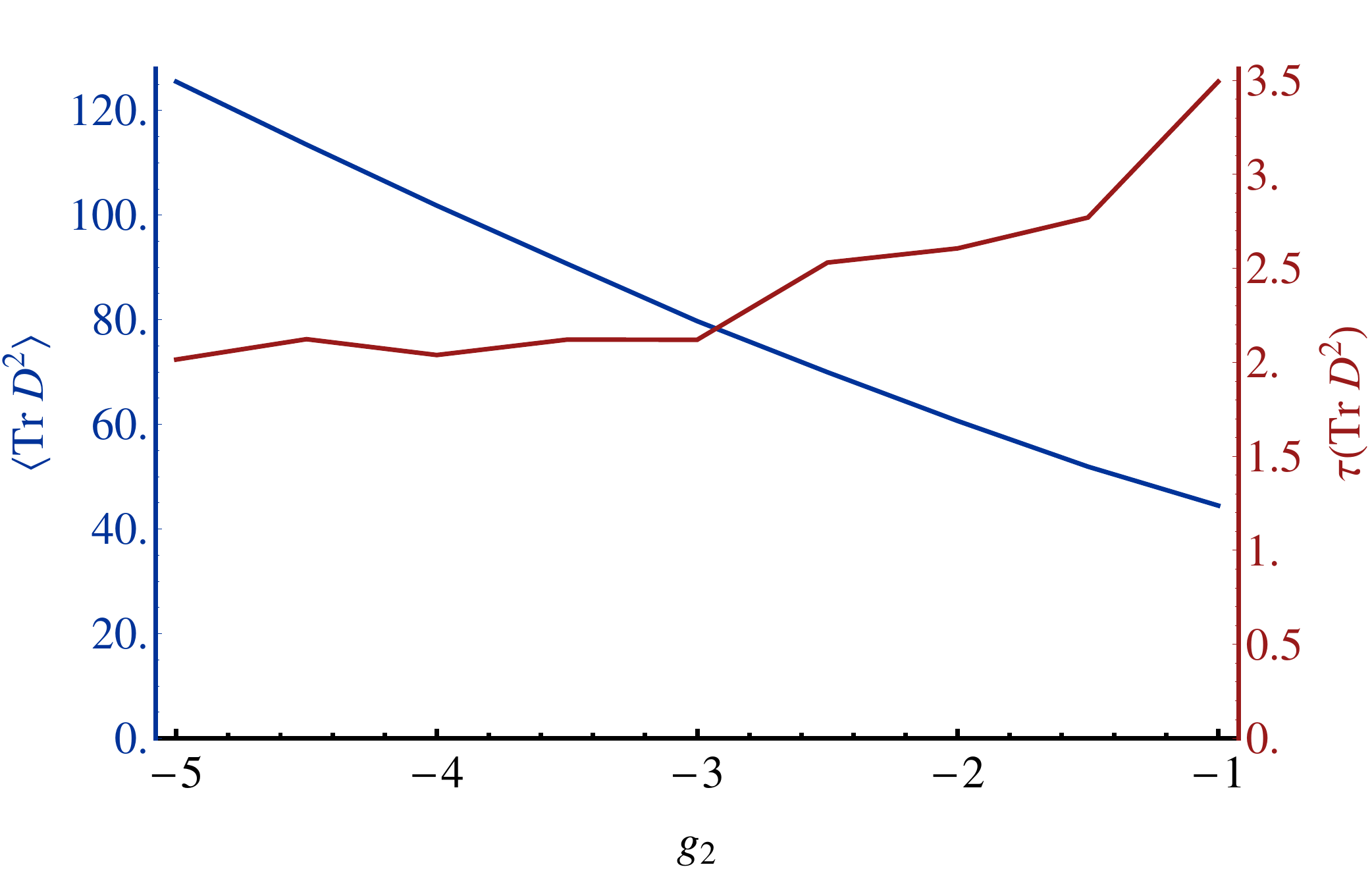}}

\subcaptionbox{Type $(2,0)$}{\includegraphics[width=0.45\textwidth]{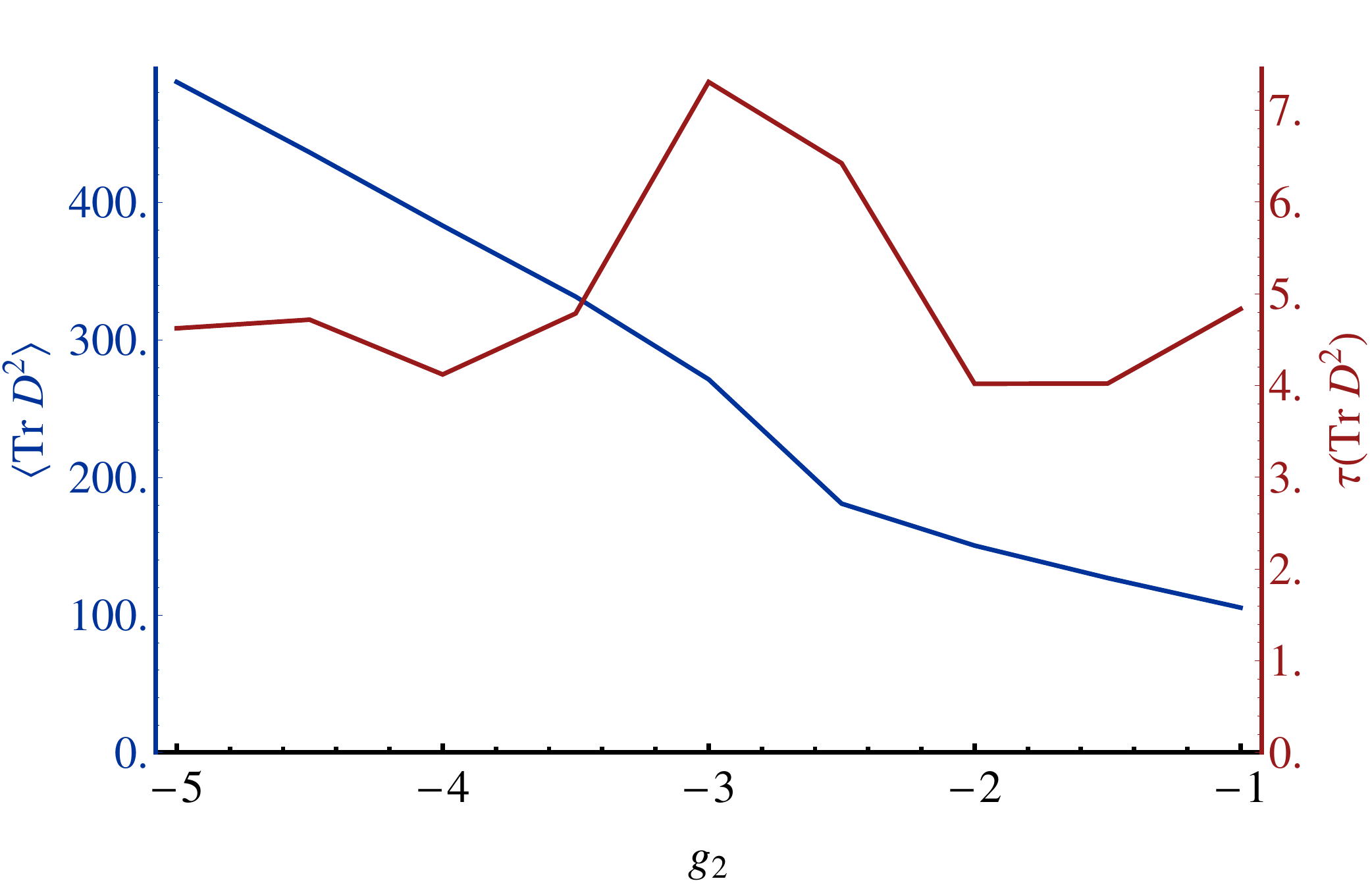}}
\subcaptionbox{Type $(1,1)$}{\includegraphics[width=0.45\textwidth]{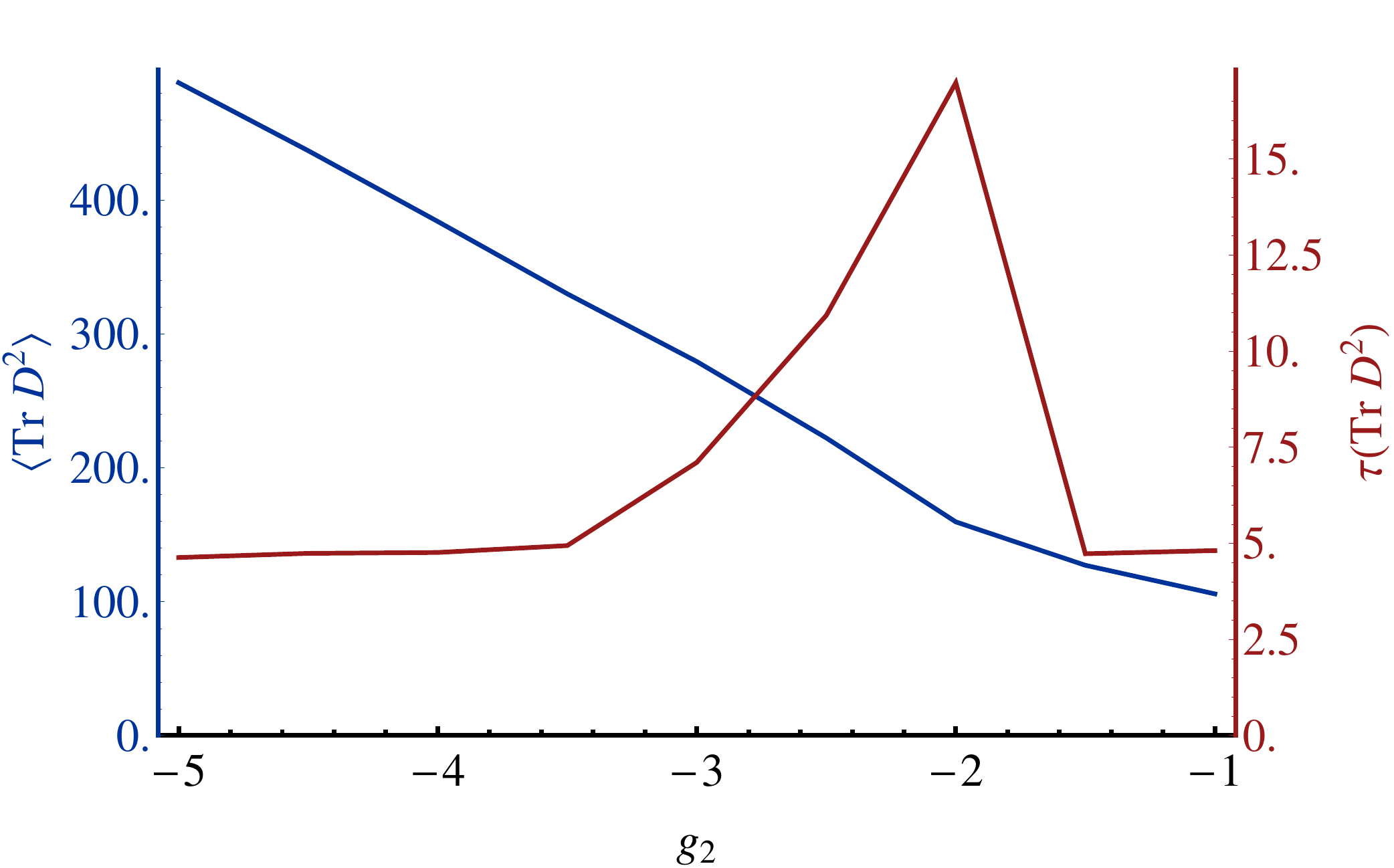}}

\subcaptionbox{Type $(0,2)$}{\includegraphics[width=0.45\textwidth]{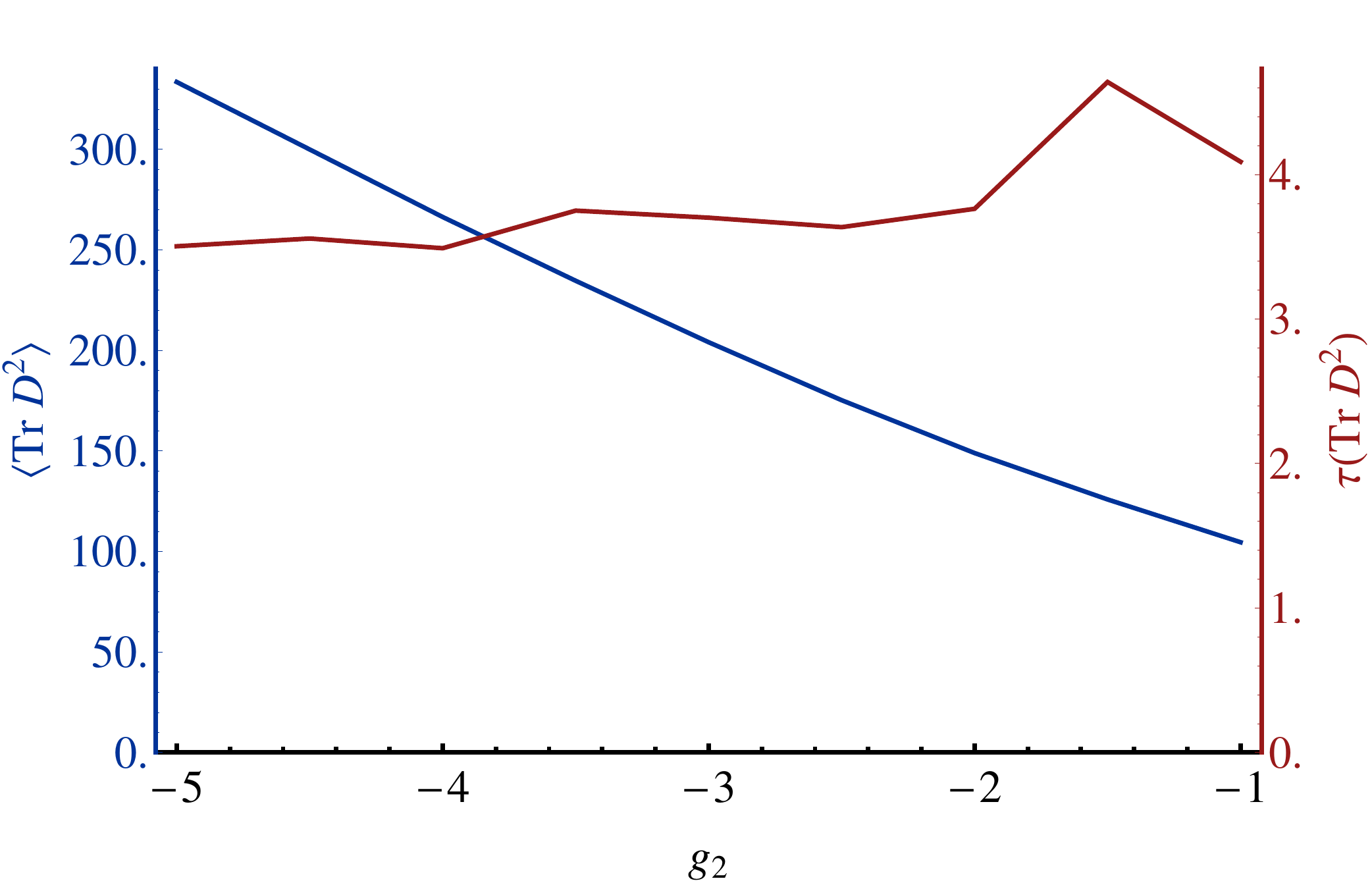}}
\subcaptionbox{Type $(0,3)$}{\includegraphics[width=0.45\textwidth]{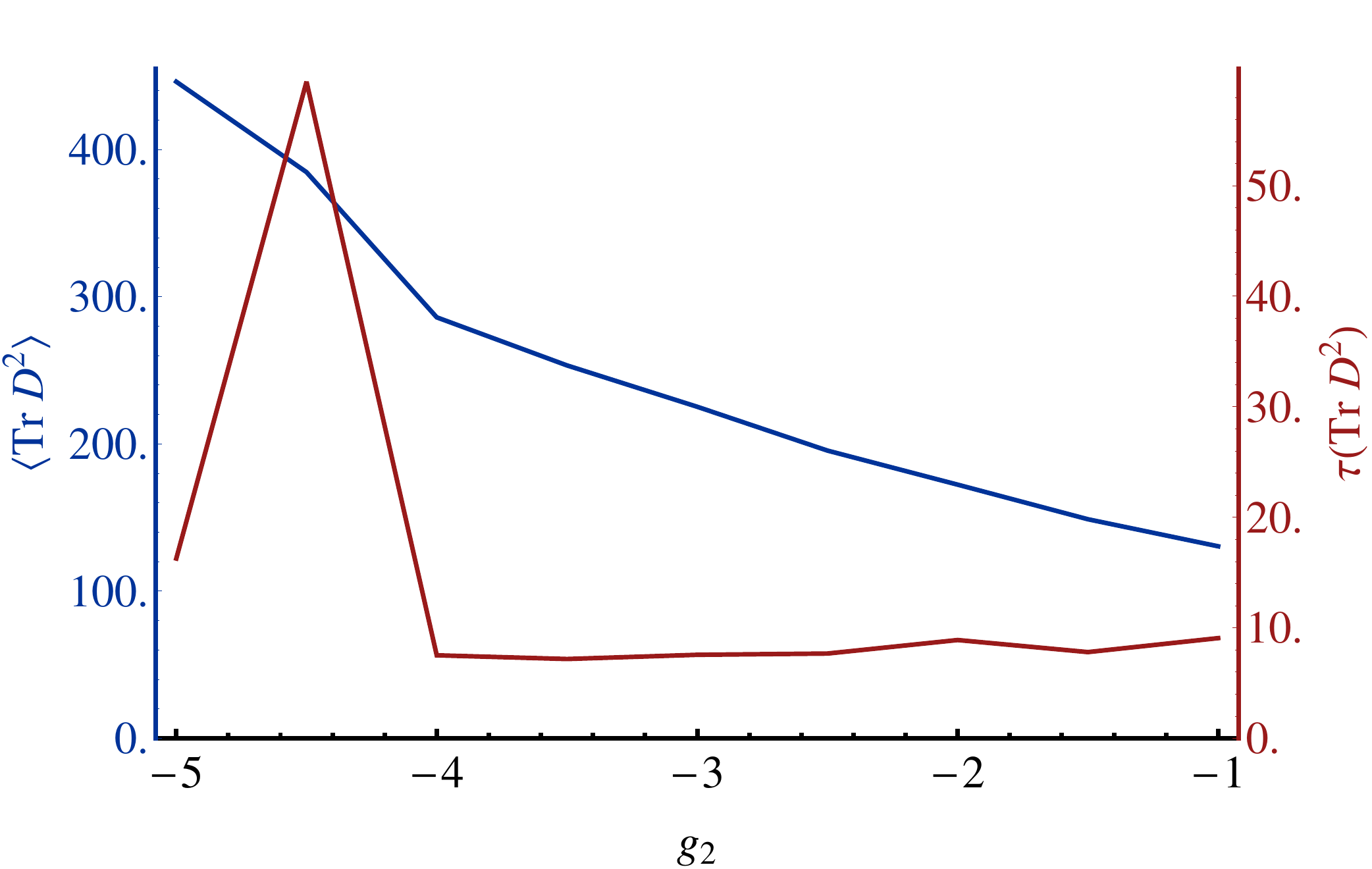}}
\caption{\label{fig:order-param} The mean of order parameter $ {\tr{ D^2}}$ and the autocorrelation time $\tau$ as $g_2$ varies.}
\end{figure}

These descriptions can be compared to the plots of the order parameter 
\begin{equation}\pd {\log Z}{g_2}=\av {\tr {D^2}}\end{equation}
and the autocorrelation time, which is usually expected to increase near a continuous phase transition due to the long-range order (`critical slowing down').
These are plotted in figure \ref{fig:order-param}. One can see that there is good evidence for a phase transition for the types $(1,0)$, $(2,0)$, $(1,1)$ and $(0,3)$. In these plots, the order parameter changes gradient at around the values of $g_2$ described above, and the autocorrelation time of $\tr D^2$ has a peak around this value also. It is difficult to see any clear signal from the plots for types $(0,1)$ and $(0,2)$.
\begin{figure}
\subcaptionbox{Type $(1,0)$}{\includegraphics[width=0.5\textwidth]{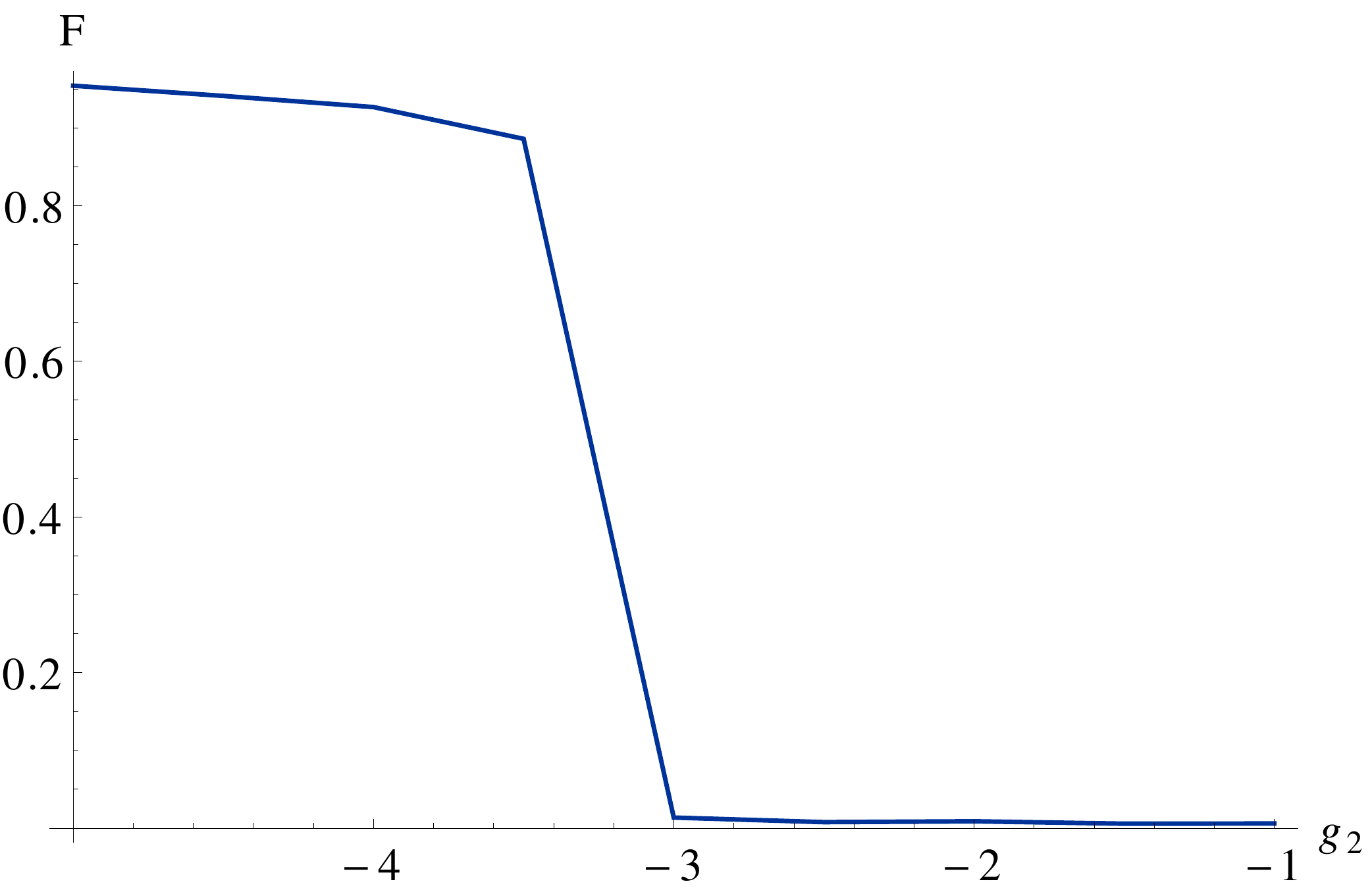}}
\subcaptionbox{Type $(1,1)$}{\includegraphics[width=0.5\textwidth]{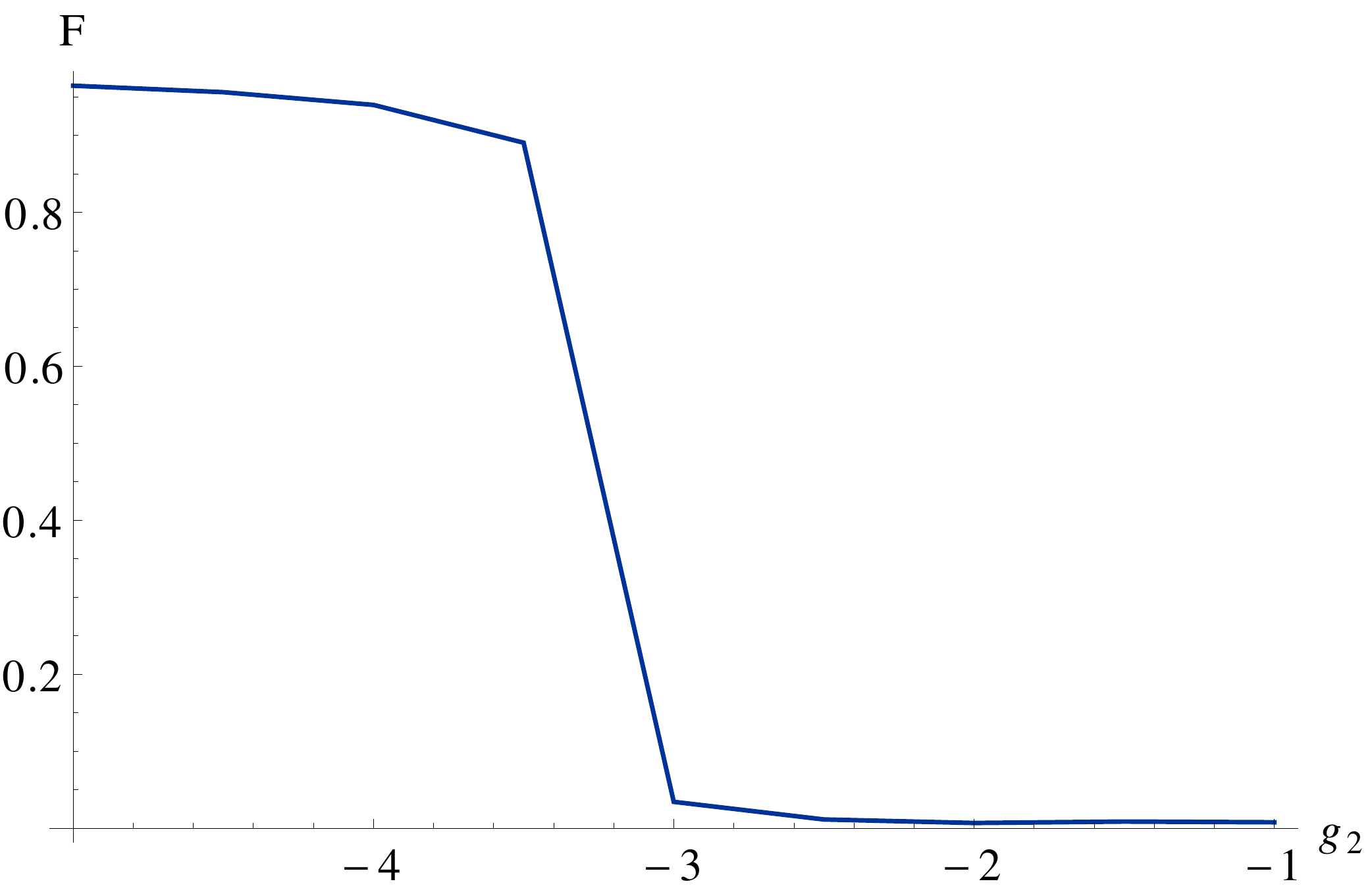}}

\subcaptionbox{Type $(2,0)$}{\includegraphics[width=0.5\textwidth]{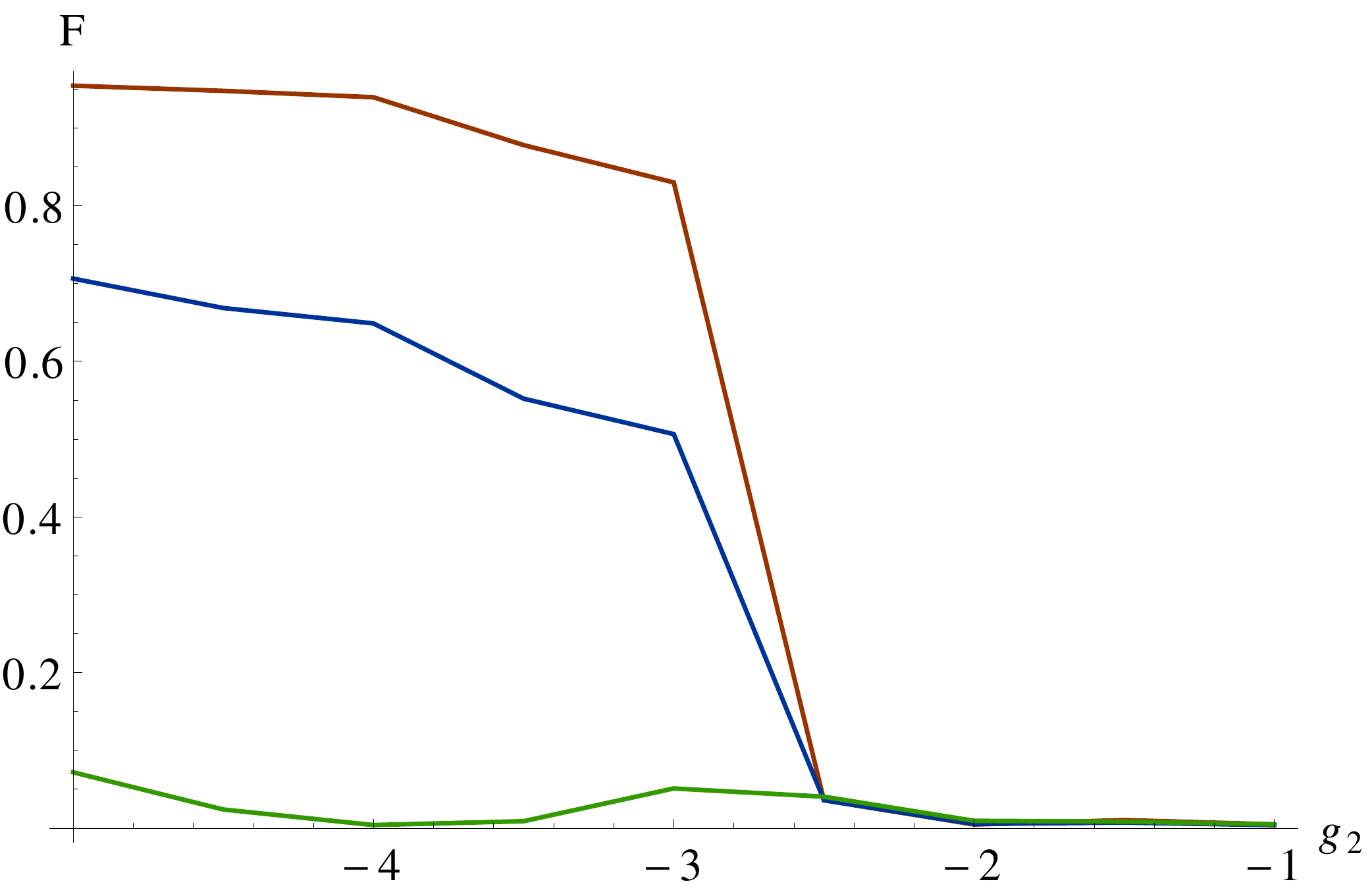}}
\subcaptionbox{Type $(0,3)$}{\includegraphics[width=0.5\textwidth]{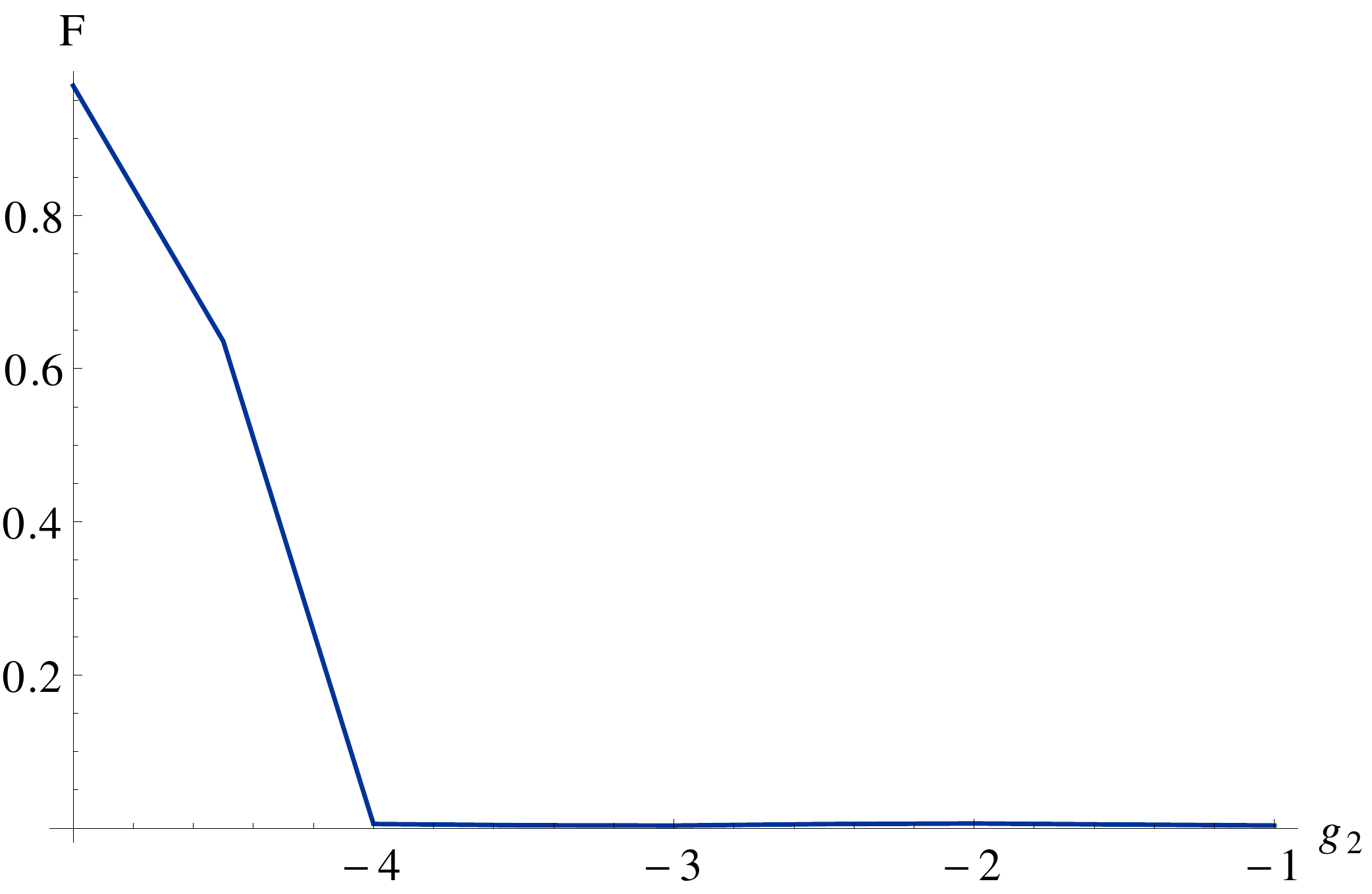}}
\caption{\label{fig:traceH}Fraction $F$ measuring the square of the proportion of $H$ that is in the trace part of $H$ as $g_2$ varies. The plot for type $(2,0)$ shows the fraction $F_1$ for $H_1$ (red), $F_2$ (green) and  combining both $F_{12}$  (blue). }
\end{figure}
It is remarkable that the types for which the evidence for a phase transition is clearest are the ones where the Dirac operator contains an anti-commutator with a Hermitian matrix $H$. Unlike the $L$ matrices, the trace of $H$ appears to play a crucial role. The observable
\begin{equation}F= \frac{(\tr H)^2}{n\tr H^2}
\end{equation}
measures the strength of $\tr H$, calculated as a square so that positive and negative values do not cancel, and as a fraction of the total strength of $H$. The averages of this are plotted in figure \ref{fig:traceH}. In the case of $(2,0)$, there are two matrices $H_1$ and $H_2$, and the $F$ for both combined  is
\begin{equation}F_{12}= \frac{(\tr H_1)^2+(\tr H_2)^2}{n\Tr{ H_1^2+ H_2^2}}
\end{equation}
In both cases, $F=1$ if the matrices are pure trace. The plots show that $\tr H$ develops a large expectation value at the phase transition. In the case of $(2,0)$ the Monte Carlo data used for the plots developed a preference for $\tr H_1$ rather than $\tr H_2$ but this is of no significance due to the rotational symmetry between the two\footnote{We have checked this with additional simulations and found that the trace degree of freedom is in general distributed randomly between the two matrices.}. The sum of squares is the correct rotationally-invariant observable.

\section{Conclusion}\label{sec:conc}
A model of random geometry has been presented here as random Dirac operators in non-commutative geometry. The integrals can be interpreted as multi-matrix models but with a new type of observables, namely the eigenvalues of the Dirac operator. 
The one-dimensional cases can be understood using theoretical results from random matrices but the higher-dimensional types are not so easy and will require further study to obtain analytic results.
Numerical results have been presented showing various phenomena that depend strongly on the type of gamma matrices used, particularly whether the spectrum of a Dirac operator for that signature is symmetric about the origin. 

From the numerical results it is clear that some features are similar to the properties of the eigenvalues of random matrices: the eigenvalue distributions appear to converge in the large $n$ limit and the dispersion of individual ordered eigenvalues decreases; also there is some evidence of a degree of eigenvalue repulsion at the origin. 

The most interesting results are for the quartic action with negative $g_2$, so that the potential is of symmetry-breaking type. For some types, the eigenvalue spectrum changes suddenly when $g_2$ reaches a critical negative value and the observable $\tr D^2$ is a good order parameter for this transition. This is taken as a strong indication that a sharp phase transition would occur in the large $n$ limit. The types where this transition is clear are those where the Dirac operator contains at least one term involving an anti-commutator with a random hermitian matrix $H$. Then the trace of $H$ develops a large expectation value, becoming the dominant contribution to $H$ after the transition.  This can't happen with commutator terms as the trace of the random matrix decouples in this instance.

\begin{figure}
\subcaptionbox{Type $(1,1)$, $g_2=-2.5$}{\includegraphics[width=0.45\textwidth]{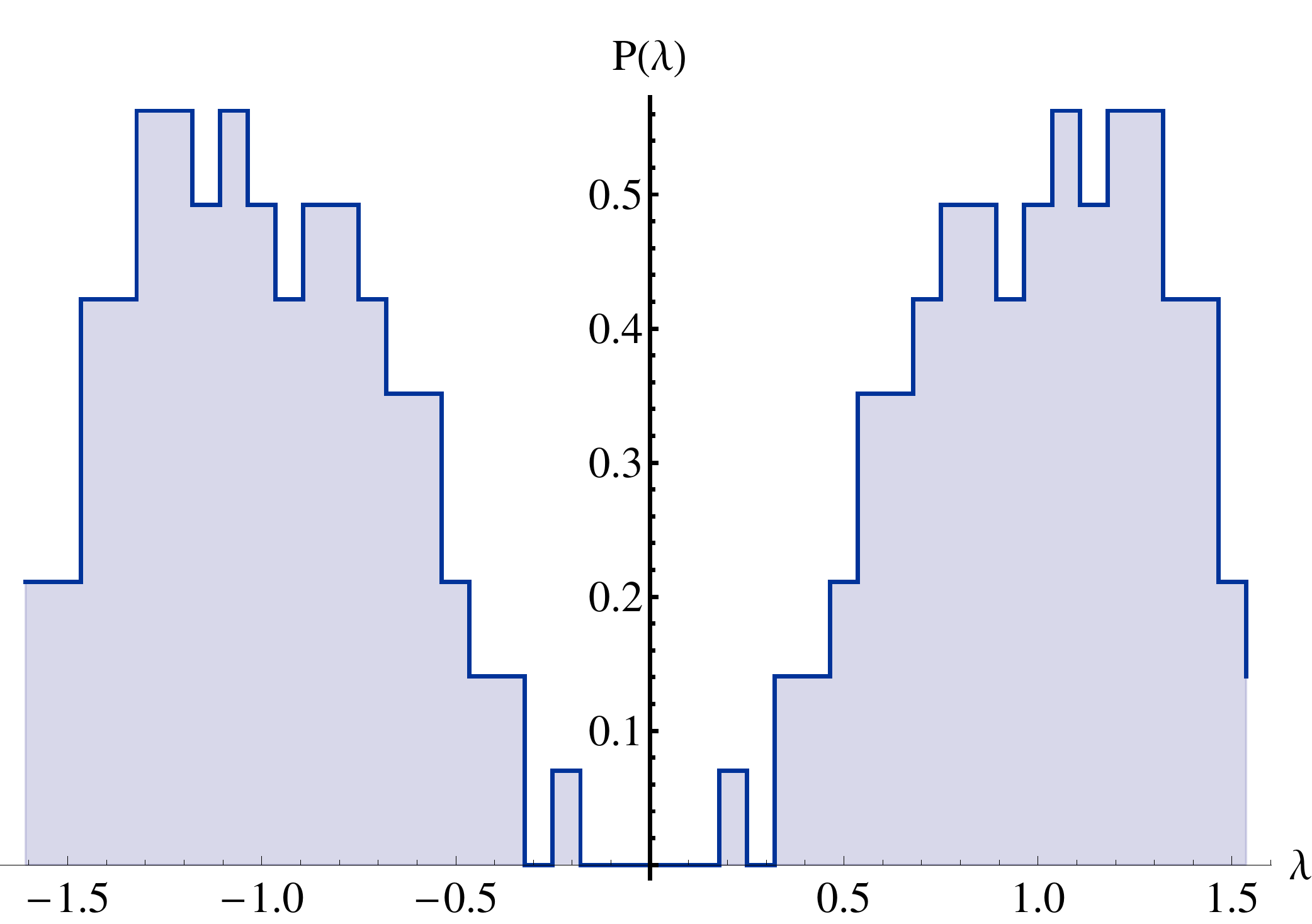}}
\subcaptionbox{Type $(2,0)$, $g_2=-3$}{\includegraphics[width=0.45\textwidth]{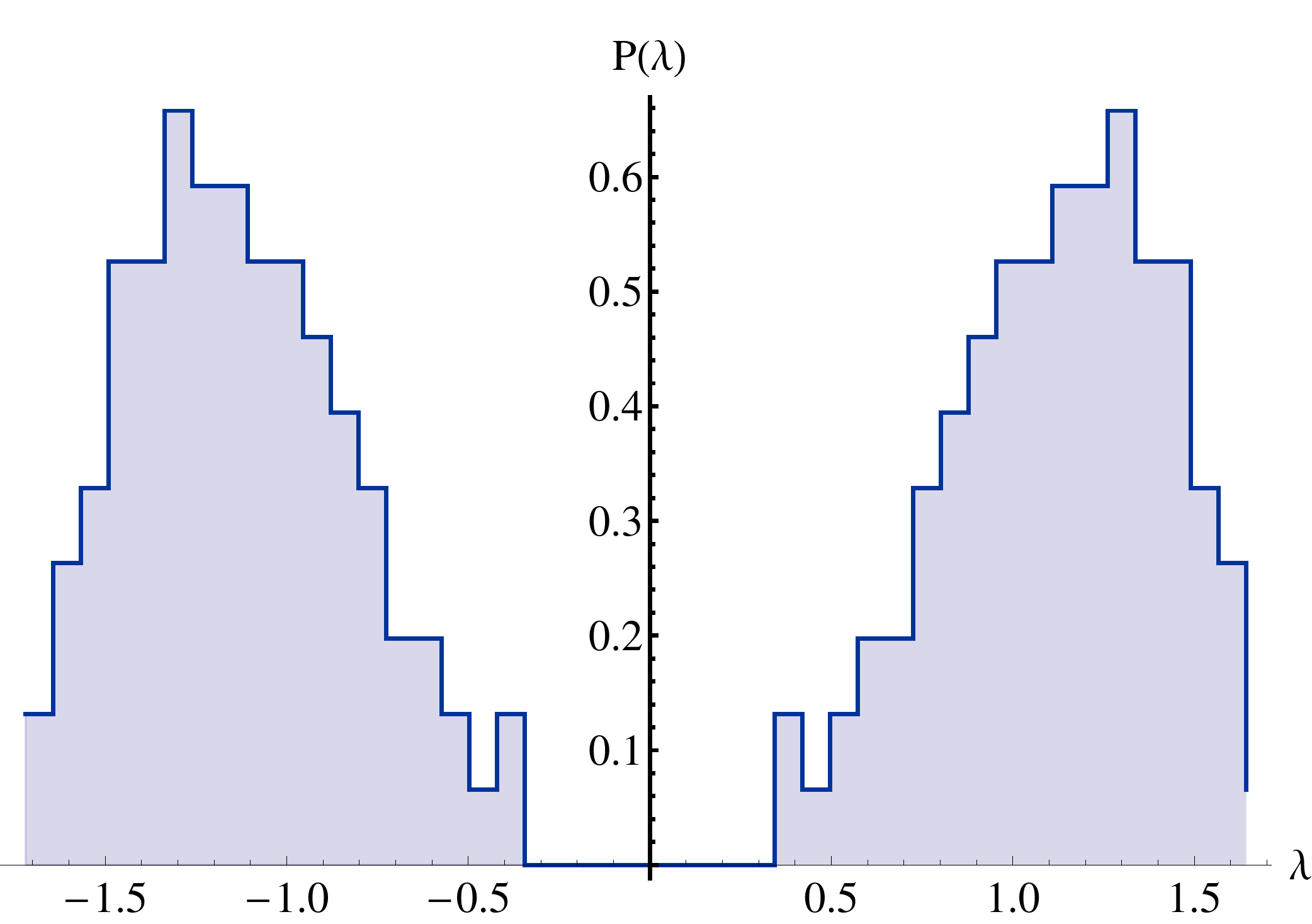}}

\subcaptionbox{Type (1,3) Fuzzy $S^2$}{\centerline{\includegraphics[width=0.45\textwidth]{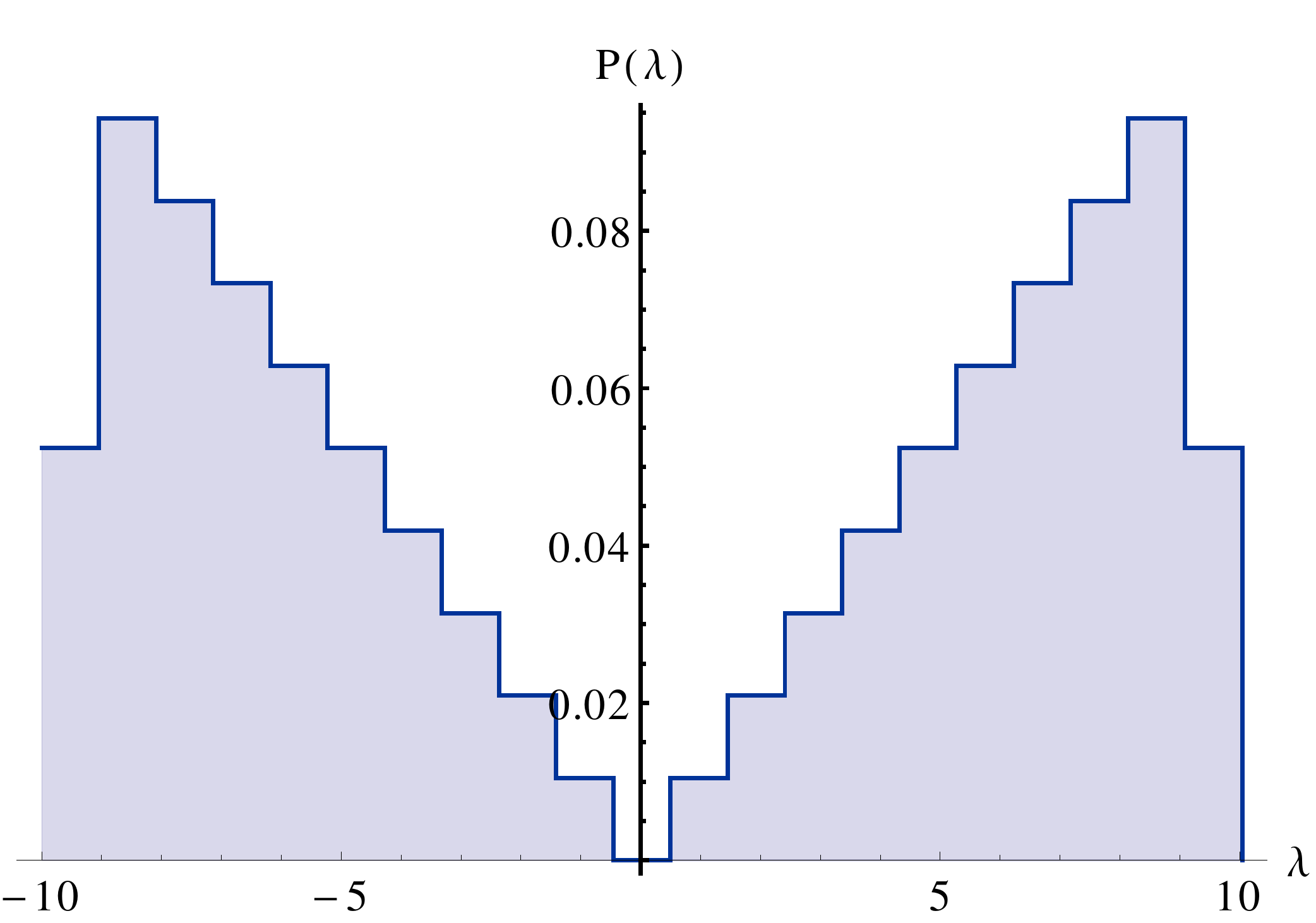}}} 
\caption{\label{fig:fuzzy}Eigenvalue distributions near the phase transition compared with the fuzzy sphere. Matrix size $n=10$.}
\end{figure}

For generic $g_2$, the eigenvalue distribution of $D$ is not a good approximation to the behaviour for the Dirac operator on any fixed (commutative) Riemannian manifold \eqref{eq:dos}, except that one could possibly argue that the distribution is approximately constant for some ranges (e.g. figure \ref{fig:TrD4}(f), which looks like a one-dimensional manifold). The exception to this is near the phase transition, where the curves in  figure \ref{fig:minima} do appear to have the right power-law behaviour. Two of these distributions are highlighted in figure \ref{fig:fuzzy}, showing the types $(1,1)$ and $(2,0)$ at values of $g_2$ just below the value for the phase transition. 

These are compared with the eigenvalue distribution for the fuzzy sphere from \cite{Barrett:2015naa}, shown in figure 
\ref{fig:fuzzy}(c). This is a type $(1,3)$ spectral triple, having signature $s=2$, and has exactly the same spectrum as the Dirac operator on the Riemannian round $S^2$ but with a maximum eigenvalue cut-off and fermion doubling.
The distributions are remarkably similar, the main differences being the gap at the origin, the size of which depends on the distance from the phase transition, and the fact that the fuzzy sphere has exactly integer eigenvalues with multiplicity, due to its rotational symmetry. The feature that is common to the plots is the approximately linear increase of the eigenvalue density with eigenvalue that is characteristic of Riemannian manifolds of dimension two, i.e., $m=2$ in \eqref{eq:dos}.  The simulations show that decreasing $g_2$ further increases the gap in the middle of the spectrum whereas the middle of the spectrum fills up for values of $g_2$ greater than the critical value.  

These results are somewhat preliminary and we have not yet carried out a systematic study of the phase transition.

The study of random non-commutative geometries gives an insight into the closely-related problem of the quantization of this fascinating theory. A quantized non-commutative space is a potential candidate for a quantum theory of gravitational interactions and will allow a better understanding of fundamental interactions.
Independent of these physical applications, it also is an interesting modification of the well-known matrix models, introducing non-trivial interactions and observables among several matrices. The results reported here will be a basis for
further investigations into the phase transition, the continuum limit, and other possible actions on this space of geometries.

\section{Acknowledgements} This work of JWB on this project was supported by STFC Particle Physics Theory Consolidated Grant ST/L000393/1, and LG was supported by  funding from the European Research Council under the European Union Seventh Framework Programme (FP7/2007-2013) / ERC Grant Agreement n.306425 ``Challenging General Relativity''. 

This research was supported in part by Perimeter Institute for Theoretical Physics. Research at Perimeter is supported by the Government of Canada through Industry Canada and by the Province of Ontario through the Ministry of Economic Development and Innovation. 

The dissemination of this work was aided by EU COST Action MP1405 `Quantum structure of spacetime'.

\begin{appendix}
\section{\label{sec:GeoAndAc}Dirac operators for the fuzzy geometries we examined}
In this appendix matrices $H$ will be Hermitian and matrices $L$ anti-Hermitian. The $L$ matrices are not assumed to be traceless, though the actions are independent of trace part of these matrices. The bracketing convention for the trace of an expression is $\tr AB\equiv \tr(AB)$ and $\tr A^n\equiv \tr(A^n)$, but $\tr A+B\equiv(\tr A)+B$.

\subsection{The $(1,0)$ geometry}
\begin{align}
\g{1}=1 \\
D=\acom{H}{\cdot}
\end{align}
\begin{align}\label{eq:S10simp}
\tr{D^2}&=2 n \tr{H^2}+ 2 (\tr{H})^2\\
\tr{D^4}&=2 n \tr{H^4}+ 8 \tr{H}\tr{H^3} + 6 (\tr{H^2})^2
\end{align}

\subsection{The $(0,1)$ geometry}
\begin{align}
\g{1}=-i \\
D=\g{1} \te \com{L_1}{\cdot}
\end{align}

\begin{align}\label{eq:S01simp}
\tr{D^2}&=-\left(2 n \tr{L^2}- 2 (\tr{L})^2\right)\\
\tr{D^4}&=2 n \tr{L^4}- 8 \tr{L}\tr{L^3} + 6 (\tr{L^2})^2
\end{align}

\subsection{The $(2,0)$ geometry}

\begin{align}
\g{1}&=\begin{pmatrix}
1 & 0 \\
0 &-1 \\
\end{pmatrix} 
&
\g{2}&=\begin{pmatrix}
0 & 1 \\
1 &0 \\
\end{pmatrix}
\end{align}
\begin{align}
D=\g{1} \te\acom{H_1}{\cdot}+\g{2} \te \acom{H_2}{\cdot}
\end{align}
The gamma matrix trace identities are $\Tr{\g{i}\g{j}}=2 \delta_{ij}$ and $\Tr{\g{i}\g{j}\g{k}}=0$ and $\Tr{\g{i}\g{j}\g{k}\g{l}}=2( \delta_{ij}\delta_{kl} -\delta_{ik}\delta_{jl}+\delta_{il}\delta_{jk})$.

\begin{align}
\tr{D^2}&=4 n (\tr{H_1^2}+\tr{H_2^2})+ 4 \left((\tr{H_1})^2+(\tr{H_2})^2\right)\\
\tr{D^4}&=2 \Tr{\acom{H_1}{\cdot}^4} +2 \Tr{\acom{H_2}{\cdot}^4}  +8 \Tr{\acom{H_1}{\cdot}^2\acom{H_2}{\cdot}^2} \\
	&-4 \Tr{\acom{H_1}{\cdot}\acom{H_2}{\cdot}\acom{H_1}{\cdot}\acom{H_2}{\cdot}}\\
		&=4 n \bigg(\tr{H_1^4}+\tr{H_2^4} +4 \tr{H_1^2H_2^2} -2 \tr{H_1H_2H_1H_2} \bigg) \\
		&+ 16 \bigg(\tr{H_1}\left(\tr{H_1^3}+\tr{H_2^2H_1} \right) \\
		&+\tr{H_2}\left(\tr{H_1^2H_2}+\tr{H_2^3}\right) +(\tr{H_1H_2})^2\bigg) \\
		&+ 12 \bigg((\tr{H_1^2})^2+(\tr{H_2^2})^2\bigg) + 8\tr{H_1^2}\tr{H_2^2}
\end{align}

\subsection{The $(1,1)$ geometry}

\begin{align}
\g{1}&=\begin{pmatrix}
1 & 0 \\
0 &-1 \\
\end{pmatrix} 
&
\g{2}&=\begin{pmatrix}
0 & 1 \\
-1 &0 \\
\end{pmatrix}
\end{align}
\begin{align}
D=\g{1} \te \acom{H}{\cdot}+\g{2} \te \com{L}{\cdot}
\end{align}
The gamma matrix trace identities are $\Tr{\g{i}\g{j}}=2 \eta_{ij}$, where $\eta_{i,j}=\mathrm{diag}(-1,1)$ and $\Tr{\g{i}\g{j}\g{k}}=0$ and $\Tr{\g{i}\g{j}\g{k}\g{l}}=2( \eta_{ij}\eta_{kl} -\eta_{ik}\eta_{jl}+\eta_{il}\eta_{jk})$.
\begin{align}
\tr{D^2}&=4 n (\tr{H^2}-\tr{L^2})+ 4 \left((\tr{H})^2+(\tr{L})^2\right)\\
\tr{D^4}&=2 \Tr{\acom{H}{\cdot}^4} +2 \Tr{\com{L}{\cdot}^4}  -8 \Tr{\acom{H}{\cdot}^2\com{L}{\cdot}^2} \\
		&+4 \Tr{\acom{H}{\cdot}\com{L}{\cdot}\acom{H}{\cdot}\com{L}{\cdot}}\\
		&=4 n \bigg(\tr{H^4}+\tr{L^4} -4 \tr{H^2L^2} +2 \tr{HLHL}\bigg) \\
		&+ 16 \bigg(\tr{H}\left(\tr{H^3}-\tr{L^2H}\right) \\
		&+\tr{L}\left(- \tr{L^3}+\tr{H^2L} \right) +(\tr{HL})^2\bigg) \\
		&+ 12 \bigg((\tr{H^2})^2+(\tr{L^2})^2\bigg) - 8\tr{H^2}\tr{L^2}
\end{align}

\subsection{The $(0,2)$ geometry}
\begin{align}
\g{1}&=\begin{pmatrix}
i & 0 \\
0 &-i \\
\end{pmatrix} 
&
\g{2}&=\begin{pmatrix}
0 & 1 \\
-1 &0 \\
\end{pmatrix}
\end{align}
\begin{align}
D=\g{1} \te \com{L_1}{\cdot}+\g{2} \te \com{L_2}{\cdot}
\end{align}
The gamma matrix trace identities are $\Tr{\g{i}\g{j}}=-2 \delta_{ij}$ and $\Tr{\g{i}\g{j}\g{k}}=0$ and $\Tr{\g{i}\g{j}\g{k}\g{l}}=2( \delta_{ij}\delta_{kl} -\delta_{ik}\delta_{jl}+\delta_{il}\delta_{jk})$.
\begin{align}
\tr{D^2}&=-4 n (\tr{L_1^2}+\tr{L_2^2})+ 4 \left((\tr{L_1})^2+(\tr{L_2})^2\right)\\
\tr{D^4}&=2 \Tr{\com{L_1}{\cdot}^4} +2 \Tr{\com{L_2}{\cdot}^4}  +8 \Tr{\com{L_1}{\cdot}^2\com{L_2}{\cdot}^2} \\
		&-4 \Tr{\com{L_1}{\cdot}\com{L_2}{\cdot}\com{L_1}{\cdot}\com{L_2}{\cdot}}\\
		&=4 n \bigg(\tr{L_1^4}+\tr{L_2^4} +4 \tr{L_1^2L_2^2} -2 \tr{L_1L_2L_1L_2}\bigg) \\
		&+ 16 \bigg(-\tr{L_1} ( \tr{L_1^3} +\tr{L_2^2L_1}) \\
		&-\tr{L_2} (\tr{L_2^3}+\tr{L_1^2L_2})  +(\tr{L_1L_2})^2 \bigg) \\
		&+ 12 \bigg((\tr{L_1^2})^2+(\tr{L_2^2})^2 \bigg) + 8\tr{L_1^2}\tr{L_2^2}
\end{align}

\subsection{The $(0,3)$ geometry}
\begin{align}
\g{1}&=\begin{pmatrix}
i & 0 \\
0 &-i \\
\end{pmatrix} 
&
\g{2}&=\begin{pmatrix}
0 & 1 \\
-1 &0 \\
\end{pmatrix} 
&
\g{3}&=\begin{pmatrix}
0 & i \\
i &0 \\
\end{pmatrix}
&
\g{}&=\begin{pmatrix}
-1 & 0 \\
0 &-1 \\
\end{pmatrix} 
\end{align}
\begin{align}
D&= \acom{H}{}+\g{1}\te\com{L_1}{\cdot}+\g{2}\te\com{L_2}{\cdot}+\g{3}\te\com{L_3}{\cdot}
\end{align}
To calculate the action, the following gamma matrix identities are used: $\Tr{\g{i}\g{j}}=-2 \delta_{ij}$ and $\Tr{\g{i}\g{j}\g{k}\g{l}}=2( \delta_{ij}\delta_{kl} -\delta_{ik}\delta_{jl}+\delta_{il}\delta_{jk})$, and $\g{i}\g{j}\g{k}=\varepsilon_{ijk}$, the totally antisymmetric tensor  with $\varepsilon_{123}=-1$.
\begin{align}
\tr{D^2}&=4 n (\tr{H^2}- \tr{L_1^2}-\tr{L_2^2}-\tr{L_2^2})\\
	&+ 4 ((\tr{H})^2+(\tr{L_1})^2+(\tr{L_2})^2+(\tr{L_3})^2) \\
\tr{D^4}&=2 \Tr{\acom{H}{\cdot}^4} +2 \Tr{\com{L_1}{\cdot}^4} +2 \Tr{\com{L_2}{\cdot}^4}+2 \Tr{\com{L_3}{\cdot}^4} \\
& +8 \Tr{\com{L_1}{\cdot}^2\com{L_2}{\cdot}^2}  +8 \Tr{\com{L_1}{\cdot}^2\com{L_3}{\cdot}^2} \\
&+8 \Tr{\com{L_2}{\cdot}^2\com{L_3}{\cdot}^2} -8 \Tr{\acom{H}{\cdot}^2\com{L_1}{\cdot}^2}\\
& -8 \Tr{\acom{H}{\cdot}^2\com{L_2}{\cdot}^2} -8 \Tr{\acom{H}{\cdot}^2\com{L_3}{\cdot}^2}\\
		&-4 \Tr{\com{L_1}{\cdot}\com{L_2}{\cdot}\com{L_1}{\cdot}\com{L_2}{\cdot}}
		-4 \Tr{\com{L_1}{\cdot}\com{L_3}{\cdot}\com{L_1}{\cdot}\com{L_3}{\cdot}}\\
		&-4 \Tr{\com{L_2}{\cdot}\com{L_3}{\cdot}\com{L_2}{\cdot}\com{L_3}{\cdot}}
		-4 \Tr{\acom{H}{\cdot}\com{L_1}{\cdot}\acom{H}{\cdot}\com{L_1}{\cdot}}\\	
		&-4 \Tr{\acom{H}{\cdot}\com{L_2}{\cdot}\acom{H}{\cdot}\com{L_2}{\cdot}}	
		-4 \Tr{\acom{H}{\cdot}\com{L_3}{\cdot}\acom{H}{\cdot}\com{L_3}{\cdot}}\\	
		&-8 \Tr{\acom{H}{\cdot} \com{L_1}{\cdot} \com{L_2}{\cdot} \com{L_3}{\cdot}}+8 \Tr{\acom{H}{\cdot} \com{L_1}{\cdot} \com{L_3}{\cdot} \com{L_2}{\cdot}}\\
		&-8 \Tr{\acom{H}{\cdot} \com{L_2}{\cdot} \com{L_3}{\cdot} \com{L_1}{\cdot}}+8 \Tr{\acom{H}{\cdot} \com{L_2}{\cdot} \com{L_1}{\cdot} \com{L_3}{\cdot}}\\
		&-8 \Tr{\acom{H}{\cdot} \com{L_3}{\cdot} \com{L_1}{\cdot} \com{L_2}{\cdot}}+8 \Tr{\acom{H}{\cdot} \com{L_3}{\cdot} \com{L_2}{\cdot} \com{L_1}{\cdot}}\\
&= 4n\bigg(\Tr{L_1^4+L_2^4+L_3^4+H^4} \\
&-2\Tr{L_1L_2L_1L_2+L_1L_3L_1L_3+L_2L_3L_2L_3}  \\
&-2\Tr{HL_1HL_1+HL_2HL_2+HL_3HL_3} \\
&+ 4 \Tr{L_1^2L_2^2+L_1^2L_3^2+L_2^2L_3^2}- 4 \Tr{H^2L_1^2+H^2L_2^2+H^2L_3^2}\\
&- 4 \Tr{HL_1L_2L_3+HL_2L_3L_1+ HL_3L_1L_2} \\
&+4 \Tr{HL_1L_3L_2+HL_2L_1L_3+HL_3L_2L_1} \bigg) \\
&- 16 \tr{L_1}\left(\tr{L_1^3}+\tr{L_2^2L_1}+\tr{L_3^2L_1}-3\tr{H^2L_1} \right) \\
&- 16 \tr{L_2}\left(\tr{L_2^3}+\tr{L_1^2L_2}+\tr{L_3^2L_2}-3\tr{H^2L_2} \right) \\
&- 16 \tr{L_3}\left(\tr{L_3^3}+\tr{L_1^2L_3}+\tr{L_2^2L_3}-3\tr{H^2L_3} \right) \\
&+ 16 \tr{H}\left(\tr{H^3}-3\tr{L_1^2H}-3\tr{L_2^2H}-3\tr{L_3^2H} \right) \\
&+ 16 \bigg(\tr{L_1}\Tr{H\com{L_2}{L_3}}-\tr{L_2}\Tr{H\com{L_1}{L_3}}\\
&+\tr{L_3}\Tr{H\com{L_1}{L_2}} -3 \tr{H}\Tr{L_1\com{L_2}{L_3}} \bigg) \\
&+ 12 \left( (\tr{L_1^2})^2+(\tr{L_2^2})^2+(\tr{L_3^2})^2+(\tr{H^2})^2\right) \\
&+ 8 \left(\tr{L_1^2}\tr{L_2^2}+\tr{L_1^2}\tr{L_3^2}+\tr{L_2^2}\tr{L_3^2}\right)\\
&- 24 \tr{H^2}\left(\tr{L_1^2}+\tr{L_2^2}+\tr{L_3^2}\right)\\
&+16 \left( (\tr{L_1L_2})^2+(\tr{L_1L_3})^2+(\tr{L_2L_3})^2\right) \\
&+48 \left( (\tr{HL_1})^2+(\tr{HL_2})^2+(\tr{HL_3})^2\right) 
\end{align}
This expression shows that the complexity of terms rises quickly with increasing $d=p+q$.


\subsection{Powers of (anti-)commutators}
For anti commutators we have
\begin{align}
\acom{H}{\cdot}&= H \te \ident{n} + \ident{n} \te H^T \\
\acom{H}{\acom{H}{\cdot}}&= \ident{n} \te H^T H^T + 2 H \te H^T + H H \te \ident{n} \\
\Tr{\acom{H}{\acom{H}{\cdot}}}&= 2 n \tr{ H^2} + 2 (\tr{H})^2 \\
\acom{H}{\acom{H}{\acom{H}{\acom{H}{\cdot}}}}&=\left( \ident{n} \te H^T H^T + 2 H \te H^T + H H \te \ident{n} \right)^2\\
		&= \ident{n} \te (H^T)^4 +4 H \te (H^T)^3 + 6 H^2 \te (H^T)^2\\
		&+ 4 H^3 \te H^T + H^4 \te \ident{n}\\
\Tr{\acom{H}{\acom{H}{\acom{H}{\acom{H}{\cdot}}}}}&= 2 n \tr{H^4} +8 \tr{H} \tr{H^3} + 6 (\tr{H^2})^2
\end{align}
And for commutators
\begin{align}
\com{L}{\cdot}&=L \te \ident{n} -\ident{n} \te L^T  \\
\com{L}{\com{L}{\cdot}}&= \ident{n} \te L^T L^T - 2 L \te L^T + L L \te \ident{n} \\
\Tr{\com{L}{\com{L}{\cdot}}}&= 2 n \tr{ L^2} - 2 (\tr{L})^2 \\
\com{L}{\com{L}{\com{L}{\com{L}{\cdot}}}}&=\left( \ident{n} \te L^T L^T - 2 L \te L^T + L L \te \ident{n} \right)^2\\
		&= \ident{n} \te (L^T)^4 -4 L \te (L^T)^3 + 6 L^2 \te (L^T)^2\\
		&- 4 L^3 \te L^T + L^4 \te \ident{n}\\
\Tr{\com{L}{\com{L}{\com{L}{\com{L}{\cdot}}}}}&= 2 n \tr{L^4} -8 \tr{L} \tr{L^3} + 6 (\tr{L^2})^2
\end{align}


\end{appendix}

\bibliographystyle{plain}

\bibliography{references}

\end{document}